\documentclass{aa}  

\usepackage{graphicx}
\usepackage{txfonts}
\usepackage{hyperref}
\usepackage{longtable}
\begin{document} 

\title{Cepheid Metallicity in the Leavitt Law (C--MetaLL) survey:}
\subtitle{IX. Metallicity dependence of period-Wesenheit relations based on a homogeneous spectroscopic sample. \thanks{Based on the European Southern Observatory programs P105.20MX; P106.2129; 108.227Z; 109.231T; 110.23WM and on the Telescopio Nazionale Galileo programmes A39TAC\_9; A40TAC\_11; A41TAC\_29; A42TAC\_15; A43TAC\_16; A44TAC\_27; A45TAC\_12; A46TAC\_15. Based on observations obtained at the Canada-France-Hawaii Telescope (CFHT), which is operated by the National Research Council of Canada, the Institut National des Sciences de l´Univers of the Centre National de la Recherche Scientique of France, and the University of Hawaii.}}
\titlerunning{C-MetaLL Survey IX}

   \author{V. Ripepi\inst{1}\thanks{E-mail: vincenzo.ripepi@inaf.it}
       \and
          E. Trentin\inst{1}
          \and
          G. Catanzaro\inst{2}
          \and 
          M. Marconi\inst{1} 
          \and
          A. Bhardwaj\inst{3}
          \and
         G. Clementini \inst{4}
          \and
          F. Cusano \inst{4}
          \and
          G. De Somma\inst{1,5}
          \and
          R. Molinaro\inst{1}
          \and
         T. Sicignano\inst{1,5,6,7}
          \and
         J. Storm\inst{8}
        }

\institute{
INAF-Osservatorio Astronomico di Capodimonte, Salita Moiariello 16, 80131, Naples, Italy
\and
INAF-Osservatorio Astrofisico di Catania, Via S.Sofia 78, 95123, Catania, Italy 
\and
Inter-University Center for Astronomy and Astrophysics (IUCAA), Post Bag 4, Ganeshkhind, Pune 411 007, India
\and
INAF-Osservatorio di Astrofisica e Scienza dello Spazio, Via Gobetti 93/3, I-40129 Bologna, Italy 
\and
Istituto Nazionale di Fisica Nucleare (INFN)-Sez. di Napoli, Via Cinthia, 80126 Napoli, Italy
\and
European Southern Observatory, Karl-Schwarzschild-Strasse 2, 85748 Garching bei München, Germany 
   \and 
Scuola Superiore Meridionale, Largo San Marcellino10 I-80138 Napoli, Italy
\and
Leibniz-Institut f\"{u}r Astrophysik Potsdam (AIP), An der Sternwarte 16, D-14482 Potsdam, Germany
}

   \date{Received September 15, 1996; accepted March 16, 1997}

 
  \abstract
{
The C-MetaLL project has provided homogeneous spectroscopic abundances of 290 Classical Cepheids (DCEPs) for which we have the intensity-averaged magnitudes in multiple optical and NIR bands, periods, pulsation modes, and $Gaia$ parallaxes corrected for individual zero-point (ZP) biases.
}
{Our goal is to derive updated period--Wesenheit--metallicity (PWZ) relations using the largest and most homogeneous metallicity sample ever used for such analyses, covering a range of $-1.3<$[Fe/H]$<+0.3$ dex, and to assess the metallicity dependence of these relations.}
{We computed several optical and NIR Wesenheit magnitudes adopting both Cardelli et al. and Fitzpatrick reddening laws, and transformed Johnson-Cousins Wesenheit magnitudes into their HST equivalents using empirical relations. Using 275 DCEPs with reliable parallaxes, we applied a robust photometric parallax technique, which simultaneously fits all parameters -- including the global ZP counter-correction to \textit{Gaia} parallaxes -- and handles outliers via a Cauchy likelihood to account for the sample's excess variance.}
{We find a stronger metallicity dependence ($\gamma \approx -0.5$ mag/dex in optical, $-0.4$ mag/dex in NIR) than recent literature reports.
\textit{Gaia} parallax ZP counter-correction ($\epsilon$) varies moderately across bands, with an average value of $\sim$10 $\mu$as, aligning with previous determinations. Applying our PWZ relations to $\sim$4500 LMC Cepheids yields distances generally consistent within $1\sigma$ with geometric estimates. The choice of reddening law has a small impact, while using only fundamental-mode pulsators significantly increases the uncertainties. Including $\alpha$ element corrections increases $|\gamma|$ and reduces $\epsilon$. However, we find 1$\sigma$ consistency $\gamma$ values with the literature, particularly for the Wesenheit magnitude in the HST bands, by restricting the sample to brighter (i.e. closer) objects, or by including only pulsators with $-0.7<$[Fe/H]$<$0.2 dex. Our results hint at a large $\gamma$ or a non-linear dependence on metallicity of DCEP luminosities at the metal-poor end, which is difficult to quantify with the precision of parallaxes of the present dataset.
}
{}


   \maketitle
%

\section{Introduction}

Classical Cepheids (DCEPs) are crucial standard candles in the extragalactic distance scale due to the Leavitt law \citep{Leavitt1912}, which defines a correlation between their period and luminosity -- a PL relation. Once calibrated using independent distances derived from geometric methods such as trigonometric parallaxes, eclipsing binaries, and water masers, these relations form the first step in constructing the cosmic distance scale. The first rung provides the basis for calibrating secondary distance indicators, such as Type Ia supernovae (SNIa), which enable us to measure distances to distant galaxies within the steady Hubble flow. The calibration of this three-step process (commonly referred to as the cosmic distance ladder) allows us to reach the Hubble flow, where the Hubble constant ($H_0$) -- the value linking the distance to the recession velocity of galaxies -- can be estimated \citep[e.g.][and references therein]{Sandage2006,Freedman2012,Riess2016}.

The value of $H_0$ is a key parameter in cosmology as it determines both the scale and age of the Universe. Consequently, achieving a 1\% precision in its measurement is one of the primary goals of modern astrophysics. However, there is an ongoing and well-known discrepancy between the $H_0$ values obtained by the SH0ES\footnote{Supernovae, H0, for the Equation of State of Dark energy} project using the cosmic distance ladder \citep[$H_0=$73.01$\pm$0.99 km s$^{-1}$ Mpc$^{-1}$,][]{Riess2022a} and those measured by the Planck mission from the cosmic microwave background under the flat $\Lambda$ cold dark matter ($\Lambda$CDM) model \citep[$H_0=$67.4$\pm$0.5 km s$^{-1}$ Mpc$^{-1}$,][]{Planck2020}. The tension seems to be confirmed by new James Webb Space Telescope observations by the SH0ES team \citep{Riess2024}, albeit \citet{Freedman2024} results, again based on JWST data but using a different sample of SNIa, appear to favour a reduction in the size of the discrepancy between the early and late estimates of $H_0$.
On the other hand, if confirmed, the 5$\sigma$ discrepancy would signal the need to revise the $\Lambda$CDM model.

One of the remaining sources of uncertainty in the cosmic distance ladder is the debated metallicity dependence of the DCEP $PL$ relations used to calibrate secondary distance indicators. A variation in metallicity is expected to influence the shape and width of the DCEP instability strip \citep[e.g.][]{Caputo2000}, which in turn alters the coefficients of the $PL$ relations \citep[][and references therein]{Marconi2005,Marconi2010,DeSomma2022}. 
Although it has been demonstrated that the metallicity dependence of the PL relations alone cannot solve the Hubble tension \citep[e.g.][]{Riess2022a}, its precise value could nevertheless be crucial to establish the actual size of the discrepancy in $\sigma$s between the cosmic ladder and the CMB+$\Lambda$CDM measurements of $H_0$. 
Furthermore, accurately knowing the metallicity dependence of PL relations allows us to constrain pulsation models and validate their physical assumptions.

Direct empirical tests of metallicity effects on the PL and PW relations, based on Galactic DCEPs with reliable [Fe/H] abundances from high-resolution (HiRes) spectroscopy, have long been limited by the lack of precise independent distances for enough Milky Way (MW) DCEPs. The $Gaia$ mission \citep[][]{Gaia2016} has radically changed this. Starting with data release 2 (DR2) \citep[][]{Gaia2018} and further improved in early data release 3 (EDR3) \citep[][]{Gaia2021}, $Gaia$ has provided accurate parallaxes. It has also led to the discovery of hundreds of new Galactic DCEPs \citep[][]{Clementini2019,Ripepi2019,Ripepi2023}, alongside other surveys such as OGLE \citep[][]{Udalski2018} and ZTF \citep[][]{Chen2020}, creating a large, valuable DCEP sample for both extragalactic distance scale and Galactic studies \citep[e.g.][]{Lemasle2022,Trentin2023,Trentin2024b}.

Yet until recently, DCEPs with HiRes-based metallicities were mostly confined to the solar neighbourhood, covering a narrow [Fe/H] range, around the solar value, with small scatter (0.2–0.3 dex) \cite[e.g.][]{Genovali2014,Luck2018,Groenewegen2018,Ripepi2019}. The limited range makes it difficult to assess metallicity effects on Galactic Cepheid PL relations with strong statistical significance.

To tackle this issue, a few years ago we launched the C-MetaLL project\footnote{https://sites.google.com/inaf.it/c-metall/home}  \citep[Cepheid — Metallicity in the Leavitt Law; see][for details]{Ripepi2021a}. Its goal is to measure chemical abundances for approximately 400 Galactic DCEPs using HiRes spectroscopy, and specifically extending the iron abundance range into the metal-poor regime, particularly [Fe/H]$<-0.4$ dex. In the project’s papers I,\,II,\,V, and VI \citep[][]{Ripepi2021a,Trentin2023,Bhardwaj2024,Trentin2024b}, we provided accurate abundances of over 25 chemical species for 290 DCEPs spanning a wide range in metallicity (+0.3$<$[Fe/H]$<-1.1$ dex; see Sect.~\ref{sect:abundances}) and with an approximately uniform number of DCEPs at all [Fe/H] values.

In \citet{Trentin2024b} we exploited this sample to study the chemical composition of the Galactic disc and spiral arms. In this paper, we investigate the metallicity dependence of the DCEPs' PW relations using our updated and homogeneous sample alone, i.e. without using literature abundances as in our previous papers, due to insufficient proprietary data. 
In our previous works \citep[][]{Ripepi2020,Ripepi2021a,Ripepi2022a,Bhardwaj2024,Trentin2024a}, we have found a generally large metallicity dependence of the DCEPs PL and PW relations, of the order of $\gamma \sim -0.3/-0.5$ mag/dex, where $\gamma$ is the usual notation for the metallicity term of the intercept in the PL and PW relations (see Sect.~\ref{sect:analysis} for details). These findings are in contrast to those of the SH0ES team, who identified a smaller dependence \citep[$\sim -0.2$ mag/dex][]{Riess2021} based on a local sample of 75 DCEPs with a narrow [Fe/H] range. Moreover, our results provide values of $\gamma$ that are larger (in the absolute sense) than those reported in the past by several authors \citep[e.g.][]{Scowcroft2009,Gieren2018,Cruz2023} and, in particular, by \citet{Breuval2021,Breuval2022,Breuval2024}, who, similarly to what was reported in SH0ES works, found a mild metallicity dependence from the analysis of DCEP samples in the MW and the Large and Small Magellanic Clouds (LMC and SMC). These systems show different metallicities, decreasing from about solar to [Fe/H]$\sim$-0.75 dex.
\citet{Bhardwaj2023} showed that the two approaches of using spectroscopic metallicities of individual Galactic Cepheids with $Gaia$ parallaxes and using the MW and Magellanic Clouds as representative samples provide different quantifications of the metallicity dependence of PL relations. 
Even more discrepant are the results from \citet{Madore2025a}, who performed five empirical tests across a broad metallicity range and various wavelength regimes (optical to mid-infrared), concluding that no statistically significant metallicity dependence could be detected in the available PL relations. In a subsequent paper, \citep{Madore2025b} reinforced their conclusions using theoretical static stellar atmosphere models to evaluate the differential impact of metallicity on Cepheid spectral energy distributions. However, a recent and detailed re-examination of the methods used by \citet{Madore2025a} has suggested potential uncertainties regarding the robustness of their conclusions \citep[][]{Breuval2025}. 
Finally, in a recent paper, \citet{Wang2025}, adopting basically the same input data and a very similar approach to that of our previous paper C-MetaLL-IV \citep[][]{Trentin2024b}, found very similar results to ours in the same filters.

Intending to shed some light on this intricate matter, in this paper we analyse the metallicity dependence of the PW relations for Galactic DCEPs. 
We do not address the PL relations because of the significant uncertainties on the interstellar extinction for a sizeable fraction of our targets, which are placed far away in the disc and can easily reach values of $E(B-V)$ larger than one magnitude. The use of Wesenheit magnitudes substantially mitigates this problem, especially in the NIR bands, where the coefficients multiplying the colour are small. 

The paper is structured as follows. In Sect.~\ref{sect:dataset} we present the sample used in our analysis, in Sect.~\ref{sect:analysis} we describe the analysis procedure, and in Sect.~\ref{sect:results} we describe our results. We discuss them and give our conclusion in Sect.~\ref{sect:conclusions}.

  \begin{figure}
   \includegraphics[width=8cm]{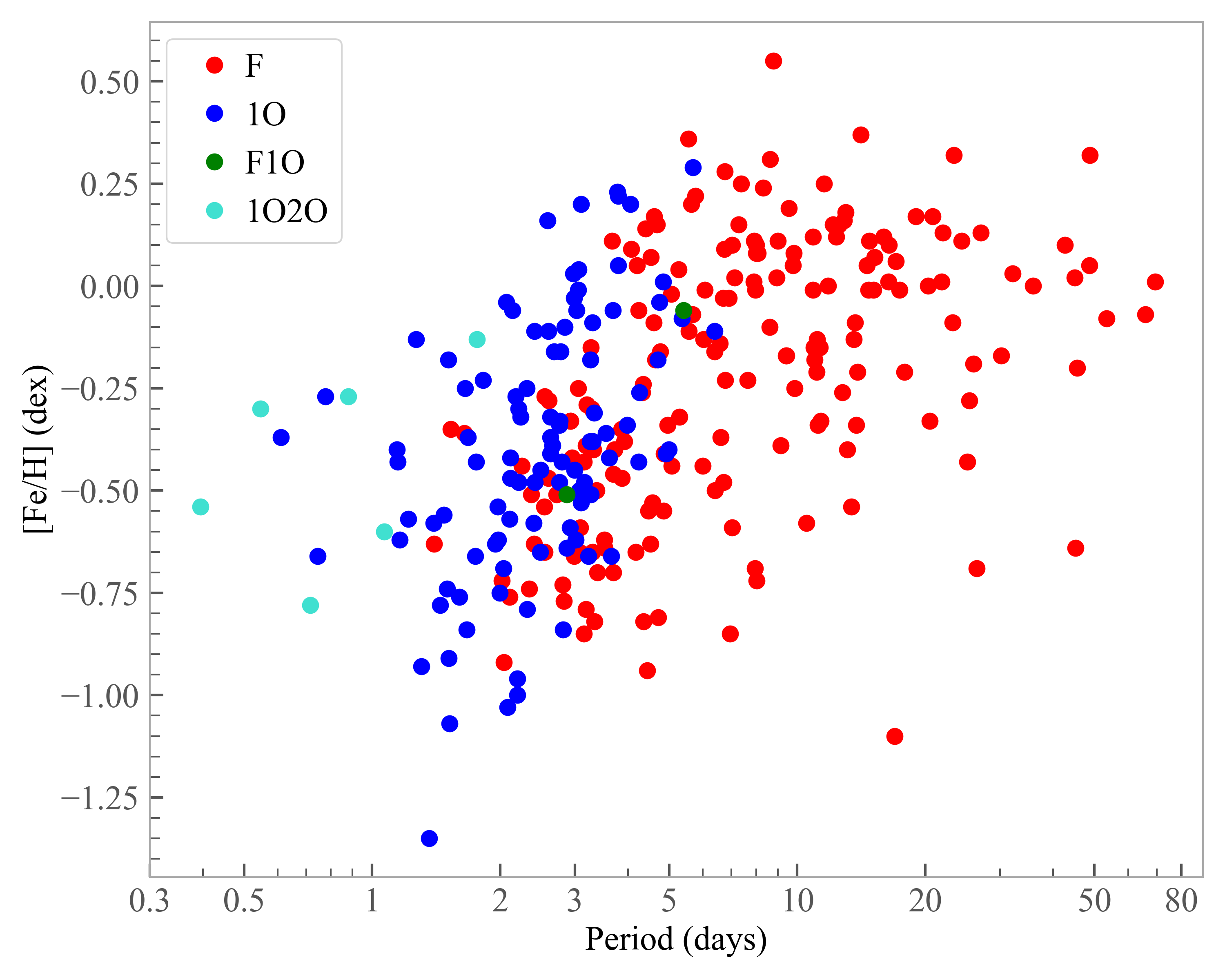}
   \caption{Periods and iron abundances spanned by the investigated sample. The different pulsation modes are labelled in the figure.}
              \label{fig:period_metallicity}
   \end{figure}

   \begin{figure}
   \includegraphics[width=9cm]{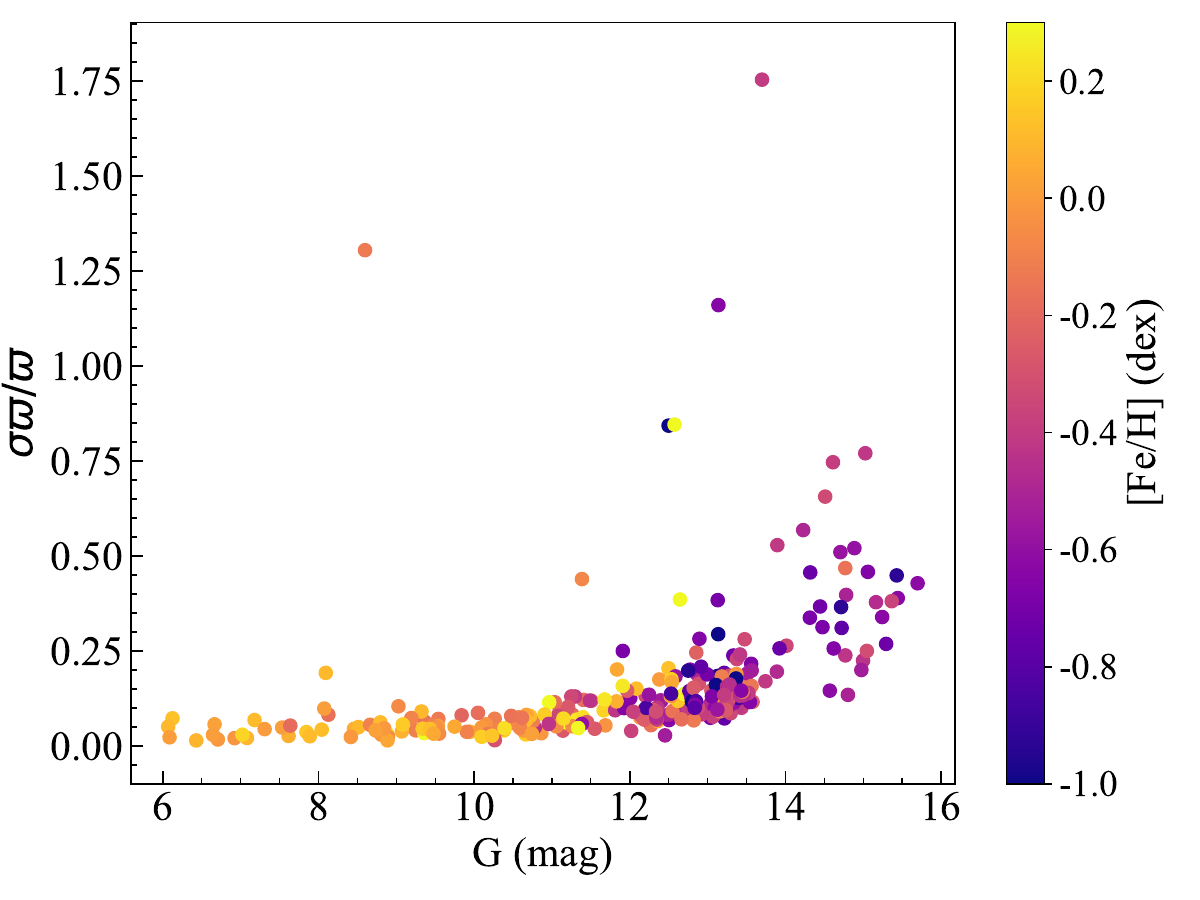}
   \caption{Relative parallax error as a function of the $Gaia$  G magnitude. The points are colour-coded according to their iron abundance.}
              \label{fig:parallaxError}
   \end{figure}

   \begin{figure}
   \includegraphics[width=9cm]{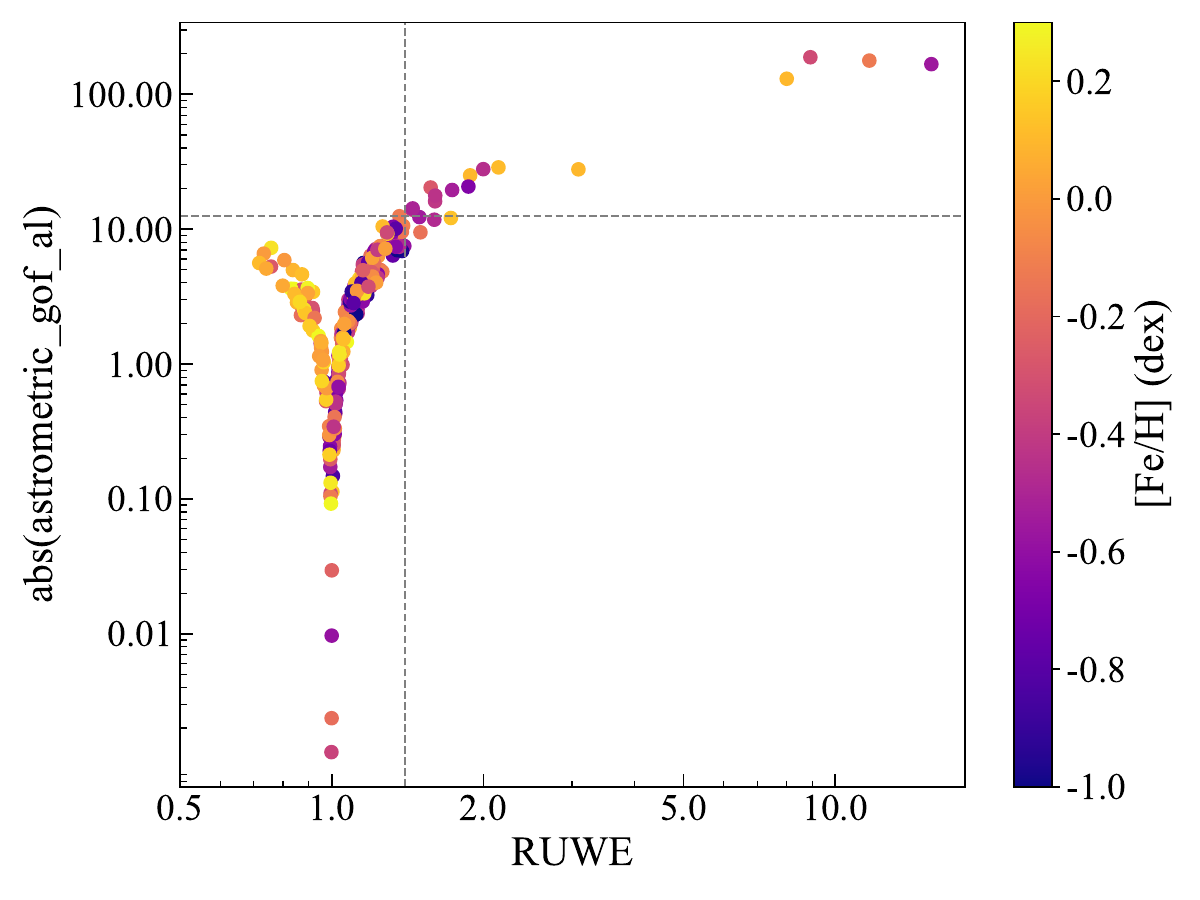}
   \caption{Reliability of parallaxes based on the two parameters {\tt RUWE} and {\tt astrometric\_gof\_al} (in absolute value). For reference, the typical limits at 1.4 and 12.5 have been reported with dashed lines.}
              \label{fig:astrometry}
   \end{figure}

\section{Dataset}\label{sect:dataset}

In this section, we describe in detail the properties of the DCEP sample used in this work. To calculate the desired period--Wesenheit--metallicity (PWZ) relations, we require additional ingredients besides the abundances, such as the multi-band photometry and parallaxes. The collection of this information is described in the following sections. All data used in this work are provided in Table~\ref{tab:data}.

\subsection{Periods of pulsation and photometry}
\label{photometry}

As in our previous works, the relevant data for the DCEP sample used in this work have mainly been taken from the list published in the $Gaia$ Data Release 3 \citep[DR3, see][]{Ripepi2023}. From the same source, we adopted periods, epochs of maximum light, amplitudes in the $G$ band, and the intensity-averaged $G, G_{BP}$, and $G_{RP}$ magnitudes, derived according to a specific treatment of Cepheid pulsating stars \citep[see][]{Clementini2016,Clementini2019,Ripepi2019,Ripepi2023}. For the 15 stars without specific treatment in $Gaia$  DR3, we adopted periods from \citet{Piet2021} and average magnitudes from the general $Gaia$  DR3 catalogue\footnote{See the discussion in \citet{Ripepi2022a} regarding the use of these magnitudes compared to the intensity-averaged ones.} \citep[see][]{GaiaVallenari}. 

The $Gaia$ photometry was used to calculate homogeneous $V,\,I$ magnitudes for 248 stars using the precise transformations by \citet{Pancino2022}, which were corrected for a small shift as is explained in \citet{Trentin2024a}. For the remaining 42 (bright) stars, we adopted the photometry by \citet{Bhardwaj2015}, which, as has been demonstrated by \citet{Bhardwaj2024}, is consistent with the $V,\,I$ magnitudes obtained with the \citet{Pancino2022} transformations corrected as mentioned above.

Photometry in the NIR bands $J,H,K_s$ was collected from the literature using intensity-averaged magnitudes from full light curves when possible. In particular, we used the data by i) \citet{Bhardwaj2015} and \citet{Groenewegen2018} for 36 and 6 bright DCEPs, respectively; ii) \citet{Bhardwaj2024} data collected in the context of the C-MetaLL survey for an additional 61 stars (paper V of the series). For the remaining 187 stars, we had to rely on the single-epoch photometry from the 2MASS survey \citep[][]{Skrutskie2006}, which was used together with template light curves to estimate average magnitudes in the NIR bands. To this aim, we proceeded exactly as is described in \citet{Ripepi2021a} and \citet{Trentin2024a}. We recall that the periods and epochs of maximum light have been taken from $Gaia$ DR3 when available and from the ASAS-SN survey \citep[All-Sky Automated Survey for Supernovae][]{Shappee2014,Christy2023} for seven stars for which epochs were missing in $Gaia$ DR3. 

To calculate the Wesenheit magnitudes in the HST bands, we proceeded as in \citet{Trentin2024a}, i.e. we adopted the \citet{Riess2021} transformations between Johnson-Cousins $H,\,V\,,I$ and the $F160W,\,F555W,\,F814W$ bands. 
In the following, to avoid confusion, we call $cF160W,\,cF555W,\,cF814W$ the HST-like magnitudes obtained starting from ground-based data, where the ‘$c$’ stands for ‘calculated’. The same prefix will be applied to the Wesenheit magnitudes obtained with these data (see Table~\ref{tab:wesenheit}). 
To verify the reliability of the calculated HST magnitudes, we carried out a test which is described in detail in Appendix~\ref{sect:appendix_comparison_HST_mag}. As a result, we find that apart from some scattering, the calculated HST magnitudes (and Wesenheit functions) do not present any relevant systematic difference with the native HST photometry. 

The adopted photometry is listed in Table~\ref{tab:data}, where different labels point to the sources of the distinct values. 
We note that, although the spectroscopy used in this paper is fully homogeneous, as is shown in this section, the photometry is not yet. The analysis of a fully homogeneous sub-sample of C-MetaLL data can be found in \citet{Bhardwaj2024}.

\subsection{Fundamentalisation of 1O pulsators}
\label{sect:fundamentalisation}

In a recent work, \citet{Pilecki2024a} discussed in detail the fundamentalisation of the periods of first-overtone (1O) pulsators. They provided new empirical relationships for different environments such as the MW, LMC, and SMC. However, they also specified that for the aim of using the fundamentalisation for the PL and PW relations, it is preferable to use the relations they had published in their previous work \citep[][]{Pilecki2024b}, in which they had provided new relations for LMC and SMC. Therefore, in this work, we decided to use for all the 1O (and also for first over second overtone modes - 1O2O) DCEPs the following relation from \citet{Pilecki2024b}:  
\begin{equation}
    P_F=P_{1O}(1.418+0.114 \log P_{1O})
,\end{equation}
which is valid for the LMC, i.e. for [Fe/H]$\sim-$0.4 dex \citep[with $\sigma=0.07$ dex][]{Romaniello2022}. This is justified by the fact that: i) this relation has been accurately derived using the best data available (the LMC DCEPs are almost all at the same distance, contrarily to those in the MW and SMC); and ii) the average metallicity of the LMC is close to the mean metallicity of our sample of 1O pulsators, which is exactly $<$[Fe/H]$>_{1O}\sim-$0.4 dex, with a dispersion of $\sim$0.3 dex.

\subsection{Abundances}\label{sect:abundances}

The dataset utilised in this work comprises DCEPs whose abundances have been published in our previous papers. In more detail, we adopted data for one star from both \citet{Catanzaro2020} and \citet{Ripepi2021a}, and 47, 65, 42, and 134 stars from \citet{Ripepi2021b}, \citet{Trentin2023}, \citet{Bhardwaj2023}, and \citet{Trentin2024b}, respectively\footnote{Even if \citet{Bhardwaj2023} is not a paper of the C-MetaLL series, the abundance analysis for the 42 stars observed with ESPADONS@CFHT has been carried out with the same methodology and codes adopted in the C-MetaLL project for all the other objects in that survey. Note also that only [Fe/H] have been published in \citet{Bhardwaj2023}, while the abundances of the different chemical species are presented in \citet{Trentin2024b}.}. This data comprises  290 DCEPs, for which we have provided abundances for the iron and more than 20 chemical species in a homogeneous way.  
The sample's period and iron abundance ranges are shown in Fig.~\ref{fig:period_metallicity}. To obtain a good fit of the PLZ and PWZ relations, the entire space of these two parameters should be filled evenly. Although a larger number of long-period metal-poor DCEPs would be desirable (we are collecting additional observations for this aim in the context of the C-MetaLL project), the sample appears already to be rather equilibrated. In particular, a comparison with the sample adopted in \citet{Trentin2024b}, displayed in Fig.~\ref{fig:comparisonPeriodMetallicity}, reveals a substantial improvement in this respect.

\subsection{Extinction law and definition of the Wesenheit magnitudes}

In this work, for homogeneity with our previous papers, we adopted the standard \citet{Cardelli1989} extinction law with $R_V=3.1$ as the baseline. In addition, we adopted the \citet{Fitzpatrick1999} results, when the difference with the previous one was significant (e.g. in the $V\,,I$ and $J,K_S$ colours). Some specific Wesenheit magnitudes were taken directly from the literature. All the quantities used in this work are listed in  Table~\ref{tab:wesenheit}.

\subsection{Parallaxes}
\label{sect:parallaxes}
All the 290 objects in our sample have a valid parallax value from $Gaia$  DR3. In more detail, three objects have negative parallaxes, while 141 and 237 objects have a relative error on the parallax of $\sigma \varpi / \varpi <$0.1 and 0.2, respectively. 
Figure~\ref{fig:parallaxError} shows the variation in $\sigma \varpi / \varpi$ with the $Gaia$ G magnitude. As expected, fainter objects have less precise parallaxes. The figure also shows that more metal-rich objects generally have more precise parallaxes, because, on average, they are closer to the Sun. The farthest objects, which are also the most metal-poor, have in general less precise parallaxes, which is why in the C-MetaLL project we aim to further enlarge the sample of metal-poor objects, i.e. to compensate for the lower precision of individual sources with statistics. A comparison with the sample adopted in \citet{Trentin2024b} (see Fig.~\ref{fig:comparisonParallaxError}) confirms the above considerations and testifies to the improvement of the present work sample in this context.

Concerning the goodness of the parallaxes, 19 objects have a $Gaia$  {\tt RUWE} parameter \citep[Renormalised Unit Weight Error][]{Lindegren2021} beyond the generally accepted good threshold of 1.4. On the other hand, we have 15 objects with {\tt astrometric\_gof\_al} below the threshold of 12.5 adopted by \citet{Riess2021} for their DCEP sample. The distribution of these parameters for our sample is shown in Fig.~\ref{fig:astrometry}\footnote{We recall that the two parameters are not independent as {\tt astrometric\_gof\_al} includes the {\tt RUWE} in its calculation}. The slight gap on the y axis between 12 and 14 seems to justify a preliminary cut at {\tt astrometric\_gof\_al}=12.5, and thus keeping four objects with {\tt RUWE} slightly exceeding the value 1.4. In any case, our fitting procedure allows for outlier rejection; hence, the occurrence of a few DCEPs with possibly incorrect parallaxes will not impact the results. 
The figure also shows that there is no dependence on metallicity among the rejected objects. The list of the 15 rejected objects is shown in Table~\ref{tab:excludedStars} in Appendix~\ref{sect:misc}.

\begin{table}
    \centering
    \caption{Photometric bands and colour coefficients adopted to calculate the Wesenheit magnitudes considered in this work.}
    \footnotesize\setlength{\tabcolsep}{3pt}
    \begin{tabular}{llc}
    \hline
    \hline
    Acronym & Wesenheit definition & Source\\
    \hline
    WG &  $G-1.90(G_{BP}-G_{RP})$ & R19 \\
    WVI(C89) & $I-1.55(V-I)$ & C89 (OGLE) \\
    WVI(F99) & $I-1.39(V-I)$ & F99 \\
    WcVI & $cF814W-1.19(cF555W-cF814W)$ & B24\\
    WcHVI & $cF160W-0.386(cF555W-cF814W)$ & R16\\
    WVK$_S$ & $K_S-1.13(V-K_S)$  & C89\\
    WJK$_S$(C89) & $K_S-0.69(J-K_S)$ & C89\\
    WJK$_S$(F99) & $K_S-0.75(J-K_S)$ & F99\\
    \hline
    \end{tabular}
    \tablefoot{Sources for the coefficient used to define the Wesenheit magnitudes: R19=\citet{Ripepi2019}; OGLE=\citep{Udalski1999}; F99=\citet{Fitzpatrick1999}; B24=\citet{Breuval2024}; R16=\citet{Riess2016}; and C89=\citet{Cardelli1989}.}
    \label{tab:wesenheit}
\end{table}

\begin{figure}
\vbox{
\includegraphics[width=9cm]{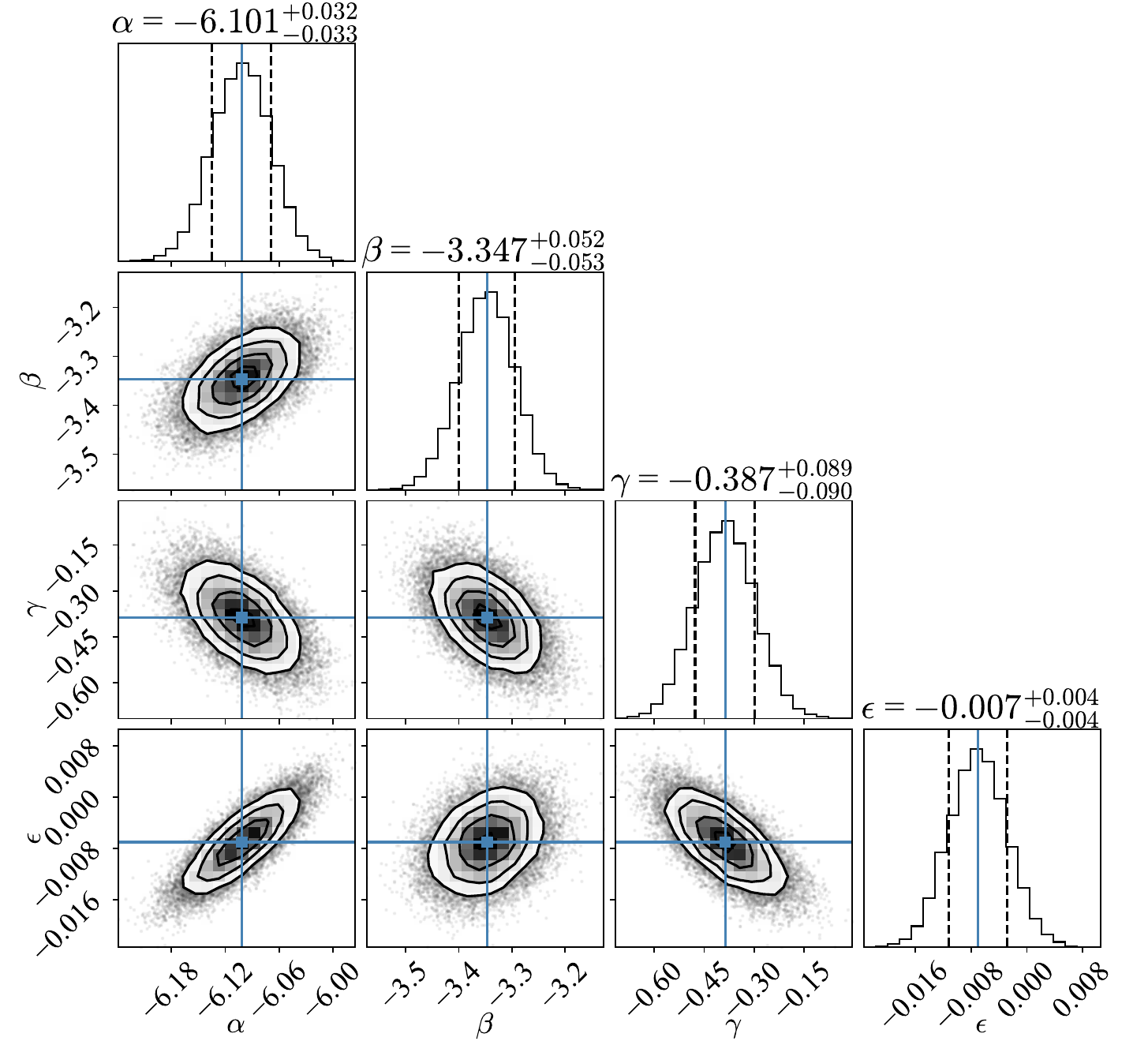}
\includegraphics[width=9cm]{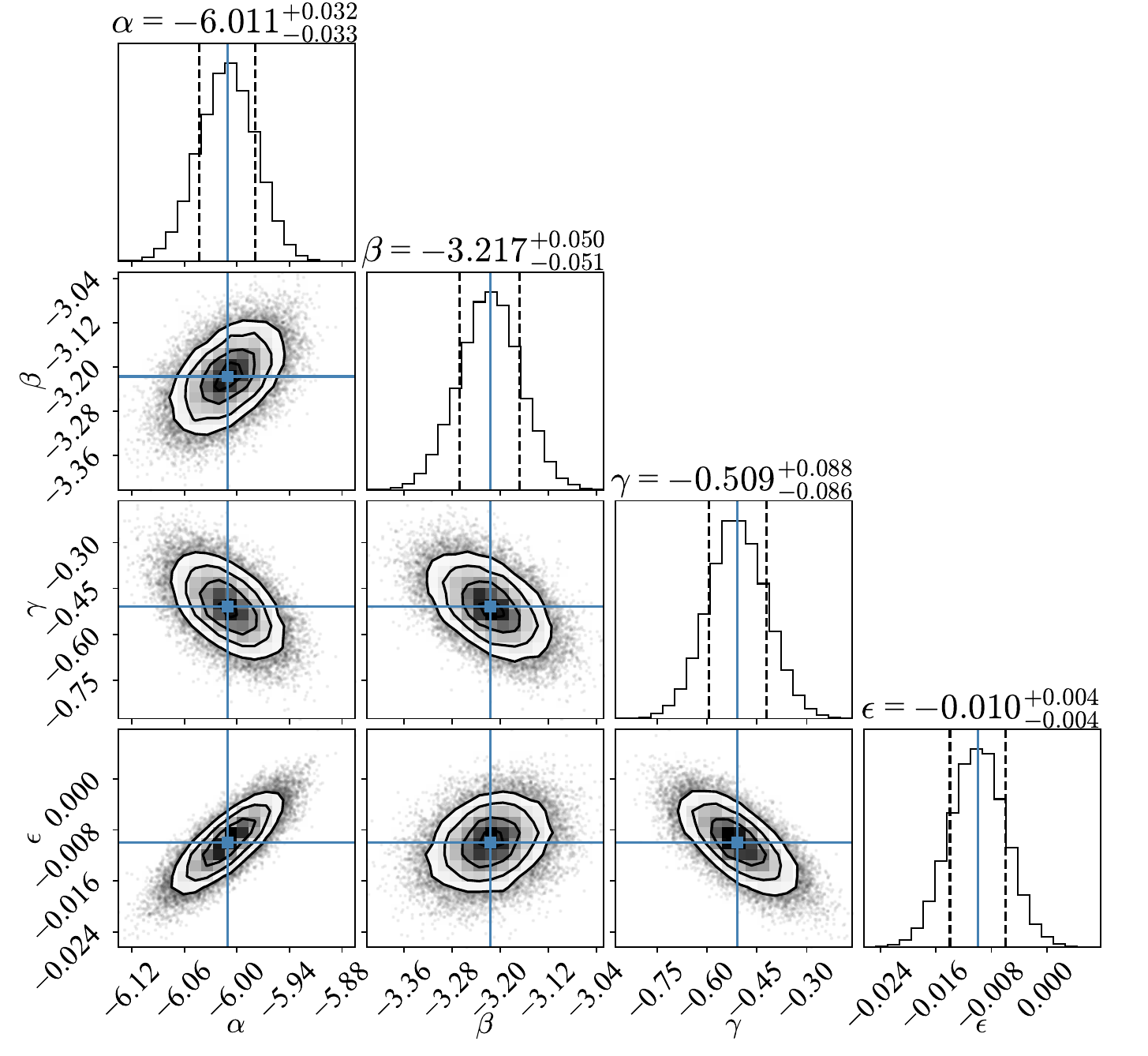}
}
\caption{Examples of corner plot with the posterior distributions for the output parameters and the best-fit solution obtained for the $WJK_S$ (top) and $WcHVI$ (bottom) Wesenheit magnitudes. The units of $\alpha,\, \beta,\, \gamma,\,\epsilon$ and of their uncertainties are mag, mag/dex, mag/dex, and mas, respectively.}
\label{fig:cornerPlotJK}
\end{figure}

\begin{figure}
\vbox{
\includegraphics[width=9cm]{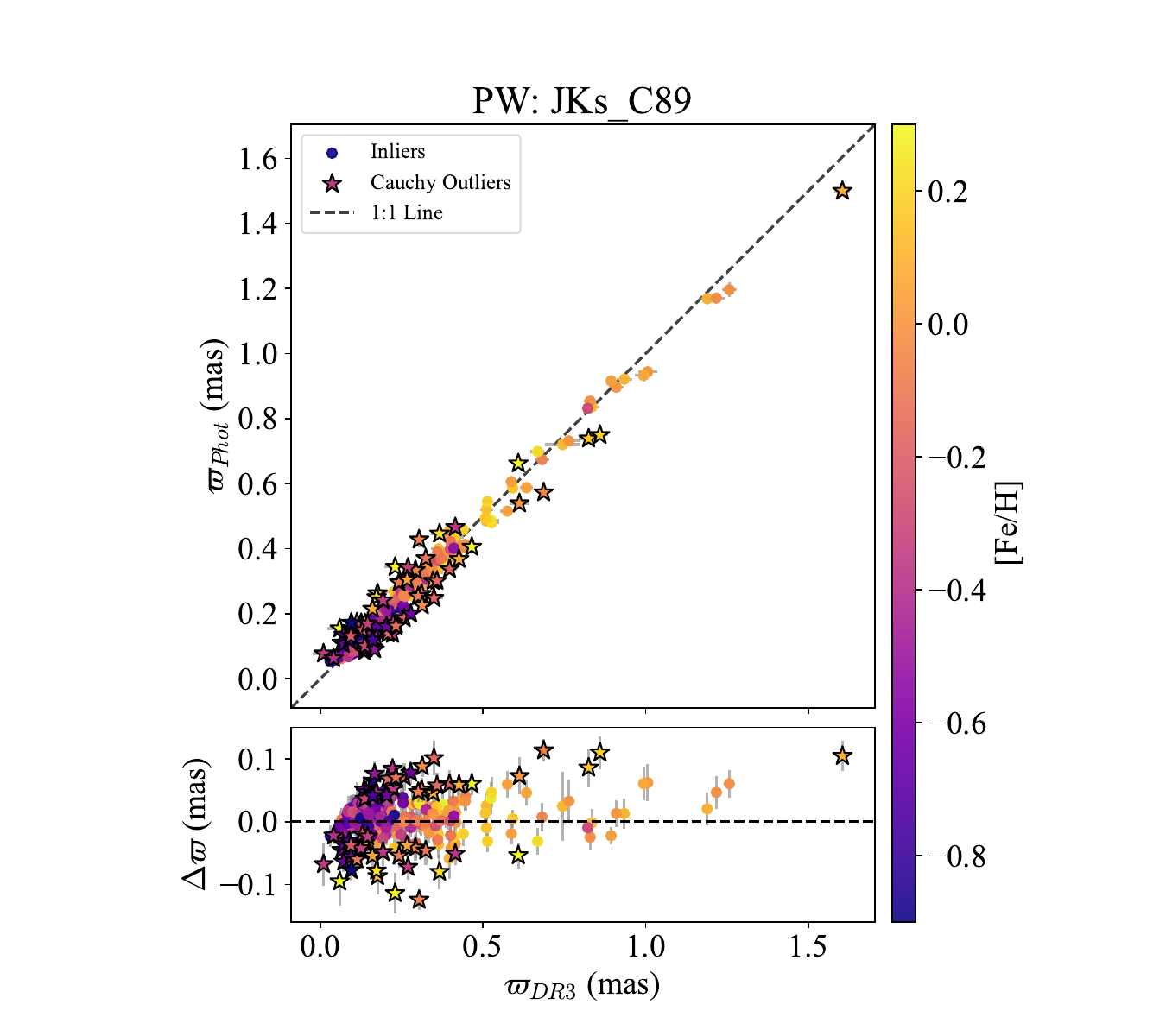}
\includegraphics[width=9cm]{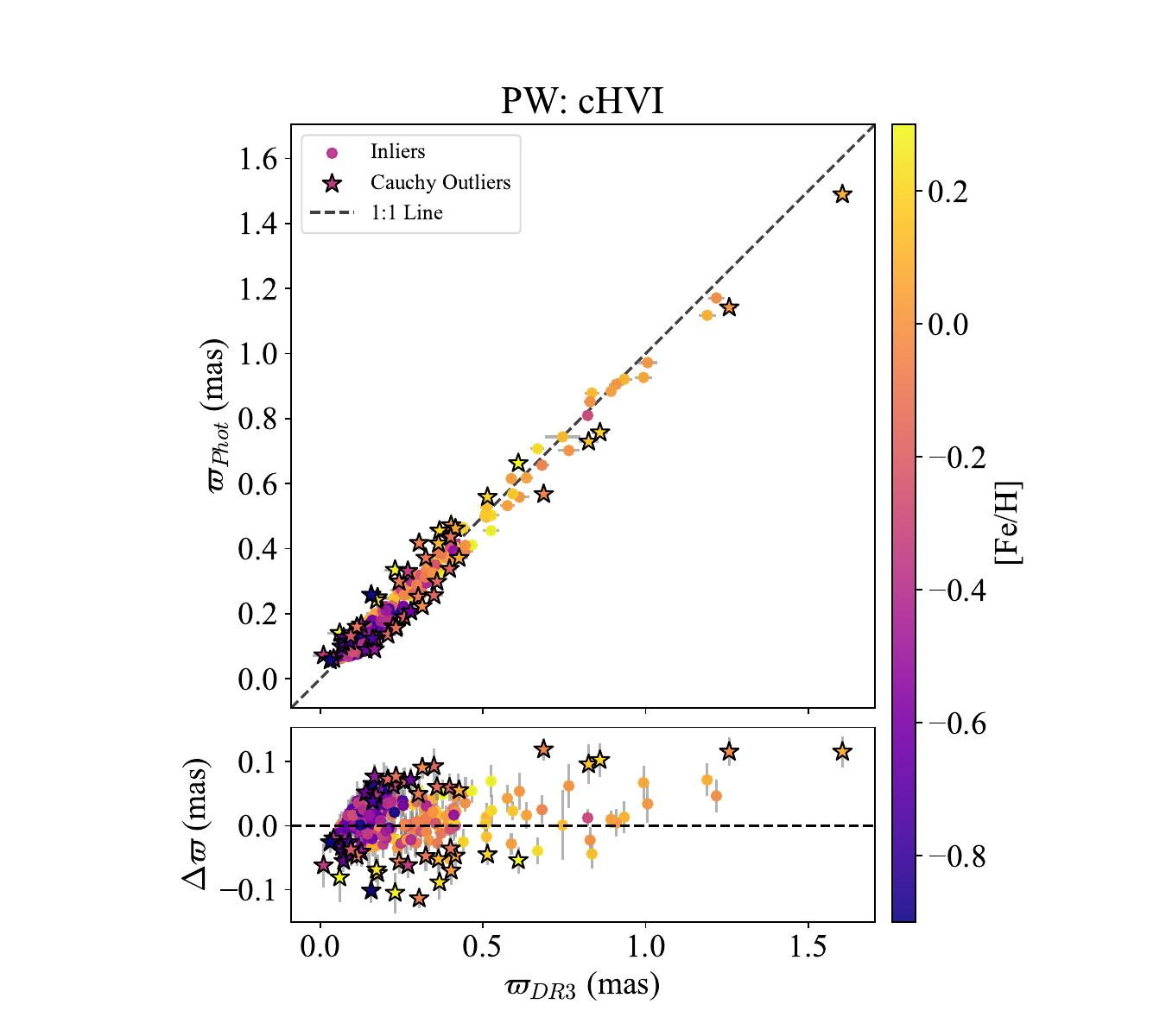}
}
\caption{Comparison between final photometric parallax (i.e. the one calculated with the final result of the fitting procedure) and that from $Gaia$. In filled circles and filled stars are the objects not considered or considered as outliers by the Cauchy loss function. All the data are colour-coded according to their [Fe/H] ratio. The top and bottom panels show the results for the $WJK_S$ and $WcHVI$ Wesenheit magnitudes, respectively. The units of $\alpha,\, \beta,\, \gamma,\,\epsilon$ and $\varpi$ are mag, mag/dex, mag/dex, mas, and mas, respectively.}
\label{fig:comparisonParallaxesJK}
\end{figure}

\section{Analysis}
\label{sect:analysis}
 
The fitting method used in this paper is the same as in \citet{Ripepi2022a}, which, in turn, is based on the approach by \citet{Riess2021}. We first defined the photometric parallax (in mas) as 

\begin{equation}
\varpi_{phot}=10^{-0.2(w-W-10)} 
,\end{equation}

\noindent where $w$ is a generic apparent Wesenheit magnitude (see Table~\ref{tab:riessComparison}), while $W$ is the absolute Wesenheit magnitude, which can be written as

\begin{equation}
W=\alpha+\beta(\log_{10} P-1.0)+\gamma {\rm {\rm [Fe/H]}}
\label{eq:PWZ}
.\end{equation}

Note that, contrary to our previous work, given the reduced number of objects compared with \citet{Trentin2024a}, we did not try to study the metallicity dependence of the slope, i.e. of the $\log P$ coefficient. This is also justified because we calculated the zero-point (ZP) counter-correction directly from the data, and thus already added a fourth parameter to the fit. In any case, the analysis of the slope dependence on metallicity is postponed to a future paper, which will include the complete C-MetaLL sample, consisting of more than 400 DCEPs. 

Indicating with $\varpi_{DR3}$ the DR3 parallax already corrected for the ZP bias (see Sect.~\ref{sect:parallaxes}), we wish to minimise the following quantity:

\begin{equation}
\chi^2=\sum \frac{(\varpi_{DR3}-\varpi_{phot}+\epsilon)^2}{\sigma^2}    
\label{eq:chi}
.\end{equation}
 
Here $\sigma$ is the total error obtained by summing up in quadrature the uncertainty on $\varpi_{DR3}$ and $\varpi_{phot}$: $\sigma=\sqrt{\sigma_{\varpi_{DR3}}^2+\sigma_{\varpi_{phot}}^2}$, while $\epsilon$ is the parallax ZP counter-correction. In more detail, the astrometric $\sigma_{\varpi_{DR3}}$ consists of two contributions: i) the standard error of the parallax as reported in the  DR3 catalogue, which we conservatively increased by 10\%; and ii) the uncertainty on the individual ZP corrections, assumed to be 5 $\mu$as \citep[see e.g.][]{Lindegren2021,Riess2021}. 
The uncertainty on $\varpi_{phot}$ is more complex to evaluate. Using the equivalence $\delta \varpi / \varpi = \delta D / D$, where $D$ denotes the distance, and recalling the definition of the distance modulus $\mu = -5 + 5 \log_{10} D$, error propagation and straightforward algebra yield
\begin{equation}
    \sigma_{\varpi_{phot}}=0.46 \times \sigma_\mu \times \varpi_{phot}, \label{eq:error}
\end{equation}
where $\sigma_\mu=\sqrt{\sigma_{w}^2+\sigma_{W}^2}$. 
The term $\sigma_w$ is readily computed by propagating the uncertainties on the single magnitudes in the Wesenheit definitions of Table~\ref {tab:wesenheit}. In contrast, estimating $\sigma_W$ is more challenging, as it requires prior knowledge of the intrinsic scatter of the adopted relation. To this end, we followed the approach by \citet{Riess2021}, adopting a scatter of 0.06 mag in the NIR bands, while in the optical bands we assumed a more conservative value of 0.1 mag. As is discussed in \citet{Ripepi2022a}, this is justified on theoretical grounds \citep[e.g.][and references therein]{Desomma2020}. 
Note also that Eq.~\ref{eq:error} requires the knowledge of $\varpi_{phot}$, and, in turn, of $w$ and $W$. While $w$ is a measured quantity, $W$ includes the unknowns $\alpha,\,\beta,\,\gamma$. Therefore, the minimisation procedure detailed below proceeded iteratively: 

\begin{table*}[h]
 \caption[]{\label{table:PWresults} Results of the photometric parallax fit to the PWZ relation in different bands.}
 \footnotesize\setlength{\tabcolsep}{3pt}
\begin{tabular}{lrrrrrrrrrrrrrrrrr}
\hline
  \multicolumn{1}{c}{Wesenheit} &
  \multicolumn{1}{c}{$\alpha$} &
  \multicolumn{1}{c}{$\sigma_\alpha^{low}$} &
  \multicolumn{1}{c}{$\sigma_\alpha^{high}$} &
  \multicolumn{1}{c}{$\beta$} &
  \multicolumn{1}{c}{$\sigma_\beta^{low}$} &
  \multicolumn{1}{c}{$\sigma_\beta^{high}$} &
  \multicolumn{1}{c}{$\gamma$} &
  \multicolumn{1}{c}{$\sigma_\gamma^{low}$} &
  \multicolumn{1}{c}{$\sigma_\gamma^{high}$} &
  \multicolumn{1}{c}{$\epsilon$} &
  \multicolumn{1}{c}{$\sigma_\epsilon^{low}$} &
  \multicolumn{1}{c}{$\sigma_\epsilon^{high}$} &
  \multicolumn{1}{c}{N} &
  \multicolumn{1}{c}{$\mu_{\rm LMC}$} &
  \multicolumn{1}{c}{$\sigma\mu_{\rm LMC}$} &
    \multicolumn{1}{c}{AIC} &
  \multicolumn{1}{c}{BIC} \\
\hline
  Gaia & -6.047 & 0.034 & 0.035 & -3.309 & 0.053 & 0.054 & -0.527 & 0.095 & 0.093 & -0.016 & 0.004 & 0.004 & 270 & 18.524 & 0.06 & 351 & 365\\
  VI\_C89 & -6.075 & 0.035 & 0.035 & -3.264 & 0.057 & 0.058 & -0.632 & 0.099 & 0.097 & -0.011 & 0.004 & 0.004 & 267 & 18.418 & 0.064 & 298 & 312\\
  VI\_F99 & -5.858 & 0.033 & 0.034 & -3.173 & 0.057 & 0.057 & -0.436 & 0.095 & 0.095 & -0.018 & 0.004 & 0.004 & 274 & 18.442 & 0.06 & 350 & 365\\
  cVI & -5.869 & 0.033 & 0.033 & -3.173 & 0.054 & 0.054 & -0.317 & 0.092 & 0.09 & -0.015 & 0.004 & 0.004 & 272 & 18.386 & 0.057 & 319 & 333\\
  cHVI & -6.011 & 0.032 & 0.033 & -3.214 & 0.05 & 0.051 & -0.509 & 0.088 & 0.086 & -0.01 & 0.004 & 0.004 & 270 & 18.417 & 0.055 & 327 & 341\\
  VKs & -6.073 & 0.032 & 0.033 & -3.289 & 0.05 & 0.051 & -0.47 & 0.088 & 0.084 & -0.008 & 0.004 & 0.004 & 271 & 18.457 & 0.056 & 350 & 365\\
  JKs\_C89 & -6.101 & 0.032 & 0.033 & -3.347 & 0.052 & 0.053 & -0.387 & 0.089 & 0.09 & -0.007 & 0.004 & 0.004 & 271 & 18.439 & 0.057 & 331 & 346\\
  JKs\_F99 & -6.145 & 0.033 & 0.034 & -3.366 & 0.052 & 0.054 & -0.397 & 0.091 & 0.089 & -0.006 & 0.004 & 0.004 & 271 & 18.446 & 0.058 & 336 & 350\\
  \hline\end{tabular}
\tablefoot{C89 and F99 refer to the \citet{Cardelli1989} and \citet{Fitzpatrick1999} extinction laws, respectively. The values of $\sigma^{low}$ and $\sigma^{high}$ correspond to the 16th and 84th percentiles of the posterior probability, respectively (see text). The LMC distance modulus and its error were calculated in the manner described in the text. The units of $\alpha,\, \beta,\, \gamma,\,\epsilon$ and of their uncertainties are mag, mag/dex, mag/dex, and mas, respectively.}
\end{table*}

\begin{itemize}
\item To obtain a reasonable starting point for the parameter estimation (priors), we first minimised the standard $\chi^2$ function described in Eq.~\ref{eq:chi}. The resulting values of $\alpha,\,\beta,\,\gamma$, and $\epsilon$ were used as inputs in the subsequent steps. During this first step, we checked the sample for large outliers, which could indicate undetected problems with the astrometry or photometry. To do this, we applied a $sigma$-clipping procedure, identifying stars whose weighted difference between the calculated and observed parallaxes exceeded 3$\sigma$. This method flagged eight problematic stars (listed in Table~\ref{tab:outliers}): four of them were deviant in almost all PW relations, likely due to incorrect parallaxes, and the other four were outliers only in the optical relations, likely due to poor photometry in those bands. These stars were subsequently removed from the derivation of any PW relation in which they were identified as outliers (see Table~\ref{table:PWresults}).

\item To further investigate the statistical properties of our dataset, we examined the residuals of the standard $\chi^2$ fit described in the previous point. Taking as an example the WHVI magnitude, we find that the residuals exhibit significant excess variance throughout the sample, with a reduced chi-squared of ${\chi}^2/\text{dof}\footnote{degree of freedom} \approx 2.6$ and $AIC \approx 700$. This indicates that the scatter in the data is approximately 1.6 times larger than the formal uncertainties.\footnote{Note that the excess variance is absent in the \citet{Riess2021} sub-sample ($\chi^2/dof\sim1.04$, see the end of this section and Appendix:\ref{sect:testPhotParallax}).} Furthermore, the distribution of residuals is non-Gaussian, displaying tails with significantly more high-significance deviations than expected. 
To address this over-dispersion from biasing our parameter estimation, we adopted a robust loss function based on the Cauchy distribution\footnote{We tested different loss functions \citep[see][for a review]{barron2019general}, particularly the Huber and Cauchy ones. In the end, we chose the Cauchy loss, which provided more reliable results.}. The Cauchy loss is defined as

\begin{equation}
    \mathcal{L}_{\text{Cauchy}} = \frac{c^2}{2} \sum_i \log \left(1 + \left( \frac{r_i}{c} \right)^2 \right),
\end{equation}
where $r_i=(\varpi_{DR3}-\varpi_{phot}+\epsilon)^2/\sigma^2$ is the normalised residual, and \( c \) is a scale parameter that controls the influence of outliers. The parameter \( c \) was estimated from the residuals of the initial chi-square fit using the median absolute deviation (MAD), scaled to be consistent with a normal distribution: $c = 1.4826 \times \text{MAD}$.

This formulation provides a smooth, non-quadratic penalty for large residuals, limiting the influence of potential outliers without requiring explicit clipping.

\item To estimate the posterior distributions of the parameters $\alpha,\,\beta,\,\gamma$, and $\epsilon$, we adopted a Bayesian approach and sampled from the posterior using the affine-invariant Markov Chain Monte Carlo (MCMC) sampler \texttt{emcee} \citep{Foreman-Mackey2013}. The log-likelihood was defined as

\begin{equation}
    \log \mathcal{L}(\theta) = -\frac{1}{2} \mathcal{L}_{\text{Cauchy}}(\theta),
\end{equation}
where $\theta = (\alpha,\,\beta,\,\gamma$,\,$\epsilon$) is the vector of model parameters. We initialised a set of 50 walkers around the initial solution and evolved them for 5000 steps, discarding the burn-in phase and thinning the chain to obtain a clean sample of the posterior. Tests using a larger number of walkers (up to 1000) showed no significant improvement, so we adopted 50 to optimise computational efficiency.
Our approach enables the estimation of parameter uncertainties and degeneracies, as well as robust inference in the presence of non-Gaussian errors.
\end{itemize}

Our best estimation of each parameter is represented by the median of the posterior distribution, sampled via the MCMC. We set the 16th and 84th percentiles as uncertainties. 

To test the soundness of the procedure outlined above, we tried to reproduce the results by \citet{Riess2021}. Indeed, we used the same method as in that work, the only difference being a different minimisation technique. As is shown in detail in 
 Appendix~\ref{sect:testPhotParallax}, our procedure gives the same numbers as in the literature.

\section{Results}
\label{sect:results}

Based on this positive test, we obtained the PWZ relations for the selected Wesenheit magnitudes defined in Table~\ref{tab:wesenheit}. Two examples of the results for the WJK$_S$ and WcHVI are shown in Fig.~\ref{fig:cornerPlotJK}, where we show the posterior distribution of the four  $(\alpha,\,\beta,\,\gamma$,\,$\epsilon$) parameters. The comparison between the $Gaia$ DR3 and the photometric parallaxes for the same selected Wesenheit magnitudes is shown in Fig.~\ref{fig:comparisonParallaxesJK}. Here, the objects considered outliers by the Cauchy-loss functions are highlighted in red. We recall that the adopted procedure assigns less weight to the outlying points but does not remove them. The figure shows the robustness of our fits, and in particular that the distribution of residuals shows no significant, unmodeled dependence on metallicity, further supporting the consistency of the baseline fit.

\begin{figure}
\includegraphics[width=9.5cm]{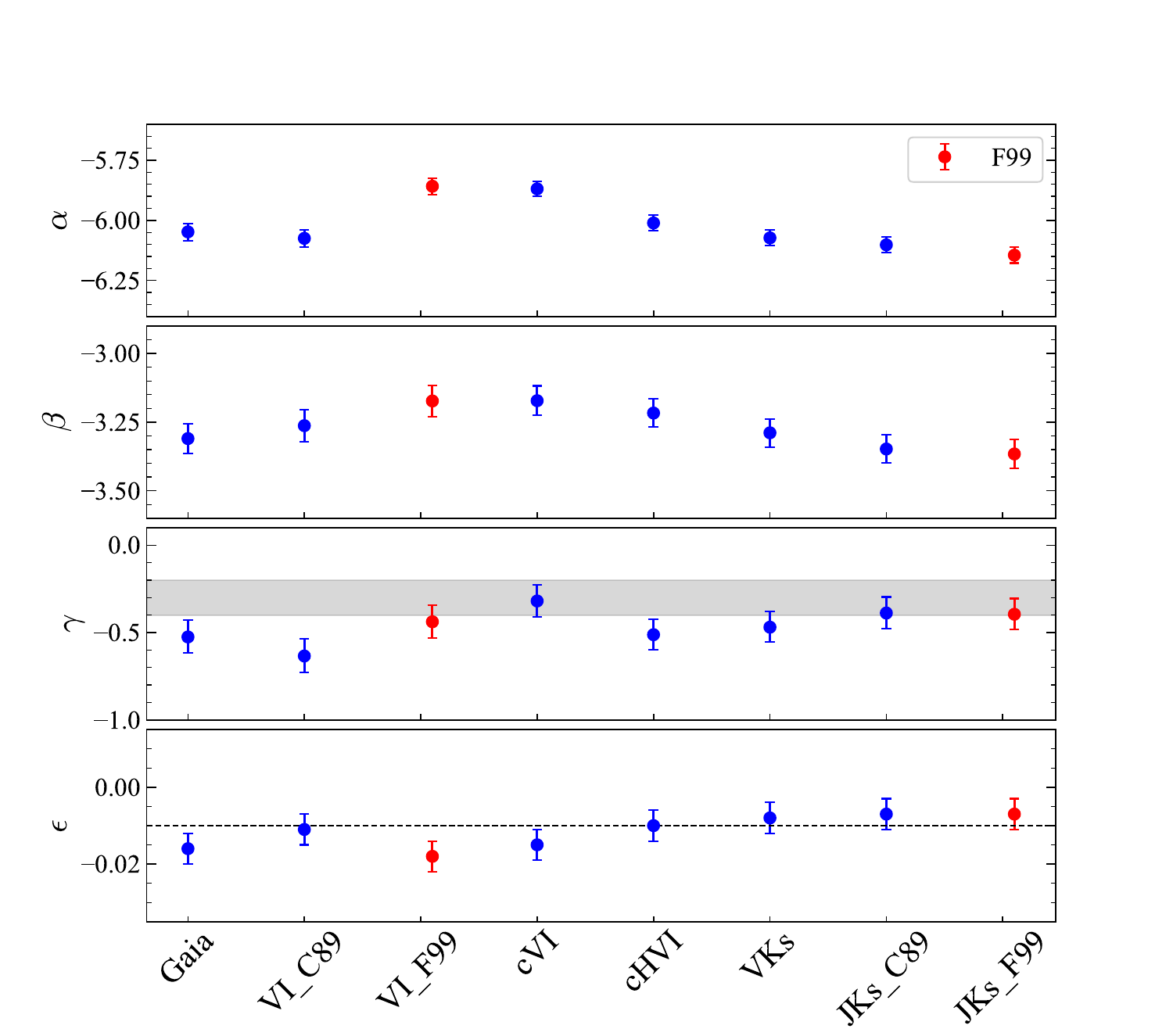}
\caption{Values of the four parameters, $\alpha,\,\beta,\,\gamma$, and $\epsilon$, for the different Wesenheit magnitudes considered in this work. Red dots show the results obtained by adopting the \citet{Fitzpatrick1999} reddening law (F99 in the label). The shaded area comprises the range $-0.4<\gamma<-0.2$ mag/dex, where most of the recent literature results are located.   
The units of $\alpha,\, \beta,\, \gamma,\,\epsilon$ and of their uncertainties are mag, mag/dex, mag/dex, and mas, respectively.}
\label{fig:results_PW_baseline}
\end{figure}

Applying the procedure to all the Wesenheit magnitudes leads to the derivation of the parameters, which are shown in Fig.~\ref{fig:results_PW_baseline} and listed in Table~\ref{table:PWresults}. The figure displays the results for Wesenheit magnitudes with increasing wavelength from left (optical) to right (NIR). We note that:
\begin{itemize}
    \item 
$\alpha$ and $\beta$ have similar behaviour. As expected, their values tend to increase (in absolute value) towards the NIR. However, the trend is reversed when considering the $Gaia$ and $VI$ (OGLE) Wesenheits. In particular, there is a significant difference between the $\alpha$ and $\beta$ values for the $VI$ Wesenheits defined by OGLE and those calculated from the \citet{Fitzpatrick1999} extinction law. This does not occur for the $JK_S$ owing to the smaller difference between the coefficients that multiply the colour in this case.

\item 
$\gamma$ does not vary significantly from the optical to the NIR, even though it approaches the literature values in WJK (see shaded region in Fig.~\ref{fig:results_PW_baseline}), but remains more negative in WVK and WcHVI.

\item
The $Gaia$ parallaxes ZP counter-correction tends to vary systematically from $\sim$16 $\mu$as to a few $\mu$as from the optical to the NIR, having the lowest value for the WJK$_S$. The average value of about 10 $\mu$as is marked in the figure for reference. For instance, this is the value obtained for WcHVI.  

\item 
The values of $\gamma$ and $\epsilon$ are correlated. This means that, for example, for the WcHVI used in the distance scale, passing from $\epsilon=10 ~\mu as$ as found from our fitting procedure to $\epsilon=14 ~\mu as$ \citep[][]{Riess2021} would increase $\gamma$ to $\sim-0.35$ mag/dex. Naturally, since other correlations are present, this would also imply $\alpha \sim -6.04$ mag, increasing the discrepancy of the ZP compared with \citet[][see also Appendix~\ref{sect:appendix_comparison_HST_mag}]{Riess2021}

\end{itemize}
  
As in our previous papers, we validated the distances derived from these PWZ relations by comparing them with the 1\% precise geometric distance of the LMC derived by \citet{Pietrzynski2019}. To this aim, we adopted a sample of about 4500 DCEPs in the LMC that have optical photometry from $Gaia$ and OGLE IV surveys as well as $J,K_S$ photometry from the Vista Magellanic Cloud survey \citet[VMC,][]{Cioni2011,Ripepi2022b} and $H$ data from \citet{Inno2016}. 
Similarly to the Galactic sample, we used these $H,\,V,\,I$ data to calculate the corresponding HST Wesenheit magnitudes. 
Then, we applied the derived PWZ relations to this LMC sample, adopting [Fe/H]=$-0.41$ dex \citep[][]{Romaniello2022}, and thus derived a distance modulus from each star. 

\begin{figure}
\includegraphics[width=9cm]{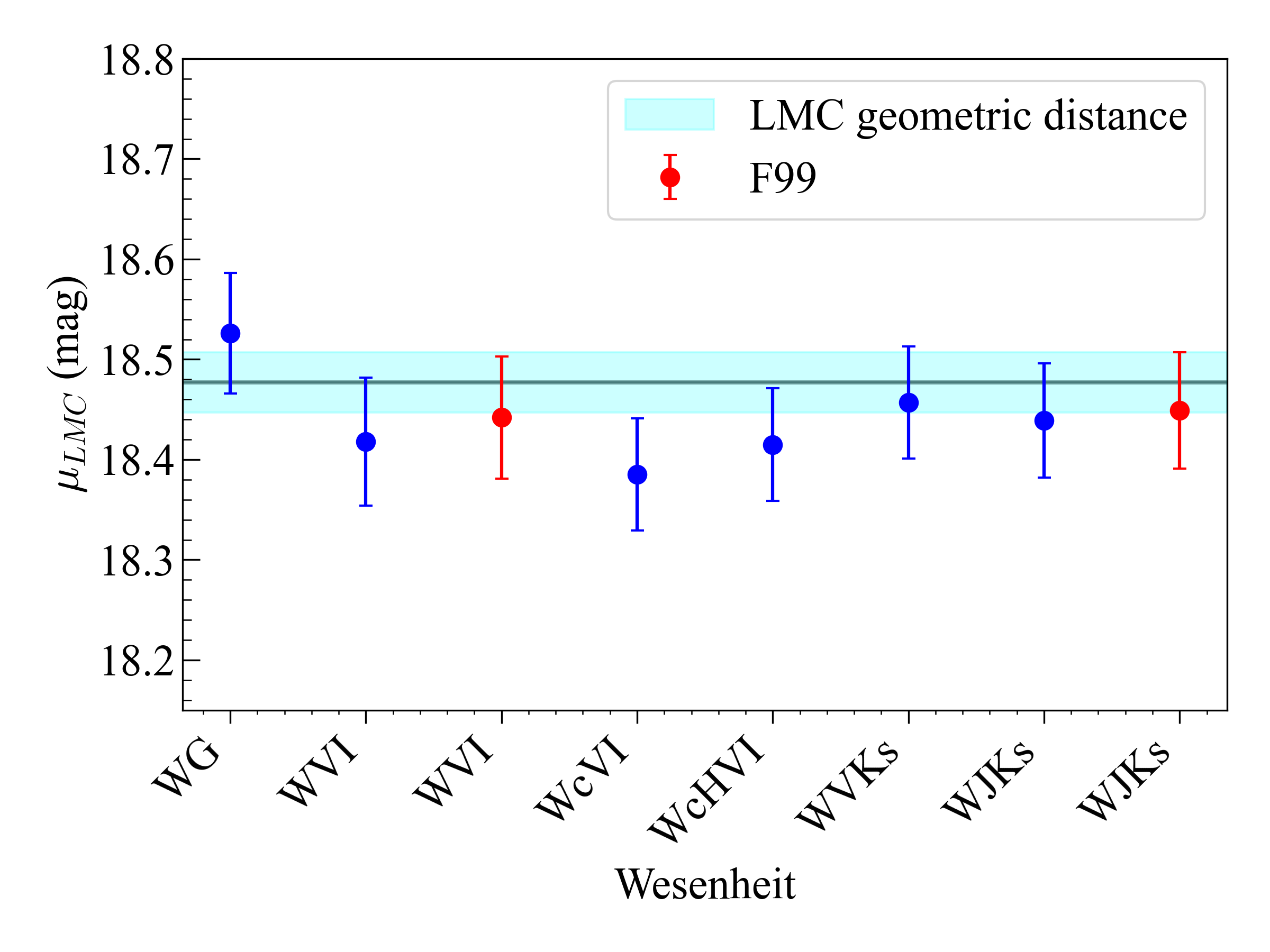}
\caption{LMC distance moduli calculated from our best-fitting PWZ relations. Red dots show results relative to the Wesenheit definition according to the 
\citet{Fitzpatrick1999} reddening law. The shaded region corresponds to the geometric distance of the LMC by \citet{Pietrzynski2019} $\pm 1 \sigma$.}
\label{fig:lmcDistance}
\end{figure}

The median of the resulting distance moduli distribution provides us with an estimate of the LMC distance. The calculation of its error is discussed in Appendix~\ref{sect:lmcDistances}. The derived LMC distances for each PWZ relation are displayed in Fig.~\ref{fig:lmcDistance} and listed in Table~\ref{table:PWresults}. The resulting $\mu_{LMC}$ values are typically about 2--5\% smaller than the geometric distance by \citet{Pietrzynski2019} except for WG, which provides a $\sim$4\% farther distance. In any case, given the typical 5.5\% uncertainty, almost all the Wesenhit magnitudes analysed in this work provide LMC distances well within $1\sigma$ compared to the geometric distance of the LMC. We note that the best results are provided by the NIR Wesenheit magnitudes, especially the WVKs one. 
Overall, this validation procedure suggests that the results presented in this section are reliable. We consider the PWZ in Table~\ref{table:PWresults} to be our baseline results. 

\section{Variations from the baseline}

To explore possible biases in our results, which could lead to the large metallicity dependence of the PWZ relations in the previous section, we re-ran the calculations by applying possible variations described in the following sections. In the next sections, we use the Akaike information criterion (AIC) and the Bayesian information criterion (BIC) to compare different model calculations  (see Sect:~\ref{sect:aic} for a detailed definition of these quantities) that involve samples of approximately the same size.

\subsection{Use of pure $\sigma$-clipping method}

To be consistent with the techniques adopted in previous works, in this variation of the baseline, instead of using the Cauchy-loss technique, we applied a pure $\sigma$-clipping method. The results of this procedure for the case with 3$\sigma$ and 5 iterations are listed in the top part of Table~\ref{tab:parameterComparison} and shown in Fig.~\ref{fig:parameterComparisonPlot}. We experimented by lowering the sigma-clipping to 2.5 and 2 $\sigma$, and/or increasing the number of iterations, without significant changes in the coefficient values, but with increasingly smaller errors, as expected. 

In general, the main difference from the baseline is a modest variation (both increasing or decreasing) in the $\gamma$ terms at different wavelengths, accompanied by a general reduction in errors. The distance of LMC is essentially the same as in the baseline, albeit with smaller errors. On the contrary, the AIC and BIC values are significantly larger. However, these goodness of fit values are not directly comparable with the baseline, because in this variant we minimised the $\chi^2$, while in the baseline we minimised the Cauchy-loss function.  Overall, this experiment shows that the introduction of the Cauchy-loss algorithm does not introduce significant differences with more standard procedures.

\subsection{Use of F-mode pulsators only}

\begin{figure}
\includegraphics[width=9.0cm]{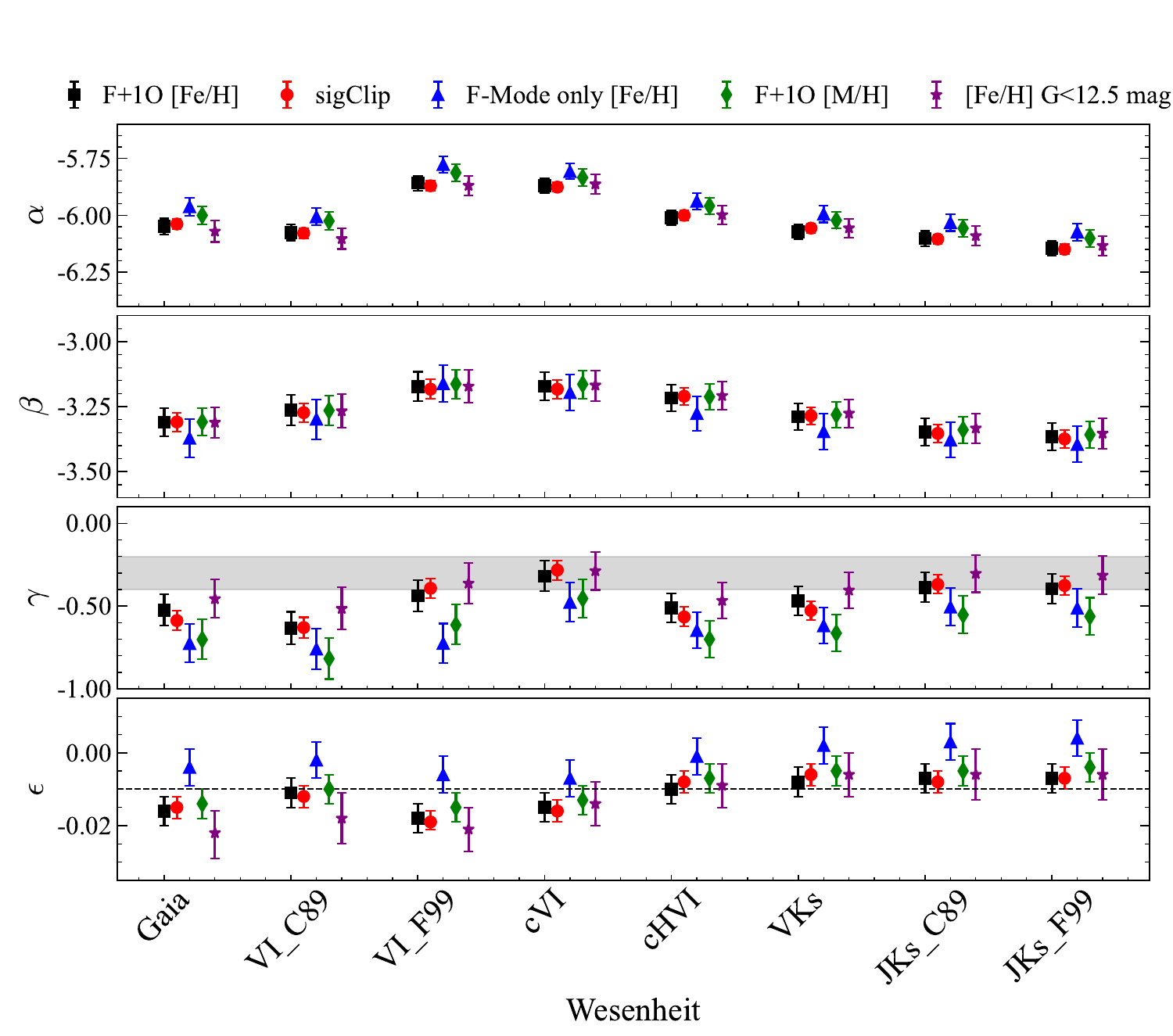}
\caption{Comparison between the PWZ parameters obtained with our baseline sample (F+1O pulsators, using [Fe/H], black squares) and those obtained:: i) using a pure $\sigma$-clipping algorithm (red circles) ii) using only F-mode pulsators with [Fe/H] as abundance indicator (blue triangles); iii) using F+1O pulsators with the total metallicity [M/H] as abundance indicator (green diamonds); and iv) using only objects with $G<12.5$ mag (violet stars). 
The shaded area comprises the range $-0.4<\gamma<-0.2$ mag/dex, where most of the recent literature results are located.   
The units of $\alpha,\, \beta,\, \gamma,\,\epsilon$ and of their uncertainties are mag, mag/dex, mag/dex, and mas, respectively.}
\label{fig:parameterComparisonPlot}
\end{figure}

To investigate the impact of using 1O-mode pulsators with fundamentalised periods together with F-mode DCEPs, we re-ran the PWZ calculations using only the latter. Of the 183 F-mode DCEPs in our sample, only 173 have usable parallaxes. The results of the calculations are shown in Fig.~\ref{fig:parameterComparisonPlot} and in the middle-top part of Table~\ref{tab:parameterComparison}. The adoption of the reduced sample essentially produces: i) errors larger by a factor of 1.5 or more on all the parameters; ii) $\gamma$ values much more negative (larger in absolute value) compared with the baseline calculation; iii) parallax ZP counter-correction much larger (even positive) than the baseline PWZs; and iv) low values for the LMC distance modulus. 
We interpret these findings as resulting from the reduced sample size. Indeed, since the overall relative error on the parallaxes for most of the C-MetaLL sample stars is of the order of 10-15\%, with an increase towards the most metal-poor objects (see Fig.~\ref{fig:parallaxError}), we need to use a larger sample to compensate for the reduced precision of the parallaxes with the statistics. In other words, currently, using the F-mode pulsators alone is not feasible.

\begin{figure}
\includegraphics[width=8cm]{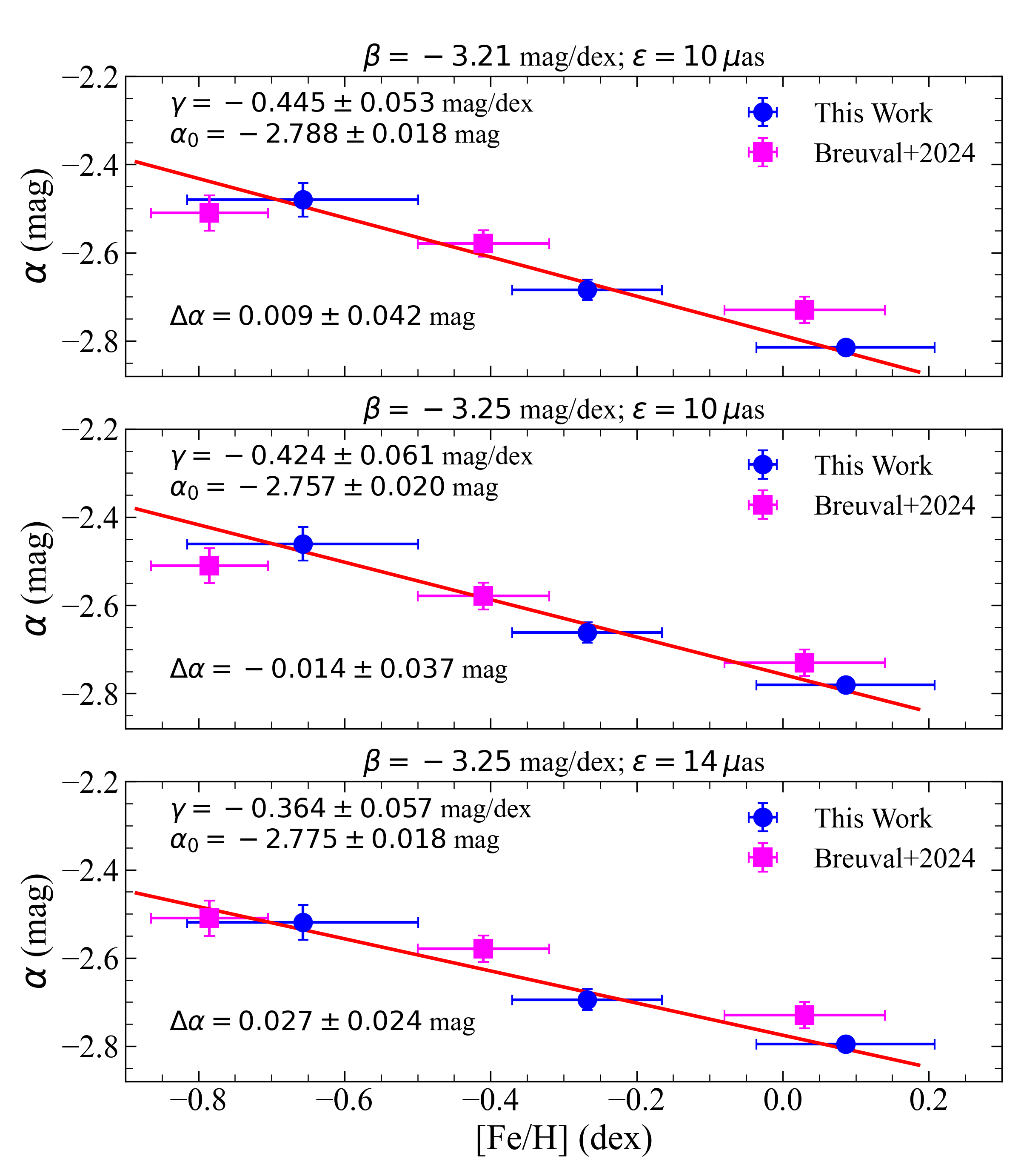}
\caption{Estimate of the $\gamma$ parameter for the WcHVI Wesenheit magnitude obtained by binning our data in three [Fe/H] intervals (filled blue circles). The top and middle panels show the results for two different choices of the $\beta$ parameter (see labels). The bottom panel shows the results for the same slope as in the middle panel but for a different value of $\epsilon$. For comparison, filled magenta circles show the results from \citet{Breuval2024}. 
}
\label{fig:binnedFit}
\end{figure}

\subsection{Use of the total metallicity [M/H] instead of [Fe/H]}

It is usual to express the metal abundance in terms of [Fe/H], i.e. the difference of the logarithm of the iron over hydrogen abundance ratios measured in the atmosphere of the target star and the Sun. 
This is based on the assumption that the solar heavy element distribution is universal. However, we have known for a long time that this is not always the case. For example, metal-poor stars in the Galactic halo show various degrees of enhancement of the so-called $\alpha$ elements (e.g. O, Ne, Mg, Si, S, Ca, and Ti) compared with the Sun \citep[see e.g. Chapter 8 of ][for a general discussion]{SalarisCassisi2005}. The amount of this enhancement can be of the order of [$\alpha$/Fe]$\sim$0.3-0.4 dex. In these cases, one should consider using the total metallicity [M/H] instead of [Fe/H]. 
As was demonstrated by \citet{Salaris1993}, the net effect of including the [$\alpha$/Fe] enhancement is to simulate a larger metallicity that, according to \citet{Salaris1993}, can be computed from the relation 

\begin{equation}
[\mathrm{M}/\mathrm{H}] \sim  [\mathrm{Fe}/\mathrm{H}] + \log_{10} \left( 0.638 \times 10^{[\alpha/\mathrm{Fe}]} + 0.362 \right)
\label{eq:alpha}
.\end{equation}

Typically, in the MW halo, the [$\alpha$/Fe] enhancement becomes significant at [Fe/H]$<-0.6$ dex \citep[this number varying in different environments, see e.g.][]{Tolstoy2009}. Since our sample includes objects more metal-poor than the threshold, we tested the effect of the [$\alpha$/Fe] enhancement on the determination of the PWZ relations for DCEPs using the [$\alpha$/Fe]  values published by \citet{Trentin2024b} for all the 290 stars in the C-MetaLL sample (their fig. 6). Although our DCEPs are disc and not halo stars, their [$\alpha$/Fe] enhancement can reach values of $\sim$+0.4 dex. We therefore calculated a new series of PWZ relations using [M/H] instead of [Fe/H] as an abundance indicator. To this aim, we converted [Fe/H] to [M/H] using Eq.~\ref{eq:alpha} and re-ran the calculations on the F+1O sample. 
The result of this test is shown in Fig.~\ref{fig:parameterComparisonPlot} and Table~\ref{tab:parameterComparison} (middle-bottom part). 
The effect of using [M/H] provides results and errors close to the baseline. However, we obtain $\gamma$ and $\epsilon$ values which are smaller (more negative) and larger than the baseline, respectively. Thus, the large (in the absolute sense) values of $\gamma$ obtained from our baseline calculations cannot be caused by the use of [Fe/H] instead of [M/H]. Since the changes of $\gamma$ and $\epsilon$ compensate for each other, the calculated LMC distance moduli are about 1\% larger than the baseline\footnote{Note that for this variation we adopted [$\alpha$/Fe]=0.3 dex, according to \citet{Mucciarelli2010}}, and thus more in agreement with the geometric value. 
The AIC and BIC values are close to those of the baseline in the optical but slightly lower for the NIR Wesenheit magnitudes. In the latter cases, therefore, the present variation suggests a slightly better fit to the data.

\subsection{Use of the brightest DCEPs only}

In this variation, we explore the impact of restricting the sample to only the brightest DCEPs, i.e. those with the most precise parallaxes. To avoid reducing the sample size excessively, we adopt a magnitude limit of $G<12.5$ mag, which yields 142 usable objects, slightly more than half of the full usable sample of 275 stars. Applying such a magnitude cut introduces well-known observational biases, notably the Malmquist bias\footnote{The Malmquist bias refers to the systematic overrepresentation of intrinsically brighter objects in flux-limited samples, leading to biased estimates of average luminosities and distances. For a comprehensive discussion, see e.g. the review by \citet{Sandage2000}.}. Therefore, this test should be regarded as a diagnostic exercise aimed at better understanding the properties of the DCEP sample used in this work.

A further consequence of the magnitude cut is a significant reduction in the metallicity range covered by the sample, as the fainter -- typically more distant -- DCEPs also tend to be more metal-poor. The result of this variation is shown in Fig.~\ref{fig:parameterComparisonPlot} and in Table~\ref{tab:parameterComparison} (bottom part). 
While large uncertainties are introduced due to the smaller sample size, it is evident that the magnitude cut leads to a general (absolute) decrease in the $\gamma$ values, whereas the other parameters remain nearly unchanged and the average distance modulus of the LMC approaches the geometric one by \citet{Pietrzynski2019}. It is unclear whether this shift in $\gamma$, bringing the values closer to those found by the SH0ES collaboration, is driven by parallax precision or the reduced metallicity range. However, a combination of both effects is likely.

We also attempted the same test on the complementary faint sample. Still, the results were inconclusive due to the low precision of the parallaxes and the limited metallicity range, this time biased towards more metal-poor stars. Some experiments in this direction for the WcHVI magnitudes are listed in Appendix~\ref{sect:testParallaxcHVI}.

\subsection{Binning of the data in [Fe/H]}

To further investigate the origin of the large metallicity dependence of our PWZ relations, we decided to estimate the $\gamma$ term similarly to that adopted by \citet{Breuval2024}. These authors assume a universal slope for the HST WHVI magnitudes and then determine the intercept of the simple PW relations in different environments using geometric distances; for example, $Gaia$ parallaxes in the MW, and eclipsing binaries in the LMC \citep[][]{Pietrzynski2019} and SMC \citep{Graczyk2020}. Finally, they plot the values of these intercepts against [Fe/H] in the three environments. The slope of the linear regression provides the value of $\gamma$ \citep[see e.g. fig. 8 in][]{Breuval2024}.

Besides the assumption about the universal slope, the subsequent approximation in this approach is to consider each environment as ‘mono-metallic’. This is justified, based on the present data, in the LMC and SMC, where the dispersion in [Fe/H] of the measured DCEPs is below 0.1 dex \citep[see][for the LMC and SMC, respectively]{Romaniello2022,Romaniello2008}. However, this is less accurate in the MW, where a wide range of abundances is present, thanks to the disc metallicity radial gradient \citep[e.g.][]{Luck2011,Genovali2014,Kovtyukh2022,Trentin2024b}.
In any case, to try to simulate this approach, we binned our C-MetaLL DCEP sample with usable parallaxes (excluding the outstanding outliers, as explained in Sect.~\ref{sect:analysis}) into three metallicity bins, including almost the same number of pulsators (about 90) with average [Fe/H]=$-0.66,\,-0.27,\,+0.0$ dex and dispersions of 0.16, 0.10, and 0.12 dex, respectively. For each bin, we then carried out the photometric parallax $\chi^2$ minimisation using the Cauchy loss function, as described above (Eq~\ref{eq:chi}), removing the $\gamma$ term and fixing both the slope, $\beta$, and the $Gaia$ parallax ZP counter-correction, $\epsilon$. In this way, only $\alpha$ is constrained by the $\chi^2$ minimisation. We carried out the calculations using the value of $\epsilon=10~\mu$as found from our baseline results in the WcHVI case (see Table~\ref{table:PWresults}) and two values for $\beta$; namely, 3.21 and 3.25. The former comes from our baseline calculation, while the latter is the one used by the SH0ES team. The result of this exercise is displayed in Fig.~\ref{fig:binnedFit}, where we plot the estimated values of $\alpha$ (which still includes the effect of metallicity) versus the binned metallicity [Fe/H] with the respective dispersions.
The three data points in the figure have been fitted with a regression line using the ODR (orthogonal distance regression) algorithm as implemented in the {\tt Scipy} package \citep[][]{2020SciPy-NMeth}. This method was chosen because it is robust, and errors on both axes can be accounted for. 

The $\gamma$ values derived this way (ranging from $\gamma \sim-0.36$ to $\sim-0.44$ mag/dex) are smaller in absolute value than those obtained by fitting all PWZ parameters simultaneously ($\sim-0.5$ mag/dex). This makes them more similar to the value from \citet{Breuval2024} ($\sim-0.234\pm0.052$ mag/dex), although ours are still more negative by about 1–2$\sigma$. Unsurprisingly, the least discrepant $\gamma$ value (a difference of slightly more than 1$\sigma$) was obtained using the $\beta$ and $\epsilon$ values adopted by the SH0ES group (bottom panel).

Using the three selected pairs of $\beta$ and $\epsilon$, the average differences between our results and those of \citet{Breuval2024} are consistent with zero within the errors for $\epsilon=10\mu$as (see top and middle panels of Fig.~\ref{fig:binnedFit}). This difference increases to a barely significant 2\% for $\epsilon=14\mu$as (bottom panel of Fig.~\ref{fig:binnedFit}).

Therefore, we seek to understand the cause of the discrepancy in $\gamma$.
Using the regression lines in the three panels of Fig.~\ref{fig:binnedFit} as references, we can evaluate the difference between our solution and the \citet{Breuval2024} data. The top and middle panels (obtained with $\epsilon=10\mu$as) show that the largest deviation between \citet{Breuval2024} and the regression line is about $\pm$0.07 mag. This occurs at the most metal-rich and metal-poor bins (representing MW and SMC galaxies), while the LMC bin aligns almost perfectly with our solution. This alignment explains the near-zero average difference but is difficult to interpret. While the difference in the metal-poor regime could plausibly be ascribed to less accurate $Gaia$ parallaxes for these fainter pulsators, the discrepancy in the MW data point is harder to explain, as parallaxes in this metallicity range should be the least affected by systematics. The result for $\epsilon=14\mu$as and $\beta=-3.25$ mag/dex \citep[the values adopted by][]{Breuval2024} is even more complex. In this case, the discrepancy in the most metal-poor (SMC) bin nearly vanishes, while the disagreement for the other two bins appears to be enhanced, even though the magnitude of the deviation is smaller than before ($\sim0.05$ mag). This accounts for the small, barely significant non-zero difference between our solution and the \citet{Breuval2024} data, and for the reduced scatter ($\sim$2\%) compared to the previous cases ($\sim$4\%).

The cause of this complex behaviour is not entirely clear. The main conclusion we can draw is that accurately measuring $\gamma$ is a difficult task. Indeed, Fig.~\ref{fig:binnedFit} demonstrates how changes of just a few hundredths of a magnitude at the extremes of the metallicity distribution are sufficient to significantly alter the value of $\gamma$. We conclude that a decisive step forward in establishing the correct value of $\gamma$ requires two things: first, the addition of more homogeneous DCEP data across the entire metallicity range (which we aim to provide in future C-MetaLL data releases), and second, and more importantly, the expected reduction in systematic errors in the upcoming $Gaia$ DR4 parallaxes.

\subsection{Variable parallax zero point counter-correction}

Some literature papers \citep[e.g.,][]{Riess2022,Khan2023} have reported a possible dependence of the $Gaia$ parallax ZP counter-correction -- $\epsilon$ in our case -- on the brightness of the specific sample. The general result is that there is a sort of linear decrease (in an absolute sense) in $\epsilon$ going towards fainter magnitudes, up to approximately zero at $G \sim 16$ mag. It is worth noting that there is no consensus regarding this occurrence, as has been discussed in other papers \citep[e.g.][]{Molinaro2023,Groenewegen2024}. In any case, to not leave anything unexplored, considering that our targets span a range of about six magnitudes, we substituted in Eq.~\ref{eq:chi} the constant $\epsilon$ with a linear equation as a function of the magnitude $G$: $z+p \times G$ and re-ran the calculation. The idea was to verify if the release of the assumption of a constant $\epsilon$ could be responsible for the large $\gamma$ values we find in the C-MetaLL survey. The results of this procedure are reported in the first part of Table~\ref{table:PW_linearZeropoint}. There is no effect on the values of $\gamma$, and the values of the slopes, $p$, are small in all the cases and barely significant at a 2$\sigma$ level in the NIR bands only. In any case, the solution appears valid, as the distance of the LMC, albeit systematically larger than the geometric one in all the cases, is consistent with this within the errors. Concerning the AIC and BIC values, we do not observe any significant decrease compared with the baseline, as expected in the case of better modelling. On the contrary, the BIC value, which tends to penalise the overfitting of the data, is slightly increased. 
We conclude that the dependence of the $Gaia$ parallaxes ZP counter-correction on brightness, if any, is small and not able to explain the large metallicity dependence we find in our work.

\subsection{Non-linear dependence on [Fe/H]}
\label{sect:quadratic}
To explore a non-linear dependence of the metallicity term ($\gamma$), we can introduce higher-order terms for the metallicity into the model. The most straightforward approach within the current polynomial-fitting framework is to include a quadratic term for the metallicity, $[Fe/H]^2$. Therefore the model is now the following:

\begin{equation}
W=\alpha+\beta(\log_{10} P-1.0)+{\gamma 1}{\rm {\rm [Fe/H]}}+{\gamma 2}{\rm {\rm [Fe/H]^2}}
\label{eq:PWZ1}
,\end{equation}

\noindent
where ${\gamma 1}$ and ${\gamma 2}$ are the linear and quadratic terms in metallicity. As in the previous case, we now have five parameters to fit. The result of this exercise is reported in the second part of Table~\ref{table:PW_linearZeropoint}.
We note that the linear term remains large and of the order of $-$0.5 mag/dex in all the cases, while the quadratic term is only significant at 1$\sigma$. The values of AIC and BIC are increased compared with the baseline, indicating that this quadratic model does not describe the data in a better fashion.

\smallskip

Further investigations about the value of $\gamma$ for the WcHVI have been extensively carried out in Appendix~\ref{sect:testParallaxcHVI} and are not reported here for brevity. The main results are discussed in the next section.

\section{Discussion and conclusions}\label{sect:conclusions}

In this work, we have exploited the homogeneous spectroscopic abundances provided for 290 DCEPs in the context of the C-MetaLL project \citep[][]{Trentin2024b}. We collected or calculated intensity-averaged magnitudes for these stars in various bands, which we used to calculate several optical and NIR Wesenheit magnitudes. We adopted both the \citet{Cardelli1989} and \citet{Fitzpatrick1999} reddening laws. Empirical relations from \citet{Riess2021} were employed to transform Johnson-Cousins $V,\,I$ and $H,\,V,\,I$ Wesenheit magnitudes in the respective HST $F555W,\,F814W$ and $F160W,\,F555W,\,F814W$ kins.
Our database was completed by: i) periods and modes of pulsation from the literature ($Gaia$ and OGLE IV surveys); and ii) parallaxes from the $Gaia$ mission corrected for the individual ZP bias \citet{Lindegren2021}. 
This wealth of data was then used to derive new PWZ relations from 275 DCEPs with usable parallaxes, utilising the photometric parallax technique and a new minimisation technique that can handle outlier measurements without removing the data. The fitting procedure also calculates the unknown global ZP counter-correction for the $Gaia$ parallaxes ($\epsilon$). 

Compared with our previous work, here we adopted a DCEP sample with homogeneous abundance determinations, an almost even population of objects over the entire metallicity range spanning approximately +0.3$<$[Fe/H]$<-1.1$ dex, the widest ever used in such calculations.   
Moreover, we adopted a new, more accurate fundamentalisation relation for the 1O pulsators and a new robust photometric parallax technique based on the MCMC algorithm, which allowed us to determine all fitting parameters simultaneously, including the value of $\epsilon$ directly from the data. In particular, we adopted a Cauchy likelihood, which enabled us to safely handle the substantial excess variance observed in our full dataset.

Our findings can be summarised as follows:

\begin{itemize}
\item
The metallicity dependence, $\gamma$, is confirmed to be larger in absolute value compared with the recent findings \citep[][]{Breuval2022,Breuval2024}, with typical values of $\gamma \sim -0.5$ mag/dex in the optical and HST Wesenheit bands and $\gamma \sim -0.4$ mag/dex in the NIR bands (e.g. using $J,\,K_S$). 
On the other hand, our results agree well in the NIR bands with those by \citet{Wang2025}, who used a DCEP sample and fitting technique similar to that of the present and our previous papers.
\item
The $Gaia$ parallax ZP counter-correction ($\epsilon$) varies almost monotonically from the $Gaia$ bands towards the NIR, spanning from about 16 to 7 $\mu$as with typical errors of about 4 $\mu$as. In particular, for the WcHVI magnitude, we obtain 10$ \pm 4 \mu$as, which agrees within 1$\sigma$ with \citet{Riess2021}.
\item 
The application of our PWZ relations to 4500 DCEPs in the LMC provides independent distances to this galaxy, which are typically within 1$\sigma$ of the geometric measurement by \citet{Pietrzynski2019}, with the remarkable exceptions of the WVI and WcVI magnitudes. Overall, this consistency demonstrates the soundness of our work.
\item 
The inclusion of the \citet{Fitzpatrick1999} reddening law in place of \citet{Cardelli1989}'s in the $V,\,I$ and $J,\,K_S$ Wesenheit definition does not impact the results significantly. 
\item
The use of only F-mode pulsators yields significantly larger errors in the parameters and unreliable results. A larger number of F-mode pulsators is needed to obtain results comparable with the entire F+1O sample. 
\item 
The adoption of the total metallicity [M/H] instead of [Fe/H] by correcting the latter with the [$\alpha$/Fe] measurements provided by \citet[][]{Trentin2024b} has the effect of increasing in an absolute sense the values of $\gamma$, while reducing those of $\epsilon$.
\item 
The selection of only the brightest DCEPs ($G<12.5$ mag), which have on average better parallax precision, provides values of $\alpha,\, \beta,\, \epsilon$ that follow the baseline's behaviour for the different Wesenheit magnitudes, while $\gamma$ generally decreases in absolute value to $\sim-0.3; -0.4$ mag/dex, i.e. closer to SHOES's team results. This occurrence may be driven by the improved parallax precision or the reduced metallicity range, though a combination of both effects is probable. This procedure, however, besides reducing the metallicity range, also introduces Malmquist bias, and therefore has to be regarded merely as an experiment. 

\item 
We investigated the potential dependence of the $Gaia$ parallax ZP counter-correction ($\epsilon$) on stellar brightness by using a linear correction ($z+p \times G$). This procedure was implemented to check if releasing the assumption of a constant $\epsilon$ could account for the large metallicity dependence ($\gamma$) found in our C-MetaLL survey. The results show that the values of $\gamma$ are unaffected, and the slope ($p$) is negligible, confirming that any brightness dependence cannot explain our findings.

\item
The introduction of a quadratic metallicity term, $\gamma_2 [\text{Fe/H}]^2$, into the period-luminosity relation (Eq.~\ref{eq:PWZ1}) did not significantly improve the model fit, as is indicated by the increased AIC and BIC values compared to the baseline linear model. While the linear metallicity term ($\gamma_1$) remains large ($\approx -0.5$ mag/dex), the quadratic term is only significant at the $1\sigma$ level. This suggests that a non-linear (at least a quadratic) dependence on metallicity is not strongly supported by the current dataset, and the simpler linear model provides a better description of the data.

\item 
The binning of our data into three metallicity intervals to simulate the data used by \citet{Breuval2024} and adoption of their fitting technique yield values of $\gamma$ ($\sim-0.36$ to $\sim-0.44$ mag/dex) that are closer to, but still 1-2$\sigma$ more negative than, the results from \citet{Breuval2024}.
This discrepancy is complex to understand, as small magnitude differences (0.05–0.07 mag) at the metallicity extremes (SMC and MW) are sufficient to significantly change the fitted $\gamma$. We conclude that accurately measuring $\gamma$ is extremely difficult and requires more homogeneous data and improved $Gaia$ DR4 parallaxes to reduce systematic errors. 
\item 

As is shown in Appendix~\ref{sect:testParallaxcHVI}, smaller (absolute) values of $\gamma$ ($\sim -0.3$ mag/dex) for the WcHVI relation are obtained for a sub-sample of DCEPs that are either brighter than $G\approx$11.5–12.5 mag, located at distances of $<$ 3/4 kpc (and thus benefitting from more precise parallaxes), or exhibit metallicities characteristic of solar neighbourhood stars (approximately $-0.5/0.7 <$ [Fe/H]$<$ +0.1/0.2 dex). If the larger value of $\gamma$ is not attributable to increased parallax uncertainties in more distant (d $>$ 3/4 kpc), metal-poor ([Fe/H] $< -0.7$ dex) stars, this could suggest either an intrinsically higher metallicity coefficient or a non-linear dependence of DCEP luminosities on metallicity at the low-metallicity end. Quantifying this effect with high precision remains challenging due to the significant covariance between the PWZ zero point, $\gamma$, and the adopted parallax ZP offset correction.
\end{itemize}

Overall, the results presented in this paper suggest that the exact metallicity dependence of PW relations for DCEPs remains uncertain. From a theoretical point of view, pulsational predictions are in favour of a mild value of $\gamma \sim -0.2$ mag/dex \citep[e.g.][]{Anderson2016b,DeSomma2022,Saniya2025}, similar to that provided by the empirical technique adopted by \citet{Breuval2021,Breuval2022,Breuval2024}. However, in this paper, we have shown that our fitting procedure is solid (see Appendix~\ref{sect:testPhotParallax}); therefore, the problem seems to be in the derivation of the $\gamma$ value using different geometric distance indicators, such as in \citet{Breuval2024} or using uniquely parallaxes from $Gaia$. Both approaches have pros and cons and are prone to different systematic uncertainties, such as the use of a universal slope and an average metallicity in each galaxy in the former approach or the bias on the $Gaia$ parallax in the latter. 

As was stressed in the introduction, the size of the metallicity dependence of DCEPs' PL and PW relations has important astrophysical consequences. Perhaps the most notable one is that a larger (absolute) value of $\gamma$ than the one currently used in the cosmic ladder would go in the direction to diminish the value of the inferred $H_0$ by 1-2\% for e.g. $\gamma \sim-0.4$ mag/dex. However, this reduction, far from solving the Hubble tension, would just reduce the discrepancy by about 1$\sigma$. 

In the context of the C-MetaLL project, we expect to improve our results shortly based on the measurement of abundances for about additional 100 DCEPs for which the analysis is still ongoing as well as the release of homogeneous $g,\,r,\,i,\,z,\,J,\,H,\,K_S$ rapid-eye-movement telescope\footnote{http://www.rem.inaf.it/} photometry for more than 120 DCEPs \citep[see][for early results]{Bhardwaj2024}. However, a significant improvement in this topic is expected with the next release of the $Gaia$ mission, Data Release 4, which will provide parallaxes that are not only more precise but also most likely less affected by systematic effects.  

\section{Data availability}

Table~\ref{tab:data} is only available in electronic form at the CDS via anonymous ftp to cdsarc.u-strasbg.fr (130.79.128.5) or via http://cdsweb.u-strasbg.fr/cgi-bin/qcat?J/A+A/.

\begin{acknowledgements}
We thank our anonymous referee for their insightful comments.
We also warmly thank A. Riess and L. Breuval for very useful discussions and suggestions, which helped us to improve our work. 
We acknowledge funding from: INAF GO-GTO grant 2023 “C-MetaLL - Cepheid metallicity in the Leavitt law” (P.I. V. Ripepi); 
PRIN MUR 2022 project (code 2022ARWP9C) 'Early Formation and Evolution of Bulge and Halo (EFEBHO),' PI: Marconi, M., funded by the European Union – Next Generation EU; 
Large Grant INAF 2023 MOVIE (P.I. M. Marconi).
AB thanks funding from the Anusandhan National Research Foundation (ANRF) under the Prime Minister Early
Career Research Grant scheme (ANRF/ECRG/2024/000675/PMS)
This research has made use of the
SIMBAD database operated at CDS, Strasbourg, France.
G.D.S. acknowledges funding from the INAF-ASTROFIT fellowship (PI G.De Somma), from $Gaia$ DPAC through INAF and ASI (PI: M.G. Lattanzi), and from INFN (Naples Section) through the QGSKY and Moonlight2 initiatives.

This work has made use of data from the European Space
Agency (ESA) mission $Gaia$ (https://www.cosmos.esa.int/gaia),
processed by the $Gaia$ Data Processing and Analysis Consortium (DPAC,
https://www.cosmos.esa.int/web/gaia/dpac/consortium). Funding
for the DPAC has been provided by national institutions, in particular, the
institutions participating in the $Gaia$ Multilateral Agreement.

This research was supported by the Munich Institute for Astro-, Particle and BioPhysics (MIAPbP), which is funded by the Deutsche Forschungsgemeinschaft (DFG, German Research Foundation) under Germany´s Excellence Strategy – EXC-2094 – 390783311.
This research was supported by the International Space Science Institute (ISSI) in Bern/Beijing through ISSI/ISSI-BJ International Team project ID \#24-603 - “EXPANDING Universe” (EXploiting Precision AstroNomical Distance INdicators in the $Gaia$ Universe).

\end{acknowledgements}

\bibliographystyle{aa} 
\bibliography{myBib} 

\begin{appendix} 

\section{Comparison between the native and ground-based HST magnitudes}
\label{sect:appendix_comparison_HST_mag}
\begin{figure}
\includegraphics[width=9cm]{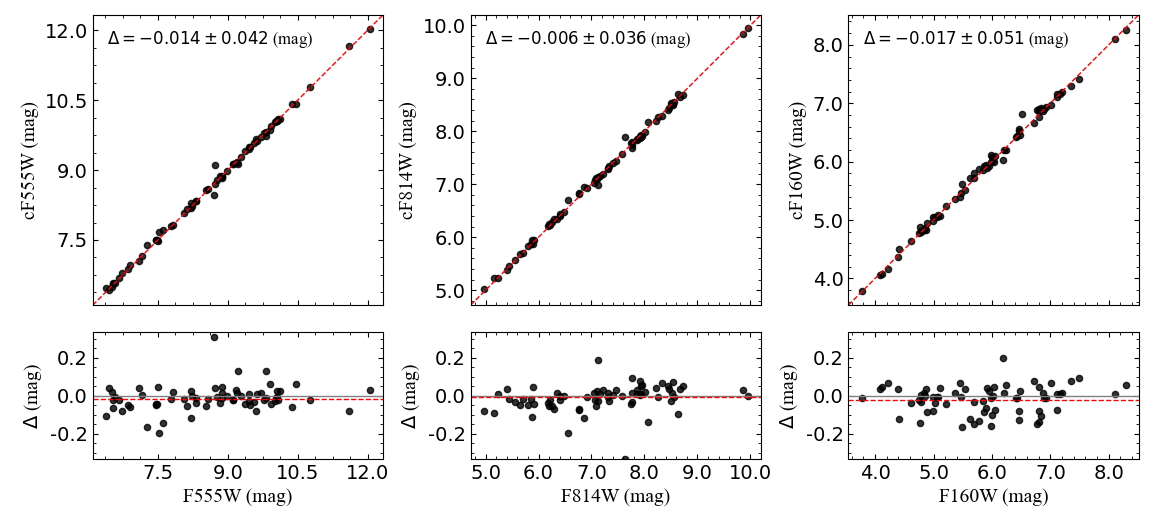}
\caption{Comparison between the F106W, F555W, and F814W magnitudes observed with HST by \citet{Riess2021} and those calculated based on ground-based $H,V,I$ magnitudes transformed into the corresponding HST magnitudes using the conversions by \citet{Riess2021}.}
\label{fig:comparisonMagnitude}
\end{figure}

\begin{figure}
\includegraphics[width=8cm]{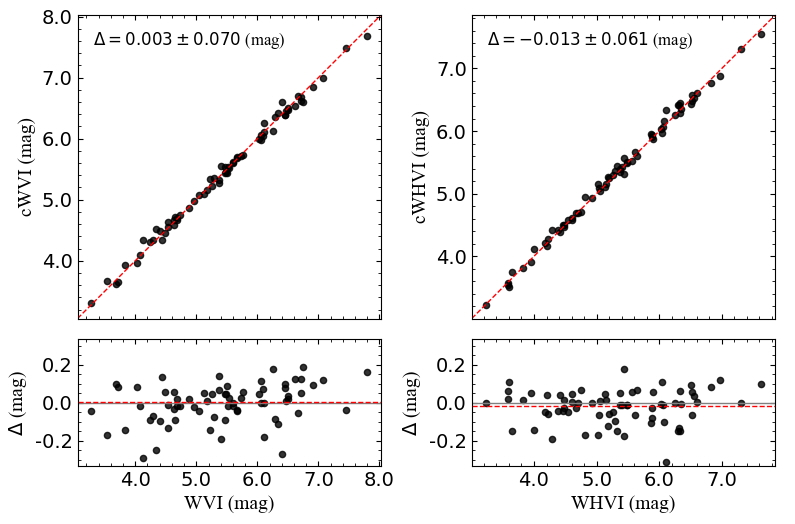}
\caption{As in Fig.~\ref{fig:comparisonMagnitude} but for the WVI and WHVI Wesenheit magnitudes.}
\label{fig:comparisonWesenheit}
\end{figure}

In this paper, we converted ground-based magnitudes in the $H, V, I$ bands into the corresponding HST ones $F106W,\, F555W,\, F814W$, adopting the transformations provided by \citet{Riess2021}. To verify the accuracy of our transformed magnitudes $cF106W, cF555W,$ and $cF814W$ and relative Wesenheit $cWVI$ and $cWHVI$, we considered a sample of 69 DCEPs with native HST photometry published by \citet{Riess2021}\footnote{The original sample was of 75 objects, but only 69 had the intensity-averaged photometry in the $Gaia$ bands} and proceeded as follows: 
1) calculated $V, I$ from the $Gaia$ DR3  $G, G_{BP}$ and $G_{RP}$ through the transformation by \citet{Pancino2022}; 2) obtained $H$ band from \citet{Trentin2024a}; 3) calculated $cF106W, cF555W$, and $cF814W$ and relative Wesenheit $cWVI$ and $cWHVI$ with the quoted transformation; 4) compared each calculated magnitudes and Wesenheit with the native HST data. 
The results of this test are reported in Fig.~\ref{fig:comparisonMagnitude} for $cF106W, cF555W$, and $cF814W$ and Fig.~\ref{fig:comparisonWesenheit} for $cWVI$ and $cWHVI$. This test shows the absence of significant discrepancies between our way of calculating the corrected HST photometry and the native HST one. 

\section{Data used in this paper}

This appendix includes an extract of the table including all the measurements used in this paper. 

\begin{table*}
\caption{Pulsational, photometric, and astrometric data used in this work.}
\label{tab:data}
\footnotesize\setlength{\tabcolsep}{3pt}
\begin{tabular}{lrrrrrlrr}
\hline
  \multicolumn{1}{c}{Star} &
  \multicolumn{1}{c}{Gaia\_source\_id} &
  \multicolumn{1}{c}{RA} &
  \multicolumn{1}{c}{DeC} &
  \multicolumn{1}{c}{P} &
  \multicolumn{1}{c}{P\_fund} &
  \multicolumn{1}{c}{Mode} &
  \multicolumn{1}{c}{G} &
  \multicolumn{1}{c}{eG} \\

  \multicolumn{1}{c}{} &
  \multicolumn{1}{c}{} &
  \multicolumn{1}{c}{(deg)} &
  \multicolumn{1}{c}{(deg)} &
  \multicolumn{1}{c}{(days)} &
  \multicolumn{1}{c}{(days)} &
  \multicolumn{1}{c}{} &
  \multicolumn{1}{c}{(mag)} &
  \multicolumn{1}{c}{(mag)} \\

  \multicolumn{1}{c}{(1)} &
  \multicolumn{1}{c}{(2)} &
  \multicolumn{1}{c}{(3)} &
  \multicolumn{1}{c}{(4)} &
  \multicolumn{1}{c}{(5)} &
  \multicolumn{1}{c}{(6)} &
  \multicolumn{1}{c}{(7)} & 
  \multicolumn{1}{c}{(8)} &
  \multicolumn{1}{c}{(9)} \\
\hline
  AA\_Gem & 3430067092837622272 & 91.64561 & 26.32922 & 11.31285 & 11.31285 & F & 9.363 & 0.005\\
  AD\_Pup & 5614312705966204288 & 117.01604 & -25.57777 & 13.59738 & 13.59738 & F & 9.541 & 0.003\\
  AP\_Sgr & 4066429066901946368 & 273.2604 & -23.11729 & 5.05808 & 5.05808 & F & 6.922 & 0.012\\
  AQ\_Pup & 5597379741549105280 & 119.59202 & -29.13008 & 30.15959 & 30.15959 & F & 8.127 & 0.011\\
  ASAS-SN\_J061713.86+022837.1 & 3124796657276655488 & 94.30773 & 2.47701 & 2.01925 & 2.01925 & F & 14.447 & 0.014\\
  ASAS-SN\_J063841.36-034927.7 & 3104095494729372032 & 99.67234 & -3.82435 & 3.86185 & 3.86185 & F & 14.014 & 0.012\\
  ASAS-SN\_J065046.50-085808.7 & 3050543819559526912 & 102.69373 & -8.9691 & 6.94564 & 6.94564 & F & 13.49 & 0.02\\
  ASAS-SN\_J072739.70-252241.1 & 5613685331497869312 & 111.91542 & -25.37809 & 2.7598 & 4.05333 & 1O & 13.891 & 0.001\\
  ASAS-SN\_J074310.73-113457.7 & 3039967784706458880 & 115.79482 & -11.58267 & 0.71534 & 1.00238 & 1O2O & 12.691 & 0.004\\
  ASAS-SN\_J074354.86-323013.7 & 5594991812757424768 & 115.97856 & -32.50378 & 3.14916 & 3.14916 & F & 14.771 & 0.007\\
\hline\end{tabular}\\
\\
\\
\begin{tabular}{rrrrrrrrrrrrrr}
\hline
  \multicolumn{1}{c}{BP} &
  \multicolumn{1}{c}{eBP} &
  \multicolumn{1}{c}{RP} &
  \multicolumn{1}{c}{eRP} &
  \multicolumn{1}{c}{V} &
  \multicolumn{1}{c}{eV} &
  \multicolumn{1}{c}{I} &
  \multicolumn{1}{c}{eI} &
  \multicolumn{1}{c}{J} &
  \multicolumn{1}{c}{eJ} &
  \multicolumn{1}{c}{H} &
  \multicolumn{1}{c}{eH} &
  \multicolumn{1}{c}{Ks} &
  \multicolumn{1}{c}{eKs} \\

  \multicolumn{1}{c}{(mag)} &
  \multicolumn{1}{c}{(mag)} &
  \multicolumn{1}{c}{(mag)} &
  \multicolumn{1}{c}{(mag)} &
  \multicolumn{1}{c}{(mag)} &
  \multicolumn{1}{c}{(mag)} &
  \multicolumn{1}{c}{(mag)} &
  \multicolumn{1}{c}{(mag)} &
  \multicolumn{1}{c}{(mag)} &
  \multicolumn{1}{c}{(mag)} &
  \multicolumn{1}{c}{(mag)} &
  \multicolumn{1}{c}{(mag)} &
  \multicolumn{1}{c}{(mag)} &
  \multicolumn{1}{c}{(mag)} \\

  \multicolumn{1}{c}{(10)} &
  \multicolumn{1}{c}{(11)} &
  \multicolumn{1}{c}{(12)} &
  \multicolumn{1}{c}{(13)} &
  \multicolumn{1}{c}{(14)} &
  \multicolumn{1}{c}{(15)} &    
  \multicolumn{1}{c}{(16)} &
  \multicolumn{1}{c}{(17)} &
  \multicolumn{1}{c}{(18)} &
  \multicolumn{1}{c}{(19)} &
  \multicolumn{1}{c}{(20)} &
  \multicolumn{1}{c}{(21)} &
  \multicolumn{1}{c}{(22)} &
  \multicolumn{1}{c}{(23)} \\

\hline
  9.956 & 0.007 & 8.62 & 0.005 & 9.714 & 0.012 & 8.583 & 0.03 & 7.647 & 0.01 & 7.191 & 0.01 & 7.086 & 0.01\\
  10.125 & 0.015 & 8.786 & 0.008 & 9.826 & 0.014 & 8.672 & 0.012 & 7.723 & 0.03 & 7.341 & 0.03 & 7.144 & 0.03\\
  7.46 & 0.039 & 6.219 & 0.035 & 6.964 & 0.013 & 6.054 & 0.011 & 5.346 & 0.01 & 4.966 & 0.01 & 4.878 & 0.01\\
  8.974 & 0.011 & 7.25 & 0.01 & 8.625 & 0.015 & 7.12 & 0.012 & 6.139 & 0.012 & 5.522 & 0.011 & 5.336 & 0.011\\
  15.102 & 0.056 & 13.639 & 0.033 & 14.856 & 0.03 & 13.572 & 0.03 & 12.669 & 0.03 & 12.08 & 0.04 & 11.931 & 0.037\\
  14.862 & 0.016 & 13.073 & 0.008 & 14.623 & 0.03 & 13.0 & 0.03 & 11.67 & 0.03 & 11.11 & 0.02 & 10.93 & 0.02\\
  14.402 & 0.029 & 12.547 & 0.017 & 14.142 & 0.03 & 12.449 & 0.03 & 11.191 & 0.023 & 10.626 & 0.023 & 10.398 & 0.023\\
  14.705 & 0.001 & 12.985 & 0.003 & 14.455 & 0.03 & 12.904 & 0.03 & 11.6 & 0.03 & 11.01 & 0.02 & 10.83 & 0.02\\
  13.02 & 0.006 & 12.19 & 0.003 & 12.84 & 0.03 & 12.147 & 0.03 & 11.484 & 0.052 & 11.248 & 0.045 & 11.16 & 0.043\\
  15.72 & 0.015 & 13.766 & 0.006 & 15.49 & 0.03 & 13.692 & 0.03 & 12.145 & 0.03 & 11.521 & 0.032 & 11.298 & 0.029\\
\hline\end{tabular}\\
\\
\\
\begin{tabular}{rrrrrrrrrlll}
\hline
  \multicolumn{1}{c}{[Fe/H]} &
  \multicolumn{1}{c}{e[Fe/H]} &
  \multicolumn{1}{c}{[$\alpha$/Fe]} &
  \multicolumn{1}{c}{e[$\alpha$/Fe]} &
  \multicolumn{1}{c}{$\varpi$} &
  \multicolumn{1}{c}{e$\varpi$} &
  \multicolumn{1}{c}{cor\_$\varpi$} &
  \multicolumn{1}{c}{gof} &
  \multicolumn{1}{c}{ruwe} &
  \multicolumn{1}{c}{flag\_source} &
  \multicolumn{1}{c}{flag\_opt\_nir} &
  \multicolumn{1}{c}{flag\_met} \\

    \multicolumn{1}{c}{(dex)} &
  \multicolumn{1}{c}{(dex)} &
  \multicolumn{1}{c}{(dex)} &
  \multicolumn{1}{c}{(dex)} &
  \multicolumn{1}{c}{(mas)} &
  \multicolumn{1}{c}{(mas)} &
  \multicolumn{1}{c}{(mas)} &
    \multicolumn{1}{c}{} &
  \multicolumn{1}{c}{} &
  \multicolumn{1}{c}{} &
  \multicolumn{1}{c}{} &
  \multicolumn{1}{c}{} \\

   \multicolumn{1}{c}{(24)} & 
  \multicolumn{1}{c}{(25)} &
  \multicolumn{1}{c}{(26)} &
  \multicolumn{1}{c}{(27)} &
  \multicolumn{1}{c}{(28)} &
  \multicolumn{1}{c}{(29)} &
  \multicolumn{1}{c}{(30)} &
  \multicolumn{1}{c}{(31)} &
  \multicolumn{1}{c}{(32)} &    
  \multicolumn{1}{c}{(33)} &
  \multicolumn{1}{c}{(34)} &    
  \multicolumn{1}{c}{(35)} \\
  
\hline
  -0.15 & 0.07 & 0.1 & 0.08 & 0.2749 & 0.0177 & -0.0365 & 4.98 & 1.249 & SOS,SOS & B15,P22,B15 & T24b\\
  -0.13 & 0.08 & 0.12 & 0.09 & 0.2331 & 0.0165 & -0.0207 & 12.48 & 1.362 & SOS,SOS & B15,B15,G18 & T24b\\
  -0.02 & 0.09 & 0.1 & 0.1 & 1.1815 & 0.024 & -0.0358 & -2.53 & 0.882 & P21,DR3 & B15,B15,B15 & T24b\\
  -0.17 & 0.08 & 0.18 & 0.09 & 0.2751 & 0.0226 & -0.0188 & 5.07 & 1.18 & SOS,SOS & B15,B15,B15 & T24b\\
  -0.72 & 0.16 & 0.24 & 0.18 & 0.0636 & 0.0233 & -0.0426 & 3.03 & 1.144 & P21,DR3 & P22,P22,2MASS & T23\\
  -0.35 & 0.14 & 0.12 & 0.15 & 0.0784 & 0.0207 & -0.0423 & 2.96 & 1.144 & SOS,SOS & P22,P22,B24 & T23\\
  -0.85 & 0.15 & 0.32 & 0.16 & 0.1198 & 0.0179 & -0.0412 & 2.27 & 1.106 & SOS,SOS & P22,P22,2MASS & T24b\\
  -0.48 & 0.16 & 0.14 & 0.17 & 0.067 & 0.0131 & -0.037 & 1.65 & 1.047 & SOS,SOS & P22,P22,B24 & T23\\
  -0.78 & 0.19 & 0.53 & 0.2 & 0.1863 & 0.0151 & -0.0249 & -0.74 & 0.966 & SOS,SOS & P22,P22,2MASS & T24b\\
  -0.43 & 0.18 & 0.16 & 0.19 & 0.0815 & 0.0194 & -0.0385 & 0.66 & 1.022 & SOS,SOS & P22,P22,2MASS & T23\\
\hline\end{tabular}
\tablefoot{The meaning of the different columns is as follows: (1) Literature name of the star; (2) $Gaia$ DR3 identification; (3) and (4) equatorial coordinates (J2000); (5) Period of pulsation; (6)  Period of pulsation where the values for 1O and 1O2O pulsators have been fundametantalised as described in the text. For mixed-mode DCEPs, the longest period is listed; (7) mode of pulsation -- F, 1O, F1O, and 1O2O indicate the fundamental, first overtone, and the mixed-mode pulsation modes, respectively; (8) to (23) magnitudes and errors for the $G$\,$G_{BP}$\,$G_{RP}$,\,$V$,\,$I$,\,$J$,\,$H$,\,$K_s$ bands, respectively; (24) and (25) iron abundance and error; (26) and (27) $\alpha$-element over iron and error; (28) and (29) original parallax value and error from $Gaia$ DR3 catalogue; (30) parallax ZP bias correction according to \citet{Lindegren2021}; (31) and (32) {\tt astrometric\_gof\_al} and {\tt RUWE} values from $Gaia$ DR3; 
(33) flag indicating the source of the identification of the DCEP and the used pulsation period (left) and source of the $Gaia$ photometry (right). The meaning of the acronyms is P21=\citet{Piet2021}; SOS=Specific Object Studies table in $Gaia$ DR3 \citep[{\it vari\_ceph}, see][]{Ripepi2023}; DR3=main $Gaia$ table \citep[{\it gaia\_source}, see][]{GaiaVallenari};  
(34) source of the $V$ (left), $I$ (middle) and $J,\,H,\,K_S$ (right) photometry: B15=\citet{Bhardwaj2015}; G18=\citet{Groenewegen2018}; P22 = magnitudes calculated on the basis of the $Gaia$ bands using the transformations to the Johnson-Cousins system by \citet{Pancino2022}; B24=\citet{Bhardwaj2024}; 2MASS=magnitudes calculated in this work based on the 2MASS survey data \citet{Skrutskie2006};
(35) source of the abundances: C20=\citet{Catanzaro2020}; R21a=\citet{Ripepi2021a}; R21b=\citet{Ripepi2021b}; T23=\citet{Trentin2023}; T24b=\citet{Trentin2024b}. A portion is shown here for guidance regarding its form and content. The full Table is available at the CDS).
}
\end{table*}

\section{Properties of the sample in comparison with literature}
\label{sect:misc}

Table~\ref{tab:excludedStars} lists the stars which have been excluded from the PW calculations due to the poor quality of the astrometric solution (see Sect.~\ref{sect:parallaxes} for details).

\begin{figure}
\includegraphics[width=8cm]{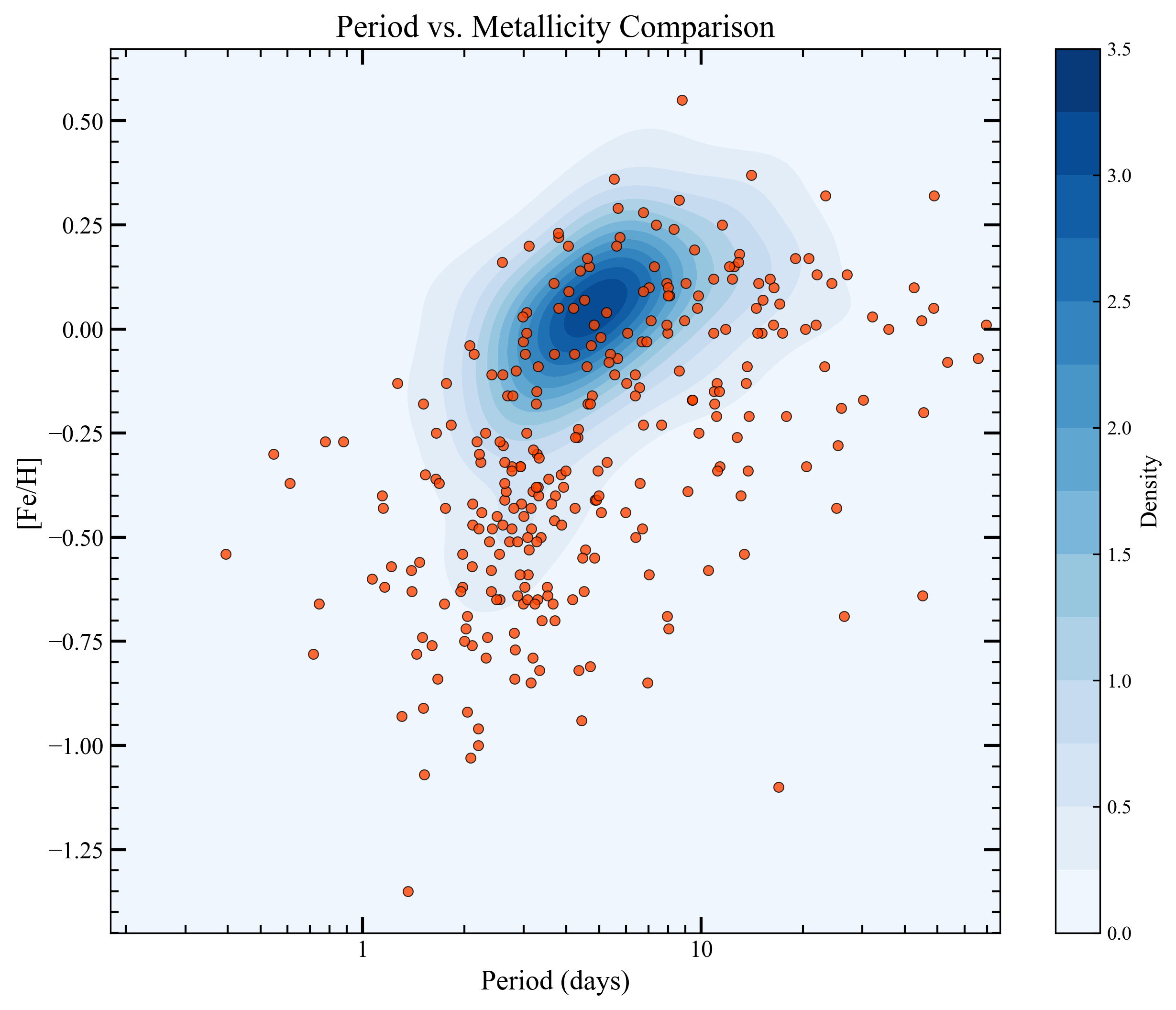}
\caption{Comparison of the period-[Fe/H] distributions between the samples adopted in C-MetaLL IV paper \citep[][]{Trentin2024b} (blue density distribution) and in the present work (red filled circles). 
}
\label{fig:comparisonPeriodMetallicity}
\end{figure}

\begin{figure}
\includegraphics[width=8cm]{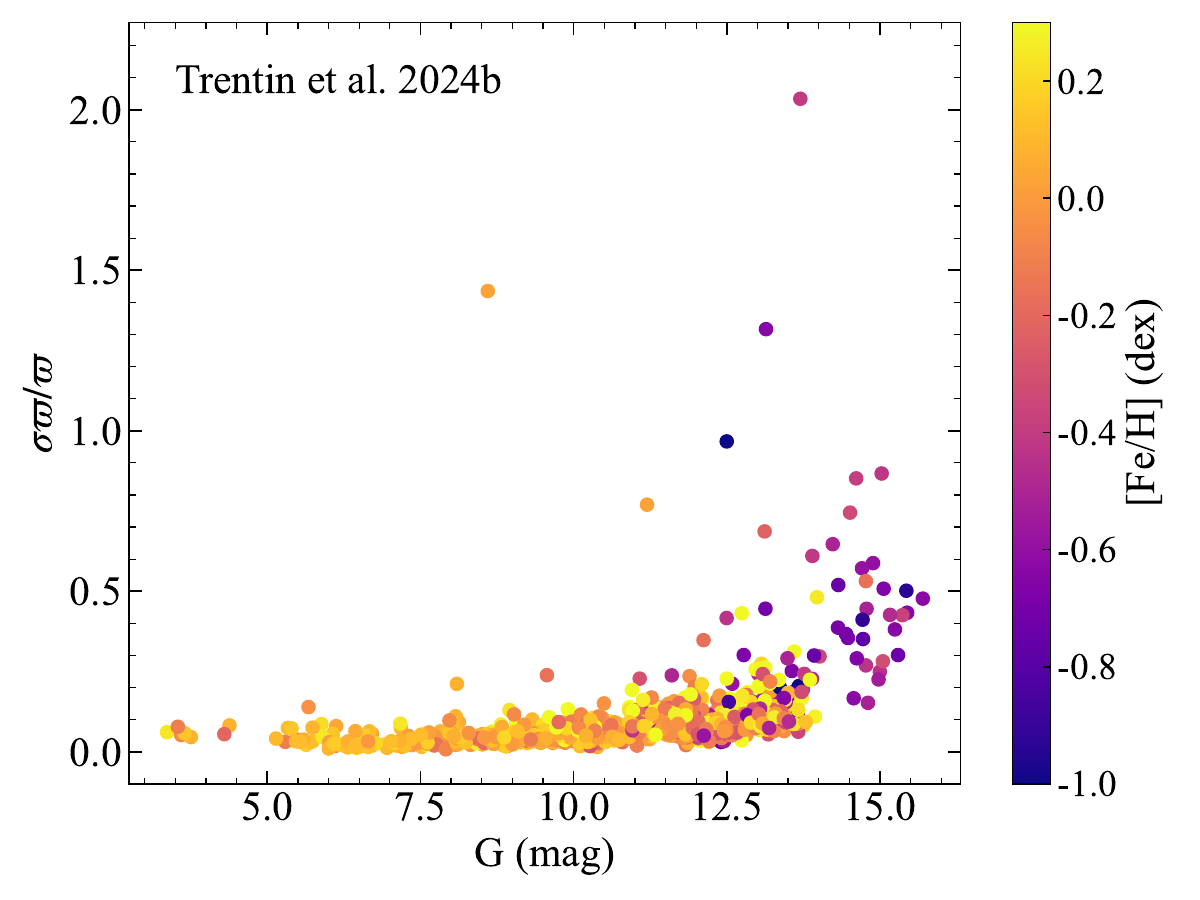}
\caption{As in Fig.~\ref{fig:parallaxError}, but for the sample adopted in C-MetaLL IV paper \citep[][]{Trentin2024b}.). 
}
\label{fig:comparisonParallaxError}
\end{figure}

Figure~\ref{fig:comparisonPeriodMetallicity} shows the different locations of the bulk data in our previous paper C-MetaLL IV \citep{Trentin2024b}, including literature values, and the present sample, only based on C-MetaLL spectroscopic data.
Similarly, Fig.~\ref{fig:comparisonParallaxError} displays the distribution of parallax error as a function of both $Gaia$ $G$ mag and [Fe/H] for the C-MetaLL IV \citep{Trentin2024b} sample. A comparison with Fig.~\ref{fig:parallaxError} reveals again the more homogeneous distribution in metallicity of the present work sample.

\begin{table}[]
    \centering
    \footnotesize\setlength{\tabcolsep}{3pt}
    \caption{List of stars excluded from the calculations due to low-quality astrometry.}
    \label{tab:excludedStars}
\begin{tabular}{lrll}
\hline
  \multicolumn{1}{c}{Star} &
  \multicolumn{1}{c}{Gaia\_source\_id} &
  \multicolumn{1}{c}{gof} &
  \multicolumn{1}{c}{ruwe} \\
\hline
  ASAS\_J060722+0834.0 & 3328542307996938496 & 167.37 & 15.53\\
  ASAS\_J074401-1707.8 & 5718755624717348096 & 13.98 & 1.44\\
  DL\_Cas & 428620663657823232 & 25.07 & 1.88\\
  NSVS\_2150508 & 266638892658936704 & 188.32 & 8.93\\
  OGLE\_GD-CEP-0104 & 2930223059544294912 & 20.37 & 1.57\\
  OGLE\_GD-CEP-0117 & 5613375028701902976 & 14.25 & 1.45\\
  OGLE\_GD-CEP-0204 & 5540101168641011840 & 19.52 & 1.73\\
  OGLE\_GD-CEP-0271 & 5318708901756763520 & 20.71 & 1.87\\
  OGLE\_GD-CEP-1313 & 2947842664653933824 & 27.87 & 2.00\\
  OGLE\_GD-CEP-1352 & 5588457861828906496 & 16.10 & 1.60\\
  RW\_Cam & 473043922712140928 & 130.48 & 8.01\\
  RX\_Cam & 470361114339849472 & 28.72 & 2.14\\
  SV\_Per & 203496585576324224 & 177.92 & 11.69\\
  U\_Aql & 4207681367143932800 & 27.78 & 3.09\\
  V432\_Ori & 3390186156826555136 & 17.74 & 1.61\\
\hline\end{tabular}
\end{table}

\begin{table}[]
    \centering
    \footnotesize\setlength{\tabcolsep}{3pt}
    \caption{List of frequent outliers.}
    \label{tab:outliers}
\begin{tabular}{lrl}
\hline
  \multicolumn{1}{c}{Star} &
  \multicolumn{1}{c}{Gaia\_source\_id} &
  \multicolumn{1}{c}{Cause} \\
\hline
  OGLE\_GD-CEP-0996 & 5878506555352293888 & opt. phot. \\
  OGLE\_GD-CEP-1111 & 5932882731154933888 & opt. phot. \\
  OGLE\_GD-CEP-0111 & 5615233993628681216 & opt. phot. \\
  OGLE\_GD-CEP-1282 & 3342468790993558528 & opt. phot. \\
  OGLE\_GD-CEP-1305 & 3050091885919345024 & Astrometry \\
  OGLE\_GD-CEP-0467 & 5258155016852594048 & Astrometry \\
  V966\_Mon & 3327474995745140608 & Astrometry \\
  ZTF\_J194453.19+251121.0 & 2020938380858128896 & Astrometry \\
\hline\end{tabular}
\end{table}

\section{Test of the photometric parallax procedure}
\label{sect:testPhotParallax}
\begin{figure}
\includegraphics[width=9cm]{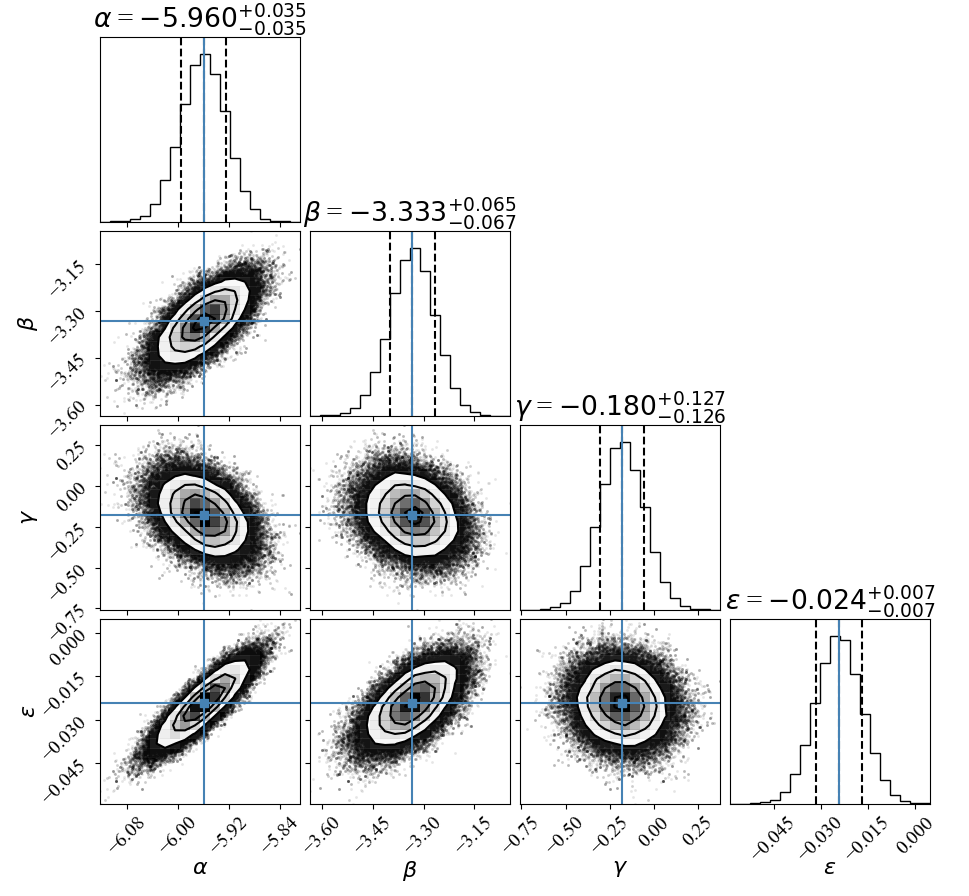}
\caption{Corner plot with the posterior distributions for the output parameters and the best-fit solution obtained for the \citet{Riess2021} sample and all parameters free to vary.}
\label{fig:cornerPlotRiess}
\end{figure}

In this section, we test the ability of our implementation of the photometric parallax method to give the correct results. To this aim, we adopted the \citet{Riess2021} sample of Galactic DCEPs observed with HST (their table 1)\footnote{As pointed out by \citet{Bhardwaj2023}, some [Fe/H] values in this table are not correct due to typos. We did not correct these metallicities on purpose, because the scope of our test is exclusively to reproduce the literature's results exactly. In any case, the use of the correct values has a minimal impact on the resulting parameters.} and applied the same selection procedures in terms of $Gaia$  astrometry (through the {\tt astrometric\_gof\_al} parameter), including the exclusion of the outliers S Vul and SV Vul. We also used the same width of the instability strip for the $WHVI$ magnitude equal to 0.06 mag. We then fitted Eq.~\ref{eq:chi} as explained in Sect.~\ref{sect:analysis}. The resulting posterior distributions for the output parameters are shown in Fig.~\ref{fig:cornerPlotRiess}. The results for the four parameters shown in the figure are also listed in Table~\ref{tab:riessComparison}, with the relative errors. They do not match exactly those by \citet{Riess2021} because these authors adopted a prior for the value of the global counter-correction, while we did not. However, our results match perfectly those obtained by \citet{Hogas2025}, which also re-analysed the data by \citet{Riess2021} but left the four fitting parameters free to vary, as we did here (see Table~\ref{tab:riessComparison}). In addition, this fit provides a $\chi^2/dof \approx 1.04$ (and AIC=71), as expected for a sample with correctly estimated errors, and in agreement with the value of $\chi^2/dof$ by \citet{Riess2021}. We conclude that our implementation of the photometric parallax fitting is reliable and free of technical errors. 

\begin{table}
\caption{Results from the Photometric parallax method with the MCMC algorithm for the $WHVI$ magnitude applied to the sample of DCEPs by  \citet{Riess2021} (see text).}             
\label{tab:riessComparison}      
\centering                          
\footnotesize\setlength{\tabcolsep}{3pt}
\begin{tabular}{c c c c c}        
\hline\hline                 
$\alpha$ & $\beta$ & $\gamma$ & $\epsilon$ & source \\    
\hline                        
-5.915$\pm$0.030 & -3.28$\pm$0.06 & -0.20$\pm$0.13 & -0.014$\pm$0.006 & R21\\
-5.960$\pm$0.035 & -3.33$\pm$0.07 & -0.18$\pm$0.13 & -0.024$\pm$0.007 & HM25 \\
-5.960$\pm$0.035 & -3.33$\pm$0.07 & -0.18$\pm$0.13 & -0.024$\pm$0.007 & TW \\
\hline                                   
\end{tabular}
\tablefoot{R21=\citet{Riess2021}; HM25=\citet{Hogas2025}; TW=This Work. 

Note that we have rounded central values and the errors for the $\beta$ and $\gamma$ parameters compared with Fig.~\ref{fig:cornerPlotRiess}, to be homogeneous with the literature results.
The units of $\alpha,\, \beta,\, \gamma,\,\epsilon$ and of their uncertainties are mag, mag/dex, mag/dex and mas, respectively.}
\end{table}

\section{Calculation of the error on the distance of LMC}
\label{sect:lmcDistances}

The total error is calculated by summing the statistical and systematic errors in quadrature: $\sigma_{\text{total}}^2 = {\sigma_{\text{stat}}^2 + \sigma_{\text{sys}}^2}$.


The statistical error represents the uncertainty on the mean distance modulus derived from the sample of $N$ DCEPs in the LMC. It is calculated as $\sigma_{\text{stat}} = \frac{\sigma_{\mu}}{\sqrt{N}}$, where: $\sigma_{\mu}$ is the standard deviation (1.48*MAD) of the distance modulus values for the $N$ stars in the sample. Since $N$ is large ($>$ 4000 stars), $\sigma_{\text{stat}}$ is of the order of 0.005 mag in all the bands, hence it is the systematic uncertainty that dominates the error budget.

The systematic error is the quadrature sum of three main components: $\sigma_{\text{sys}}^2 = \sigma_{\text{PW}}^2 + \sigma_{\text{$\epsilon$}}^2+\sigma_{\langle[\text{Fe/H}]\rangle}^2$, where the different contributions are: 
i) the uncertainty from the PW relation's calibrated coefficients errors ($\sigma_{\alpha}$, $\sigma_{\beta}$, $\sigma_{\gamma}$), propagated in the usual way; ii) the uncertainty on the $W$ absolute magnitude due to the error on the value of $\epsilon$ ($\sigma_{\epsilon}$), evaluated at the distance of the LMC and iii) the uncertainty from the LMC's assumed mean metallicity, which is however very small \citep{Romaniello2022} and can be safely neglected.

\section{Model Comparison Metrics}
\label{sect:aic}

We utilise the AIC and the BIC to formally compare the statistical performance of different PWZ models. 
These criteria balance the goodness-of-fit against 
the complexity (number of free parameters, $k$).

For a model where the maximum log-likelihood, $\ln \mathcal{L}_{\max}$, is estimated by the maximum of the log-probability function $\ln \mathcal{P}_{\max}$ from the MCMC sampler, the criteria are defined as:

\begin{align}
\label{eq:aic}
\text{AIC} &= 2k - 2 \ln \mathcal{L}_{\max} \\
\label{eq:bic}
\text{BIC} &= k \ln(N) - 2 \ln \mathcal{L}_{\max}
\end{align}
where $k$ is the number of free parameters in the model (e.g., $k=4$ for the baseline model), $N$ is the number of DCEPs used in the fit and $\ln \mathcal{L}_{\max}$ is the maximum log-likelihood, calculated using the median parameters from the MCMC chain: $\ln \mathcal{L}_{\max} \approx -0.5 \sum (\text{Cauchy Loss}) $
        
The model with the lowest AIC or BIC value is preferred. We use the result derived from the MCMC sampling as the basis for these metrics, as it is generally more robust than a simple minimisation result.

\section{Summary of the main variations of the baseline}

In this section, we report the main variations of the baseline carried out for all the 
PW relations investigated in this work.

\begin{table*}[h]
\footnotesize\setlength{\tabcolsep}{3pt}
 \caption[]{\label{tab:parameterComparison} Results of the photometric parallax fit to the PWZ relations with variations with respect to the baseline (see text).}
 
\begin{tabular}{lrrrrrrrrrrrrrrrrr}
\hline
\hline
  \multicolumn{1}{c}{Wesenheit} &
  \multicolumn{1}{c}{$\alpha$} &
  \multicolumn{1}{c}{$\sigma_\alpha^{low}$} &
  \multicolumn{1}{c}{$\sigma_\alpha^{high}$} &
  \multicolumn{1}{c}{$\beta$} &
  \multicolumn{1}{c}{$\sigma_\beta^{low}$} &
  \multicolumn{1}{c}{$\sigma_\beta^{high}$} &
  \multicolumn{1}{c}{$\gamma$} &
  \multicolumn{1}{c}{$\sigma_\gamma^{low}$} &
  \multicolumn{1}{c}{$\sigma_\gamma^{high}$} &
  \multicolumn{1}{c}{$\epsilon$} &
  \multicolumn{1}{c}{$\sigma_\epsilon^{low}$} &
  \multicolumn{1}{c}{$\sigma_\epsilon^{high}$} &
    \multicolumn{1}{c}{N} &
  \multicolumn{1}{c}{$\mu_{\rm LMC}$} &
  \multicolumn{1}{c}{$\sigma\mu_{\rm LMC}$} &
   \multicolumn{1}{c}{AIC} &
  \multicolumn{1}{c}{BIC} \\
\hline
\multicolumn{18}{c}{PWZ relations using sigma Clipping} \\
\hline
  Gaia & -6.038 & 0.022 & 0.022 & -3.309 & 0.034 & 0.035 & -0.585 & 0.061 & 0.058 & -0.015 & 0.003 & 0.003 & 268 & 18.491 & 0.042 & 716 & 730\\
  VI\_C89 & -6.078 & 0.022 & 0.023 & -3.273 & 0.036 & 0.037 & -0.628 & 0.063 & 0.062 & -0.012 & 0.003 & 0.003 & 267 & 18.418 & 0.043 & 639 & 654\\
  VI\_F99 & -5.871 & 0.022 & 0.021 & -3.183 & 0.037 & 0.037 & -0.393 & 0.059 & 0.058 & -0.019 & 0.003 & 0.002 & 274 & 18.467 & 0.039 & 766 & 781\\
  cVI & -5.877 & 0.021 & 0.021 & -3.183 & 0.036 & 0.035 & -0.278 & 0.059 & 0.059 & -0.016 & 0.003 & 0.003 & 272 & 18.406 & 0.037 & 684 & 699\\
  cHVI & -6.001 & 0.021 & 0.021 & -3.212 & 0.033 & 0.033 & -0.564 & 0.058 & 0.057 & -0.008 & 0.003 & 0.003 & 269 & 18.387 & 0.039 & 671 & 685\\
  VKs & -6.056 & 0.021 & 0.022 & -3.286 & 0.033 & 0.034 & -0.527 & 0.058 & 0.056 & -0.006 & 0.003 & 0.003 & 268 & 18.418 & 0.039 & 679 & 694\\
  JKs\_C89 & -6.105 & 0.022 & 0.021 & -3.354 & 0.034 & 0.035 & -0.369 & 0.058 & 0.056 & -0.008 & 0.003 & 0.003 & 271 & 18.447 & 0.038 & 717 & 732\\
  JKs\_F99 & -6.149 & 0.022 & 0.022 & -3.375 & 0.035 & 0.034 & -0.374 & 0.057 & 0.057 & -0.007 & 0.003 & 0.003 & 271 & 18.456 & 0.038 & 728 & 742\\
\hline
\multicolumn{18}{c}{PWZ relations using F-mode pulsators only} \\
\hline
 Gaia & -5.963 & 0.038 & 0.039 & -3.37 & 0.071 & 0.073 & -0.724 & 0.123 & 0.115 & -0.005 & 0.005 & 0.005 & 170 & 18.329 & 0.076 & 240 & 252\\
  VI\_C89 & -6.006 & 0.038 & 0.038 & -3.3 & 0.074 & 0.075 & -0.758 & 0.124 & 0.121 & -0.002 & 0.005 & 0.005 & 168 & 18.278 & 0.078 & 204 & 216\\
  VI\_F99 & -5.775 & 0.036 & 0.037 & -3.162 & 0.07 & 0.071 & -0.729 & 0.121 & 0.119 & -0.005 & 0.005 & 0.005 & 172 & 18.244 & 0.075 & 233 & 245\\
  cVI & -5.808 & 0.035 & 0.036 & -3.195 & 0.068 & 0.069 & -0.478 & 0.118 & 0.114 & -0.006 & 0.005 & 0.005 & 171 & 18.249 & 0.069 & 197 & 209\\
  cHVI & -5.937 & 0.035 & 0.036 & -3.276 & 0.067 & 0.068 & -0.648 & 0.112 & 0.108 & -0.001 & 0.005 & 0.005 & 170 & 18.26 & 0.068 & 208 & 220\\
  VKs & -5.995 & 0.036 & 0.036 & -3.346 & 0.068 & 0.068 & -0.613 & 0.115 & 0.111 & 0.002 & 0.005 & 0.005 & 171 & 18.291 & 0.071 & 224 & 237\\
  JKs\_C89 & -6.031 & 0.036 & 0.037 & -3.374 & 0.069 & 0.069 & -0.506 & 0.116 & 0.115 & 0.003 & 0.005 & 0.005 & 171 & 18.307 & 0.071 & 220 & 232\\
  JKs\_F99 & -6.073 & 0.037 & 0.038 & -3.395 & 0.069 & 0.069 & -0.513 & 0.117 & 0.113 & 0.004 & 0.005 & 0.005 & 171 & 18.313 & 0.071 & 222 & 235\\
\hline
\multicolumn{18}{c}{PWZ relations using the total metallicity [M/H]} \\
\hline
  Gaia & -6.0 & 0.038 & 0.039 & -3.31 & 0.052 & 0.054 & -0.701 & 0.121 & 0.117 & -0.014 & 0.004 & 0.004 & 270 & 18.555 & 0.056 & 352 & 366\\
  VI\_C89 & -6.024 & 0.039 & 0.04 & -3.268 & 0.056 & 0.057 & -0.815 & 0.126 & 0.123 & -0.01 & 0.005 & 0.004 & 267 & 18.463 & 0.059 & 298 & 313\\
  VI\_F99 & -5.813 & 0.037 & 0.038 & -3.163 & 0.056 & 0.055 & -0.61 & 0.122 & 0.122 & -0.015 & 0.004 & 0.004 & 274 & 18.461 & 0.056 & 350 & 364\\
  cVI & -5.834 & 0.036 & 0.037 & -3.165 & 0.053 & 0.053 & -0.452 & 0.117 & 0.114 & -0.013 & 0.004 & 0.004 & 272 & 18.397 & 0.051 & 317 & 331\\
  cHVI & -5.96 & 0.036 & 0.036 & -3.213 & 0.049 & 0.051 & -0.699 & 0.111 & 0.109 & -0.008 & 0.004 & 0.004 & 270 & 18.439 & 0.052 & 323 & 337\\
  VKs & -6.021 & 0.036 & 0.038 & -3.28 & 0.049 & 0.05 & -0.666 & 0.112 & 0.109 & -0.005 & 0.004 & 0.004 & 271 & 18.47 & 0.053 & 344 & 359\\
  JKs\_C89 & -6.057 & 0.037 & 0.037 & -3.337 & 0.051 & 0.051 & -0.554 & 0.115 & 0.113 & -0.004 & 0.004 & 0.004 & 271 & 18.45 & 0.053 & 327 & 341\\
  JKs\_F99 & -6.1 & 0.037 & 0.038 & -3.357 & 0.051 & 0.052 & -0.564 & 0.116 & 0.111 & -0.004 & 0.004 & 0.004 & 271 & 18.458 & 0.053 & 331 & 345\\
  \hline
\multicolumn{18}{c}{PWZ relations using DCEPs with $G<12.5$ mag only.} \\
\hline
 Gaia & -6.072 & 0.045 & 0.046 & -3.313 & 0.059 & 0.06 & -0.453 & 0.12 & 0.117 & -0.022 & 0.007 & 0.006 & 140 & 18.577 & 0.074 & 184 & 196\\
  VI\_C89 & -6.102 & 0.045 & 0.045 & -3.268 & 0.064 & 0.062 & -0.518 & 0.127 & 0.123 & -0.018 & 0.007 & 0.007 & 139 & 18.49 & 0.078 & 163 & 175\\
  VI\_F99 & -5.869 & 0.044 & 0.043 & -3.17 & 0.062 & 0.063 & -0.367 & 0.127 & 0.121 & -0.021 & 0.006 & 0.006 & 141 & 18.482 & 0.076 & 169 & 180\\
  cVI & -5.86 & 0.042 & 0.042 & -3.166 & 0.058 & 0.059 & -0.293 & 0.12 & 0.116 & -0.014 & 0.006 & 0.006 & 140 & 18.391 & 0.07 & 154 & 166\\
  cHVI & -5.999 & 0.041 & 0.042 & -3.207 & 0.055 & 0.056 & -0.469 & 0.114 & 0.107 & -0.009 & 0.006 & 0.006 & 140 & 18.426 & 0.068 & 173 & 185\\
  VKs & -6.057 & 0.042 & 0.042 & -3.276 & 0.054 & 0.055 & -0.401 & 0.109 & 0.109 & -0.006 & 0.007 & 0.006 & 140 & 18.475 & 0.068 & 179 & 191\\
  JKs\_C89 & -6.091 & 0.044 & 0.044 & -3.333 & 0.056 & 0.057 & -0.302 & 0.115 & 0.113 & -0.006 & 0.007 & 0.007 & 140 & 18.471 & 0.071 & 159 & 170\\
  JKs\_F99 & -6.134 & 0.043 & 0.045 & -3.353 & 0.057 & 0.059 & -0.314 & 0.117 & 0.114 & -0.006 & 0.007 & 0.007 & 140 & 18.476 & 0.071 & 163 & 175\\
\hline\end{tabular}
\tablefoot{C89 and F99 refer to the \citet{Cardelli1989} and \citet{Fitzpatrick1999} extinction laws, respectively. The values of $\sigma^{low}$ and $\sigma^{high}$ list the 16th and 84th percentiles of the posterior probability, respectively (see text). The units of $\alpha,\, \beta,\, \gamma,\,\epsilon$ and of their uncertainties are mag, mag/dex, mag/dex and mas, respectively.}
\end{table*}

\section{Test of variable parallax zero point counter-correction}
\label{sect:linearCountercorrection}

To test the possible dependence on the $Gaia$ $G$ magnitude of the parallax counter-correction, 
we modified Eq.~\ref{eq:chi}, substituting $\epsilon$ with the linear equation $z+p \times G$, where $z$ and $p$ are 
the new two unknowns which replace $\epsilon$. The introduction of an additional unknown impacts the stability and robustness of the solution, given that the number of data points remains the same as before. The result of this test is shown in Table~\ref{table:PW_linearZeropoint}.

\begin{table*}[h]
 \caption[]{\label{table:PW_linearZeropoint} Results of the photometric parallax fit to the PWZ relation in different bands using a linear dependence on the $G$ mag for the $Gaia$ parallax counter-correction.}
 \footnotesize\setlength{\tabcolsep}{2pt}
\begin{tabular}{lrrrrrrrrrrrrrrrrrrrr}
\hline
  \multicolumn{1}{c}{Wesenheit} &
  \multicolumn{1}{c}{$\alpha$} &
  \multicolumn{1}{c}{$\sigma_\alpha^{low}$} &
  \multicolumn{1}{c}{$\sigma_\alpha^{high}$} &
  \multicolumn{1}{c}{$\beta$} &
  \multicolumn{1}{c}{$\sigma_\beta^{low}$} &
  \multicolumn{1}{c}{$\sigma_\beta^{high}$} &
  \multicolumn{1}{c}{$\gamma$} &
  \multicolumn{1}{c}{$\sigma_\gamma^{low}$} &
  \multicolumn{1}{c}{$\sigma_\gamma^{high}$} &
  \multicolumn{1}{c}{$z$} &
  \multicolumn{1}{c}{$\sigma_z^{low}$} &
  \multicolumn{1}{c}{$\sigma_z^{high}$} &
    \multicolumn{1}{c}{$p$} &
  \multicolumn{1}{c}{$\sigma_p^{low}$} &
  \multicolumn{1}{c}{$\sigma_p^{high}$} &
  \multicolumn{1}{c}{N} &
  \multicolumn{1}{c}{$\mu_{\rm LMC}$} &
  \multicolumn{1}{c}{$\sigma\mu_{\rm LMC}$} & 
    \multicolumn{1}{c}{AIC} &
  \multicolumn{1}{c}{BIC} \\
\hline
   \multicolumn{21}{c}{Linear dependence of $Gaia$ parallax ZP correction on $G$} \\
  \hline
 Gaia & -6.201 & 0.061 & 0.061 & -3.41 & 0.065 & 0.069 & -0.494 & 0.103 & 0.1 & -0.105 & 0.027 & 0.027 & 0.006 & 0.002 & 0.002 & 270 & 18.641 & 0.082 & 342 & 360\\
  VI\_C89 & -6.21 & 0.058 & 0.062 & -3.357 & 0.069 & 0.071 & -0.613 & 0.106 & 0.101 & -0.094 & 0.028 & 0.028 & 0.006 & 0.002 & 0.002 & 267 & 18.514 & 0.083 & 290 & 308\\
  VI\_F99 & -5.934 & 0.058 & 0.058 & -3.227 & 0.068 & 0.068 & -0.417 & 0.1 & 0.097 & -0.063 & 0.028 & 0.028 & 0.003 & 0.002 & 0.002 & 274 & 18.498 & 0.08 & 349 & 367\\
  cVI & -5.828 & 0.056 & 0.057 & -3.143 & 0.061 & 0.063 & -0.33 & 0.093 & 0.09 & 0.011 & 0.029 & 0.029 & -0.002 & 0.002 & 0.002 & 272 & 18.354 & 0.074 & 320 & 338\\
  cHVI & -6.058 & 0.054 & 0.057 & -3.247 & 0.061 & 0.061 & -0.506 & 0.094 & 0.089 & -0.04 & 0.028 & 0.028 & 0.002 & 0.002 & 0.002 & 270 & 18.451 & 0.074 & 327 & 345\\
  VKs & -6.174 & 0.057 & 0.058 & -3.356 & 0.062 & 0.063 & -0.447 & 0.093 & 0.089 & -0.069 & 0.028 & 0.028 & 0.004 & 0.002 & 0.002 & 271 & 18.533 & 0.076 & 349 & 367\\
  JKs\_C89 & -6.201 & 0.056 & 0.057 & -3.415 & 0.062 & 0.064 & -0.366 & 0.092 & 0.092 & -0.069 & 0.027 & 0.027 & 0.004 & 0.002 & 0.002 & 271 & 18.514 & 0.075 & 331 & 349\\
  JKs\_F99 & -6.266 & 0.056 & 0.059 & -3.45 & 0.062 & 0.064 & -0.367 & 0.095 & 0.092 & -0.081 & 0.027 & 0.027 & 0.005 & 0.002 & 0.002 & 271 & 18.538 & 0.076 & 333 & 351\\
\hline
  \multicolumn{21}{c}{Quadratic dependence on [Fe/H]} \\
  \hline
  \multicolumn{1}{c}{Wesenheit} &
  \multicolumn{1}{c}{$\alpha$} &
  \multicolumn{1}{c}{$\sigma_\alpha^{low}$} &
  \multicolumn{1}{c}{$\sigma_\alpha^{high}$} &
  \multicolumn{1}{c}{$\beta$} &
  \multicolumn{1}{c}{$\sigma_\beta^{low}$} &
  \multicolumn{1}{c}{$\sigma_\beta^{high}$} &
  \multicolumn{1}{c}{$\gamma 1$} &
  \multicolumn{1}{c}{$\sigma_{\gamma 1}^{low}$} &
  \multicolumn{1}{c}{$\sigma_{\gamma 1}^{high}$} &
  \multicolumn{1}{c}{$\gamma 2$} &
  \multicolumn{1}{c}{$\sigma_{\gamma 2}^{low}$} &
  \multicolumn{1}{c}{$\sigma_{\gamma 2}^{high}$} &
    \multicolumn{1}{c}{$\epsilon$} &
  \multicolumn{1}{c}{$\sigma_\epsilon^{low}$} &
  \multicolumn{1}{c}{$\sigma_\epsilon^{high}$} &
  \multicolumn{1}{c}{N} &
  \multicolumn{1}{c}{$\mu_{\rm LMC}$} &
  \multicolumn{1}{c}{$\sigma\mu_{\rm LMC}$} & 
    \multicolumn{1}{c}{AIC} &
  \multicolumn{1}{c}{BIC} \\
\hline
  Gaia & -6.042 & 0.035 & 0.035 & -3.31 & 0.053 & 0.056 & -0.646 & 0.123 & 0.123 & -0.28 & 0.18 & 0.181 & -0.017 & 0.004 & 0.004 & 270 & 18.47 & 0.07 & 354 & 372\\
  VI\_C89 & -6.069 & 0.035 & 0.036 & -3.265 & 0.058 & 0.06 & -0.734 & 0.138 & 0.136 & -0.214 & 0.189 & 0.194 & -0.011 & 0.004 & 0.004 & 267 & 18.369 & 0.076 & 296 & 314\\
  VI\_F99 & -5.845 & 0.033 & 0.034 & -3.173 & 0.055 & 0.057 & -0.649 & 0.129 & 0.128 & -0.438 & 0.171 & 0.175 & -0.018 & 0.004 & 0.004 & 274 & 18.342 & 0.071 & 346 & 364\\
  cVI & -5.861 & 0.033 & 0.034 & -3.175 & 0.054 & 0.053 & -0.437 & 0.119 & 0.123 & -0.264 & 0.165 & 0.164 & -0.015 & 0.004 & 0.004 & 272 & 18.329 & 0.065 & 318 & 336\\
  cHVI & -6.007 & 0.032 & 0.033 & -3.219 & 0.051 & 0.052 & -0.6 & 0.117 & 0.116 & -0.212 & 0.171 & 0.173 & -0.01 & 0.004 & 0.004 & 270 & 18.374 & 0.065 & 328 & 346\\
  VKs & -6.068 & 0.033 & 0.032 & -3.292 & 0.05 & 0.051 & -0.578 & 0.113 & 0.112 & -0.254 & 0.155 & 0.158 & -0.009 & 0.004 & 0.004 & 271 & 18.406 & 0.064 & 352 & 370\\
  JKs\_C89 & -6.096 & 0.033 & 0.033 & -3.346 & 0.051 & 0.052 & -0.473 & 0.116 & 0.116 & -0.189 & 0.162 & 0.164 & -0.007 & 0.004 & 0.004 & 271 & 18.399 & 0.065 & 331 & 349\\
  JKs\_F99 & -6.14 & 0.033 & 0.034 & -3.369 & 0.051 & 0.052 & -0.489 & 0.116 & 0.114 & -0.217 & 0.158 & 0.159 & -0.007 & 0.004 & 0.004 & 271 & 18.403 & 0.065 & 335 & 353\\
  \hline\end{tabular}
\tablefoot{Note that $z$ and $p$ are correlated by $z+p \times G$, where $G$ is the $Gaia$ $G$ magnitude. Their units are mas and mas/mag, respectively. Concerning $\gamma1$ and $\gamma1$, their units are mag/dex and mag/dex$^2$, respectively.}
\end{table*}



\section{Further variations for the cHVI Wesenheit magnitude}
\label{sect:testParallaxcHVI}

In this appendix, we present the results of the photometric parallax fit for the cHVI Wesenheit magnitude, including additional variations of the baseline. This procedure is applied only to the cHVI Wesenheit because this is the Wesenheit magnitude used in the extragalactic distance scale. Thus, it is crucial to understand the origin of the larger $\gamma$ value in our baseline compared with the values used by the SH0ES team. Specifically, we carried out the photometric parallax solution making selections as follows: i) in $G$ magnitude, using only stars brighter than 10.5, 11.5, 12.5, 13.5, 14.5 and 15.7 mag (i.e. the whole sample); ii) in $G$ magnitude, using only stars fainter than 10.5, 11.5 and 12.5 mag (fainter thresholds would comprise too few stars)  iii) in parallax signal-to-noise, using 10 and 15 as thresholds; iv) in distance, using 3 and 4 Kpc as thresholds; v) in metallicity, using minimum values of -0.7 and -0.5 dex in [Fe/H] and +0.1, +0.2 and +0.3 dex for the maximum; vi) in period, SNR and metallicity for F-mode pulsators only. In this way, we cut the extreme values of metallicity that may cause extreme values of $\gamma$. 
The results of the entire procedure are reported in the first part of Table~\ref{tab:testParallax1}. 
The variations involving only 
low-quality parallaxes, namely SNR$(\varpi)<$10; SNR$(\varpi)<$15 and Dist$>$4Kpc, or $G>11.5$ produce non-realistic results, with intercepts $\sim$0.3-0.5 mag larger than the acceptable range of values. Values of $\gamma > -0.40$ mag/dex, i.e. significantly smaller in an absolute sense than the baseline, can only be obtained with samples including the brightest objects (i.e. $G<$12.5 mag) or significantly restricting the metallicity range, particularly using objects with [Fe/H]$<+0.1$ dex. 

However, due to the high correlation among the four parameters used in the calculations, it is difficult to determine which sample characteristic produces the large values of $\gamma$. Therefore, we decided to fix a couple of parameters, leaving the other two free to vary. We started by fixing $\alpha=-5.95$ mag and $\beta=-3.25$ mag/dex, according to the typical values obtained by the SH0ES team for the LMC. The result of the calculations in this case is listed in the second part of Table~\ref{tab:testParallax1}.
All the selections give high values of $\gamma$ (with large errors) apart from the same three cases mentioned above. In almost all cases, the values of $\epsilon$ are significantly smaller than when all four parameters are free to vary, and compatible with zero in most cases. Remarkably, the resulting LMC distance moduli are consistently smaller than the geometric value, indicating that this choice of alpha and beta is inadequate to describe the data.

The second test consisted of fixing $\beta$ as before and $\epsilon$ instead of $\alpha$. We choose two values for $\epsilon$, namely $-$10 $\mu$as, as resulted from our baseline calculation, and $-$14 $\mu$as, according to \citet{Riess2021}. The results of the test in the two cases are reported in the top and bottom parts of Table~\ref{tab:testParallax2}, respectively. 
We note that only the options including low-quality parallaxes give unreliable values of $\alpha$. Apart from this, both fixed values of $\epsilon$ provide a large number of cases with $\gamma<-0.4$ mag/dex, though the case of $-$14 $\mu$as provides generally lower (in absolute sense) values of $\gamma$, up to $\sim-0.2$ mag/dex, albeit often with significances only slightly larger than 1 $\sigma$ owing to the reduced samples (about 100 objects or fewer). Notably, the large majority of the solutions give reliable values for the distance to LMC. Concerning the goodness of fit, the values of AIC and BIC are generally moderately larger in the case of the four free-to-vary parameter case. The other three cases show comparable results with a slightly better fitting for $\beta$=3.25 and $\epsilon=0.014$ mas.

Overall, the experiments discussed above tell us that utilising DCEPs close to the Sun, or, equivalently, bright, or with a reduced range in metallicity, produces values of $\gamma$ generally significantly lower than those obtained using the entire sample. There are multiple explanations for this occurrence: i) the value of $\gamma$ is actually as high as -0.5 mag/dex; ii) $\gamma$ has a non-linear dependence on metallicity (but see Sect:~\ref{sect:quadratic}); iii) the high values of $\gamma$ are introduced by unknown systematic errors linked to the low-quality parallaxes. With the present dataset, we are not able to determine which scenario is the right one. To this end, increasing the sample of low-metallicity (i.e. fainter objects, with lower-quality parallaxes) DCEPs could help to increase the statistics. Alternatively, the improved $Gaia$ DR4 parallaxes may help in reducing the systematics.   

\begin{table*}[h]
\footnotesize\setlength{\tabcolsep}{2pt}
 \caption[]{\label{tab:testParallax1} Results of the photometric parallax fit to the cHVI PWZ relation with variations compared to the baseline (see text).}
\begin{tabular}{lrrrrrrrrrrrrrrrrr}
\hline
  \multicolumn{1}{c}{Variation} &
  \multicolumn{1}{c}{$\alpha$} &
  \multicolumn{1}{c}{$\sigma_\alpha^{low}$} &
  \multicolumn{1}{c}{$\sigma_\alpha^{high}$} &
  \multicolumn{1}{c}{$\beta$} &
  \multicolumn{1}{c}{$\sigma_\beta^{low}$} &
  \multicolumn{1}{c}{$\sigma_\beta^{high}$} &
  \multicolumn{1}{c}{$\gamma$} &
  \multicolumn{1}{c}{$\sigma_\gamma^{low}$} &
  \multicolumn{1}{c}{$\sigma_\gamma^{high}$} &
  \multicolumn{1}{c}{$\epsilon$} &
  \multicolumn{1}{c}{$\sigma_\epsilon^{low}$} &
  \multicolumn{1}{c}{$\sigma_\epsilon^{high}$} &
    \multicolumn{1}{c}{N} &
  \multicolumn{1}{c}{$\mu_{\rm LMC}$} &
  \multicolumn{1}{c}{$\sigma\mu_{\rm LMC}$} &
   \multicolumn{1}{c}{AIC} &
  \multicolumn{1}{c}{BIC} \\
\hline
G<10.5 mag           &   -5.956 & 0.089 & 0.092 & -3.224 & 0.136 & 0.135 & -0.496 & 0.385 & 0.403 & -0.001 & 0.017 & 0.017 & 64 & 18.365 & 0.192 & 49 & 58\\
G<11.5 mag           &   -5.988 & 0.077 & 0.081 & -3.158 & 0.103 & 0.103 & -0.403 & 0.253 & 0.254 & -0.009 & 0.013 & 0.014 & 100 & 18.463 & 0.138 & 82 & 92\\
G<12.5 mag           &   -6.006 & 0.067 & 0.067 & -3.183 & 0.091 & 0.092 & -0.39 & 0.194 & 0.187 & -0.011 & 0.011 & 0.01 & 137 & 18.476 & 0.113 & 112 & 124\\
G<13.5 mag           &   -6.013 & 0.053 & 0.053 & -3.191 & 0.088 & 0.088 & -0.542 & 0.153 & 0.148 & -0.012 & 0.007 & 0.007 & 224 & 18.417 & 0.093 & 183 & 196\\
G<14.5 mag           &   -5.994 & 0.05 & 0.051 & -3.18 & 0.086 & 0.087 & -0.552 & 0.146 & 0.143 & -0.009 & 0.007 & 0.006 & 243 & 18.398 & 0.089 & 192 & 206\\
G>10.5 mag           &   -6.065 & 0.093 & 0.104 & -3.135 & 0.133 & 0.143 & -0.611 & 0.18 & 0.17 & -0.014 & 0.008 & 0.007 & 203 & 18.466 & 0.135 & 151 & 165\\
G>11.5 mag           &   -6.34 & 0.19 & 0.218 & -3.281 & 0.22 & 0.232 & -0.768 & 0.249 & 0.245 & -0.02 & 0.01 & 0.01 & 167 & 18.611 & 0.239 & 117 & 130\\
G>12.5 mag           &   -6.469 & 0.312 & 0.333 & -3.178 & 0.341 & 0.367 & -0.975 & 0.353 & 0.324 & -0.023 & 0.013 & 0.013 & 130 & 18.7 & 0.38 & 87 & 99\\
SNR<10               &   -6.721 & 0.303 & 0.356 & -3.2 & 0.318 & 0.346 & -0.747 & 0.378 & 0.354 & -0.035 & 0.015 & 0.015 & 133 & 19.037 & 0.371 & 76 & 87\\
SNR>10               &   -6.067 & 0.067 & 0.072 & -3.165 & 0.098 & 0.1 & -0.439 & 0.202 & 0.191 & -0.026 & 0.01 & 0.01 & 133 & 18.524 & 0.116 & 110 & 121\\
SNR<15               &   -6.284 & 0.128 & 0.129 & -3.131 & 0.163 & 0.164 & -0.741 & 0.218 & 0.209 & -0.022 & 0.009 & 0.008 & 192 & 18.633 & 0.175 & 135 & 148\\
SNR>15               &   -5.944 & 0.081 & 0.085 & -3.106 & 0.107 & 0.108 & -0.441 & 0.233 & 0.243 & -0.004 & 0.015 & 0.016 & 74 & 18.427 & 0.135 & 63 & 73\\
D<3 Kpc              &   -5.988 & 0.091 & 0.098 & -3.182 & 0.117 & 0.121 & -0.532 & 0.263 & 0.295 & -0.014 & 0.02 & 0.02 & 65 & 18.399 & 0.152 & 59 & 67\\
D>3 Kpc              &   -6.203 & 0.109 & 0.117 & -3.17 & 0.151 & 0.153 & -0.671 & 0.206 & 0.193 & -0.019 & 0.009 & 0.008 & 202 & 18.564 & 0.156 & 137 & 150\\
D<4 Kpc              &   -6.001 & 0.082 & 0.089 & -3.156 & 0.098 & 0.099 & -0.431 & 0.233 & 0.235 & -0.013 & 0.015 & 0.016 & 100 & 18.466 & 0.135 & 86 & 96\\
D>4 Kpc              &   -6.506 & 0.218 & 0.236 & -3.2 & 0.23 & 0.243 & -0.824 & 0.277 & 0.282 & -0.029 & 0.011 & 0.011 & 167 & 18.79 & 0.269 & 105 & 117\\
-0.7<[Fe/H]<+0.1     &   -6.0 & 0.052 & 0.052 & -3.222 & 0.103 & 0.102 & -0.365 & 0.245 & 0.235 & -0.013 & 0.007 & 0.007 & 199 & 18.463 & 0.123 & 150 & 163\\
-0.7<[Fe/H]<+0.2     &   -6.001 & 0.052 & 0.053 & -3.188 & 0.091 & 0.093 & -0.53 & 0.195 & 0.183 & -0.01 & 0.007 & 0.007 & 223 & 18.411 & 0.105 & 160 & 174\\
-0.7<[Fe/H]<+0.3     &   -6.001 & 0.05 & 0.053 & -3.175 & 0.091 & 0.09 & -0.552 & 0.187 & 0.18 & -0.01 & 0.007 & 0.007 & 230 & 18.408 & 0.102 & 172 & 185\\
-0.5<[Fe/H]<+0.1     &   -5.999 & 0.053 & 0.053 & -3.199 & 0.105 & 0.104 & -0.224 & 0.3 & 0.292 & -0.015 & 0.007 & 0.008 & 156 & 18.53 & 0.143 & 120 & 132\\
-0.5<[Fe/H]<+0.2     &   -5.998 & 0.052 & 0.054 & -3.162 & 0.089 & 0.091 & -0.486 & 0.212 & 0.213 & -0.01 & 0.007 & 0.007 & 180 & 18.438 & 0.11 & 135 & 147\\
-0.5<[Fe/H]<+0.3     &   -6.0 & 0.054 & 0.055 & -3.149 & 0.091 & 0.091 & -0.511 & 0.209 & 0.214 & -0.01 & 0.007 & 0.007 & 187 & 18.436 & 0.111 & 142 & 155\\
P>5d                 &    -5.967 & 0.058 & 0.059 & -3.317 & 0.147 & 0.143 & -0.716 & 0.257 & 0.249 & -0.006 & 0.008 & 0.008 & 125 & 18.244 & 0.139 & 97 & 108\\
P>10d                &    -5.89 & 0.126 & 0.12 & -3.606 & 0.305 & 0.298 & -0.477 & 0.417 & 0.414 & -0.006 & 0.015 & 0.014 & 61 & 18.132 & 0.255 & 39 & 47\\
P>5d  SNR>10         &    -6.02 & 0.075 & 0.077 & -3.274 & 0.174 & 0.176 & -0.504 & 0.281 & 0.302 & -0.019 & 0.012 & 0.012 & 75 & 18.402 & 0.16 & 58 & 67\\
P>10d SNR>10         &    -6.083 & 0.197 & 0.227 & -3.618 & 0.512 & 0.449 & -0.579 & 0.623 & 0.64 & -0.042 & 0.026 & 0.03 & 34 & 18.276 & 0.401 & 20 & 26\\
P>5d [Fe/H]>$-$0.5   &   -5.975 & 0.064 & 0.063 & -3.271 & 0.155 & 0.155 & -0.749 & 0.322 & 0.335 & -0.008 & 0.009 & 0.009 & 108 & 18.258 & 0.165 & 77 & 88\\
P>10d [Fe/H]>$-$0.5  &   -5.943 & 0.136 & 0.133 & -3.594 & 0.343 & 0.319 & -0.737 & 0.524 & 0.526 & -0.015 & 0.017 & 0.015 & 53 & 18.082 & 0.301 & 30 & 38\\
\hline
  \multicolumn{18}{c}{Fixed $\alpha$ and $\beta$} \\
  \hline
G<10.5 mag           &  -5.95 & 0.0 & 0.0 & -3.25 & 0.0 & 0.0 & -0.511 & 0.348 & 0.356 & 0.001 & 0.007 & 0.007 & 64 & 18.341 & 0.144 & 45 & 50\\
 G<11.5 mag           &  -5.95 & 0.0 & 0.0 & -3.25 & 0.0 & 0.0 & -0.439 & 0.229 & 0.234 & -0.001 & 0.005 & 0.005 & 100 & 18.37 & 0.095 & 77 & 82\\
 G<12.5 mag           &  -5.95 & 0.0 & 0.0 & -3.25 & 0.0 & 0.0 & -0.449 & 0.155 & 0.155 & -0.002 & 0.005 & 0.005 & 137 & 18.366 & 0.065 & 108 & 114\\
 G<13.5 mag           &  -5.95 & 0.0 & 0.0 & -3.25 & 0.0 & 0.0 & -0.593 & 0.117 & 0.111 & -0.003 & 0.004 & 0.004 & 224 & 18.307 & 0.051 & 181 & 188\\
 G<14.5 mag           &  -5.95 & 0.0 & 0.0 & -3.25 & 0.0 & 0.0 & -0.572 & 0.116 & 0.115 & -0.003 & 0.004 & 0.004 & 243 & 18.315 & 0.051 & 184 & 191\\
 G>10.5 mag           &  -5.95 & 0.0 & 0.0 & -3.25 & 0.0 & 0.0 & -0.54 & 0.147 & 0.139 & -0.005 & 0.005 & 0.005 & 203 & 18.328 & 0.062 & 154 & 161\\
 G>11.5 mag           &  -5.95 & 0.0 & 0.0 & -3.25 & 0.0 & 0.0 & -0.591 & 0.174 & 0.164 & -0.004 & 0.006 & 0.006 & 167 & 18.308 & 0.074 & 121 & 127\\
 G>12.5 mag           &  -5.95 & 0.0 & 0.0 & -3.25 & 0.0 & 0.0 & -0.676 & 0.198 & 0.189 & -0.002 & 0.007 & 0.006 & 130 & 18.273 & 0.084 & 87 & 93\\
 SNR<10               &  -5.95 & 0.0 & 0.0 & -3.25 & 0.0 & 0.0 & -0.676 & 0.209 & 0.194 & 0.002 & 0.006 & 0.006 & 133 & 18.273 & 0.088 & 86 & 91\\
 SNR>10               &  -5.95 & 0.0 & 0.0 & -3.25 & 0.0 & 0.0 & -0.567 & 0.14 & 0.135 & -0.007 & 0.005 & 0.005 & 133 & 18.317 & 0.06 & 113 & 119\\
 SNR<15               &  -5.95 & 0.0 & 0.0 & -3.25 & 0.0 & 0.0 & -0.636 & 0.155 & 0.147 & -0.003 & 0.005 & 0.005 & 192 & 18.289 & 0.066 & 138 & 144\\
 SNR>15               &  -5.95 & 0.0 & 0.0 & -3.25 & 0.0 & 0.0 & -0.353 & 0.207 & 0.224 & -0.001 & 0.006 & 0.006 & 74 & 18.405 & 0.085 & 57 & 62\\
 D<3 Kpc              &  -5.95 & 0.0 & 0.0 & -3.25 & 0.0 & 0.0 & -0.533 & 0.253 & 0.302 & -0.004 & 0.008 & 0.008 & 65 & 18.331 & 0.105 & 54 & 58\\
 D>3 Kpc              &  -5.95 & 0.0 & 0.0 & -3.25 & 0.0 & 0.0 & -0.622 & 0.15 & 0.141 & -0.002 & 0.005 & 0.005 & 202 & 18.295 & 0.064 & 147 & 153\\
 D<4 Kpc              &  -5.95 & 0.0 & 0.0 & -3.25 & 0.0 & 0.0 & -0.432 & 0.2 & 0.206 & -0.002 & 0.005 & 0.005 & 100 & 18.373 & 0.083 & 82 & 87\\
 D>4 Kpc              &  -5.95 & 0.0 & 0.0 & -3.25 & 0.0 & 0.0 & -0.632 & 0.167 & 0.156 & -0.002 & 0.005 & 0.005 & 167 & 18.291 & 0.071 & 118 & 124\\
 -0.7<[Fe/H]<+0.1     &  -5.95 & 0.0 & 0.0 & -3.25 & 0.0 & 0.0 & -0.401 & 0.214 & 0.206 & -0.008 & 0.005 & 0.005 & 199 & 18.385 & 0.089 & 140 & 147\\
 -0.7<[Fe/H]<+0.2     &  -5.95 & 0.0 & 0.0 & -3.25 & 0.0 & 0.0 & -0.582 & 0.153 & 0.148 & -0.004 & 0.004 & 0.004 & 223 & 18.311 & 0.065 & 157 & 163\\
 -0.7<[Fe/H]<+0.3     &  -5.95 & 0.0 & 0.0 & -3.25 & 0.0 & 0.0 & -0.602 & 0.153 & 0.143 & -0.003 & 0.004 & 0.004 & 230 & 18.303 & 0.065 & 168 & 175\\
 -0.5<[Fe/H]<+0.1     &  -5.95 & 0.0 & 0.0 & -3.25 & 0.0 & 0.0 & -0.235 & 0.249 & 0.253 & -0.009 & 0.005 & 0.005 & 156 & 18.454 & 0.102 & 112 & 118\\
 -0.5<[Fe/H]<+0.2     &  -5.95 & 0.0 & 0.0 & -3.25 & 0.0 & 0.0 & -0.521 & 0.187 & 0.186 & -0.004 & 0.004 & 0.004 & 180 & 18.336 & 0.078 & 130 & 136\\
 -0.5<[Fe/H]<+0.3     &  -5.95 & 0.0 & 0.0 & -3.25 & 0.0 & 0.0 & -0.557 & 0.175 & 0.177 & -0.003 & 0.004 & 0.004 & 187 & 18.322 & 0.074 & 143 & 149\\
 P>5d                 &   -5.95 & 0.0 & 0.0 & -3.25 & 0.0 & 0.0 & -0.772 & 0.251 & 0.23 & -0.004 & 0.005 & 0.005 & 125 & 18.233 & 0.105 & 87 & 93\\
 P>10d                &   -5.95 & 0.0 & 0.0 & -3.25 & 0.0 & 0.0 & -0.436 & 0.336 & 0.354 & -0.004 & 0.006 & 0.006 & 61 & 18.371 & 0.139 & 38 & 42\\
 P>5d  SNR>10         &   -5.95 & 0.0 & 0.0 & -3.25 & 0.0 & 0.0 & -0.605 & 0.283 & 0.311 & -0.009 & 0.006 & 0.006 & 75 & 18.302 & 0.118 & 50 & 55\\
 P>10d SNR>10         &   -5.95 & 0.0 & 0.0 & -3.25 & 0.0 & 0.0 & -0.843 & 0.543 & 0.526 & -0.015 & 0.01 & 0.009 & 34 & 18.204 & 0.224 & 17 & 20\\
 P>5d [Fe/H]>$-$0.5   &  -5.95 & 0.0 & 0.0 & -3.25 & 0.0 & 0.0 & -0.811 & 0.315 & 0.313 & -0.005 & 0.005 & 0.005 & 108 & 18.217 & 0.131 & 70 & 76\\
 P>10d [Fe/H]>$-$0.5  &  -5.95 & 0.0 & 0.0 & -3.25 & 0.0 & 0.0 & -0.841 & 0.439 & 0.435 & -0.008 & 0.007 & 0.007 & 53 & 18.205 & 0.182 & 28 & 32\\
\hline\end{tabular}
\tablefoot{The units of $\alpha,\, \beta,\, \gamma,\,\epsilon$ and of their uncertainties are mag, mag/dex, mag/dex and mas, respectively.}
\end{table*}

\begin{table*}[h]
\footnotesize\setlength{\tabcolsep}{2pt}
 \caption[]{\label{tab:testParallax2} As for Table~\ref{tab:testParallax1} but for fixed $\beta$ and $\epsilon$, adopting for the latter two representative values: 10 $\mu$as and 14 $\mu$as.}
\begin{tabular}{lrrrrrrrrrrrrrrrrr}
\hline
  \multicolumn{1}{c}{Variation} &
  \multicolumn{1}{c}{$\alpha$} &
  \multicolumn{1}{c}{$\sigma_\alpha^{low}$} &
  \multicolumn{1}{c}{$\sigma_\alpha^{high}$} &
  \multicolumn{1}{c}{$\beta$} &
  \multicolumn{1}{c}{$\sigma_\beta^{low}$} &
  \multicolumn{1}{c}{$\sigma_\beta^{high}$} &
  \multicolumn{1}{c}{$\gamma$} &
  \multicolumn{1}{c}{$\sigma_\gamma^{low}$} &
  \multicolumn{1}{c}{$\sigma_\gamma^{high}$} &
  \multicolumn{1}{c}{$\epsilon$} &
  \multicolumn{1}{c}{$\sigma_\epsilon^{low}$} &
  \multicolumn{1}{c}{$\sigma_\epsilon^{high}$} & 
\multicolumn{1}{c}{N} &
  \multicolumn{1}{c}{$\mu_{\rm LMC}$} &
  \multicolumn{1}{c}{$\sigma\mu_{\rm LMC}$} &
   \multicolumn{1}{c}{AIC} &
  \multicolumn{1}{c}{BIC} \\
\hline
  \multicolumn{18}{c}{Fixed $\beta$ and $\epsilon$(=0.010 mas)  } \\
\hline
G<10.5 mag           &  -6.002 & 0.035 & 0.035 & -3.25 & 0.0 & 0.0 & -0.439 & 0.34 & 0.359 & -0.01 & 0.0 & 0.0 & 64 & 18.421 & 0.144 & 47 & 51\\
 G<11.5 mag           &  -6.006 & 0.029 & 0.028 & -3.25 & 0.0 & 0.0 & -0.362 & 0.225 & 0.232 & -0.01 & 0.0 & 0.0 & 100 & 18.457 & 0.097 & 78 & 83\\
 G<12.5 mag           &  -6.007 & 0.026 & 0.027 & -3.25 & 0.0 & 0.0 & -0.381 & 0.143 & 0.143 & -0.01 & 0.0 & 0.0 & 137 & 18.451 & 0.066 & 108 & 114\\
 G<13.5 mag           &  -6.01 & 0.027 & 0.028 & -3.25 & 0.0 & 0.0 & -0.538 & 0.103 & 0.098 & -0.01 & 0.0 & 0.0 & 224 & 18.39 & 0.053 & 176 & 182\\
 G<14.5 mag           &  -6.012 & 0.026 & 0.026 & -3.25 & 0.0 & 0.0 & -0.499 & 0.096 & 0.094 & -0.01 & 0.0 & 0.0 & 243 & 18.408 & 0.049 & 189 & 196\\
 G>10.5 mag           &  -6.058 & 0.055 & 0.057 & -3.25 & 0.0 & 0.0 & -0.595 & 0.129 & 0.127 & -0.01 & 0.0 & 0.0 & 203 & 18.414 & 0.079 & 150 & 156\\
 G>11.5 mag           &  -6.172 & 0.1 & 0.098 & -3.25 & 0.0 & 0.0 & -0.797 & 0.186 & 0.183 & -0.01 & 0.0 & 0.0 & 167 & 18.445 & 0.128 & 117 & 123\\
 G>12.5 mag           &  -6.219 & 0.137 & 0.137 & -3.25 & 0.0 & 0.0 & -0.896 & 0.244 & 0.239 & -0.01 & 0.0 & 0.0 & 130 & 18.452 & 0.172 & 88 & 94\\
 SNR<10               &  -6.263 & 0.113 & 0.121 & -3.25 & 0.0 & 0.0 & -0.831 & 0.215 & 0.229 & -0.01 & 0.0 & 0.0 & 133 & 18.522 & 0.146 & 77 & 83\\
 SNR>10               &  -5.992 & 0.027 & 0.028 & -3.25 & 0.0 & 0.0 & -0.563 & 0.137 & 0.132 & -0.01 & 0.0 & 0.0 & 133 & 18.361 & 0.065 & 105 & 111\\
 SNR<15               &  -6.132 & 0.07 & 0.072 & -3.25 & 0.0 & 0.0 & -0.736 & 0.145 & 0.146 & -0.01 & 0.0 & 0.0 & 192 & 18.431 & 0.095 & 133 & 139\\
 SNR>15               &  -5.998 & 0.03 & 0.03 & -3.25 & 0.0 & 0.0 & -0.31 & 0.216 & 0.228 & -0.01 & 0.0 & 0.0 & 74 & 18.47 & 0.094 & 56 & 61\\
 D<3 Kpc              &  -5.981 & 0.033 & 0.033 & -3.25 & 0.0 & 0.0 & -0.491 & 0.257 & 0.299 & -0.01 & 0.0 & 0.0 & 65 & 18.38 & 0.111 & 53 & 58\\
 D>3 Kpc              &  -6.118 & 0.057 & 0.058 & -3.25 & 0.0 & 0.0 & -0.703 & 0.139 & 0.138 & -0.01 & 0.0 & 0.0 & 202 & 18.429 & 0.083 & 137 & 144\\
 D<4 Kpc              &  -5.999 & 0.028 & 0.028 & -3.25 & 0.0 & 0.0 & -0.38 & 0.207 & 0.213 & -0.01 & 0.0 & 0.0 & 100 & 18.444 & 0.09 & 79 & 84\\
 D>4 Kpc              &  -6.189 & 0.092 & 0.096 & -3.25 & 0.0 & 0.0 & -0.813 & 0.175 & 0.177 & -0.01 & 0.0 & 0.0 & 167 & 18.456 & 0.119 & 109 & 116\\
 -0.7<[Fe/H]<+0.1     &  -5.988 & 0.034 & 0.035 & -3.25 & 0.0 & 0.0 & -0.427 & 0.149 & 0.148 & -0.01 & 0.0 & 0.0 & 199 & 18.413 & 0.071 & 143 & 149\\
 -0.7<[Fe/H]<+0.2     &  -6.008 & 0.027 & 0.028 & -3.25 & 0.0 & 0.0 & -0.503 & 0.127 & 0.124 & -0.01 & 0.0 & 0.0 & 223 & 18.402 & 0.061 & 154 & 161\\
 -0.7<[Fe/H]<+0.3     &  -6.011 & 0.027 & 0.028 & -3.25 & 0.0 & 0.0 & -0.513 & 0.125 & 0.122 & -0.01 & 0.0 & 0.0 & 230 & 18.4 & 0.06 & 166 & 173\\
 -0.5<[Fe/H]<+0.1     &  -5.982 & 0.033 & 0.035 & -3.25 & 0.0 & 0.0 & -0.294 & 0.205 & 0.203 & -0.01 & 0.0 & 0.0 & 156 & 18.462 & 0.091 & 110 & 116\\
 -0.5<[Fe/H]<+0.2     &  -6.009 & 0.027 & 0.028 & -3.25 & 0.0 & 0.0 & -0.438 & 0.165 & 0.161 & -0.01 & 0.0 & 0.0 & 180 & 18.429 & 0.074 & 127 & 134\\
 -0.5<[Fe/H]<+0.3     &  -6.012 & 0.027 & 0.028 & -3.25 & 0.0 & 0.0 & -0.459 & 0.162 & 0.16 & -0.01 & 0.0 & 0.0 & 187 & 18.423 & 0.073 & 136 & 143\\
 P>5d                 &   -5.994 & 0.033 & 0.033 & -3.25 & 0.0 & 0.0 & -0.661 & 0.222 & 0.208 & -0.01 & 0.0 & 0.0 & 125 & 18.323 & 0.099 & 92 & 98\\
 P>10d                &   -5.992 & 0.05 & 0.05 & -3.25 & 0.0 & 0.0 & -0.354 & 0.349 & 0.348 & -0.01 & 0.0 & 0.0 & 61 & 18.446 & 0.152 & 38 & 43\\
 P>5d  SNR>10         &   -5.969 & 0.036 & 0.035 & -3.25 & 0.0 & 0.0 & -0.609 & 0.284 & 0.294 & -0.01 & 0.0 & 0.0 & 75 & 18.319 & 0.123 & 51 & 56\\
 P>10d SNR>10         &   -5.939 & 0.061 & 0.065 & -3.25 & 0.0 & 0.0 & -0.854 & 0.589 & 0.555 & -0.01 & 0.0 & 0.0 & 34 & 18.189 & 0.25 & 17 & 20\\
 P>5d [Fe/H]>$-$0.5   &  -5.99 & 0.033 & 0.033 & -3.25 & 0.0 & 0.0 & -0.71 & 0.271 & 0.267 & -0.01 & 0.0 & 0.0 & 108 & 18.298 & 0.118 & 74 & 80\\
 P>10d [Fe/H]>$-$0.5  &  -5.972 & 0.058 & 0.06 & -3.25 & 0.0 & 0.0 & -0.793 & 0.495 & 0.458 & -0.01 & 0.0 & 0.0 & 53 & 18.247 & 0.212 & 25 & 29\\
 \hline
  \multicolumn{18}{c}{Fixed $\beta$ and $\epsilon$(=0.014 mas)  } \\
  \hline
G<10.5 mag           &  -6.019 & 0.035 & 0.036 & -3.25 & 0.0 & 0.0 & -0.416 & 0.338 & 0.358 & -0.014 & 0.0 & 0.0 & 64 & 18.448 & 0.143 & 48 & 52\\
 G<11.5 mag           &  -6.026 & 0.028 & 0.029 & -3.25 & 0.0 & 0.0 & -0.341 & 0.219 & 0.231 & -0.014 & 0.0 & 0.0 & 100 & 18.486 & 0.095 & 79 & 84\\
 G<12.5 mag           &  -6.029 & 0.027 & 0.028 & -3.25 & 0.0 & 0.0 & -0.338 & 0.148 & 0.148 & -0.014 & 0.0 & 0.0 & 137 & 18.49 & 0.067 & 105 & 110\\
 G<13.5 mag           &  -6.034 & 0.027 & 0.027 & -3.25 & 0.0 & 0.0 & -0.482 & 0.103 & 0.103 & -0.014 & 0.0 & 0.0 & 224 & 18.436 & 0.052 & 176 & 183\\
 G<14.5 mag           &  -6.037 & 0.027 & 0.028 & -3.25 & 0.0 & 0.0 & -0.44 & 0.102 & 0.099 & -0.014 & 0.0 & 0.0 & 243 & 18.456 & 0.051 & 185 & 192\\
 G>10.5 mag           &  -6.095 & 0.057 & 0.059 & -3.25 & 0.0 & 0.0 & -0.553 & 0.132 & 0.133 & -0.014 & 0.0 & 0.0 & 203 & 18.468 & 0.08 & 148 & 154\\
 G>11.5 mag           &  -6.23 & 0.103 & 0.099 & -3.25 & 0.0 & 0.0 & -0.794 & 0.194 & 0.189 & -0.014 & 0.0 & 0.0 & 167 & 18.504 & 0.132 & 116 & 122\\
 G>12.5 mag           &  -6.3 & 0.142 & 0.141 & -3.25 & 0.0 & 0.0 & -0.917 & 0.249 & 0.245 & -0.014 & 0.0 & 0.0 & 130 & 18.524 & 0.177 & 85 & 91\\
 SNR<10               &  -6.331 & 0.119 & 0.127 & -3.25 & 0.0 & 0.0 & -0.819 & 0.228 & 0.24 & -0.014 & 0.0 & 0.0 & 133 & 18.595 & 0.153 & 73 & 79\\
 SNR>10               &  -6.013 & 0.027 & 0.027 & -3.25 & 0.0 & 0.0 & -0.522 & 0.131 & 0.129 & -0.014 & 0.0 & 0.0 & 133 & 18.399 & 0.062 & 108 & 114\\
 SNR<15               &  -6.181 & 0.073 & 0.075 & -3.25 & 0.0 & 0.0 & -0.71 & 0.152 & 0.151 & -0.014 & 0.0 & 0.0 & 192 & 18.49 & 0.098 & 131 & 137\\
 SNR>15               &  -6.015 & 0.029 & 0.029 & -3.25 & 0.0 & 0.0 & -0.3 & 0.205 & 0.217 & -0.014 & 0.0 & 0.0 & 74 & 18.492 & 0.089 & 60 & 64\\
 D<3 Kpc              &  -5.997 & 0.034 & 0.033 & -3.25 & 0.0 & 0.0 & -0.496 & 0.262 & 0.304 & -0.014 & 0.0 & 0.0 & 65 & 18.394 & 0.114 & 53 & 58\\
 D>3 Kpc              &  -6.156 & 0.059 & 0.061 & -3.25 & 0.0 & 0.0 & -0.666 & 0.141 & 0.142 & -0.014 & 0.0 & 0.0 & 202 & 18.483 & 0.085 & 136 & 142\\
 D<4 Kpc              &  -6.019 & 0.028 & 0.029 & -3.25 & 0.0 & 0.0 & -0.369 & 0.204 & 0.206 & -0.014 & 0.0 & 0.0 & 100 & 18.467 & 0.089 & 81 & 86\\
 D>4 Kpc              &  -6.256 & 0.095 & 0.098 & -3.25 & 0.0 & 0.0 & -0.811 & 0.178 & 0.18 & -0.014 & 0.0 & 0.0 & 167 & 18.523 & 0.123 & 107 & 114\\
 -0.7<[Fe/H]<+0.1     &  -6.007 & 0.034 & 0.036 & -3.25 & 0.0 & 0.0 & -0.338 & 0.156 & 0.151 & -0.014 & 0.0 & 0.0 & 199 & 18.468 & 0.073 & 142 & 148\\
 -0.7<[Fe/H]<+0.2     &  -6.032 & 0.029 & 0.029 & -3.25 & 0.0 & 0.0 & -0.433 & 0.13 & 0.125 & -0.014 & 0.0 & 0.0 & 223 & 18.455 & 0.062 & 153 & 160\\
 -0.7<[Fe/H]<+0.3     &  -6.035 & 0.028 & 0.028 & -3.25 & 0.0 & 0.0 & -0.445 & 0.129 & 0.124 & -0.014 & 0.0 & 0.0 & 230 & 18.453 & 0.061 & 164 & 171\\
 -0.5<[Fe/H]<+0.1     &  -6.001 & 0.033 & 0.035 & -3.25 & 0.0 & 0.0 & -0.202 & 0.196 & 0.2 & -0.014 & 0.0 & 0.0 & 156 & 18.518 & 0.087 & 113 & 119\\
 -0.5<[Fe/H]<+0.2     &  -6.032 & 0.028 & 0.029 & -3.25 & 0.0 & 0.0 & -0.37 & 0.167 & 0.161 & -0.014 & 0.0 & 0.0 & 180 & 18.48 & 0.075 & 125 & 132\\
 -0.5<[Fe/H]<+0.3     &  -6.035 & 0.028 & 0.028 & -3.25 & 0.0 & 0.0 & -0.392 & 0.162 & 0.159 & -0.014 & 0.0 & 0.0 & 187 & 18.474 & 0.073 & 137 & 143\\
 P>5d                 &   -6.018 & 0.033 & 0.034 & -3.25 & 0.0 & 0.0 & -0.587 & 0.212 & 0.207 & -0.014 & 0.0 & 0.0 & 125 & 18.378 & 0.095 & 94 & 100\\
 P>10d                &   -6.021 & 0.047 & 0.049 & -3.25 & 0.0 & 0.0 & -0.276 & 0.336 & 0.349 & -0.014 & 0.0 & 0.0 & 61 & 18.507 & 0.146 & 40 & 44\\
 P>5d  SNR>10         &   -5.989 & 0.034 & 0.035 & -3.25 & 0.0 & 0.0 & -0.557 & 0.266 & 0.281 & -0.014 & 0.0 & 0.0 & 75 & 18.361 & 0.116 & 52 & 56\\
 P>10d SNR>10         &   -5.964 & 0.063 & 0.068 & -3.25 & 0.0 & 0.0 & -0.805 & 0.589 & 0.575 & -0.014 & 0.0 & 0.0 & 34 & 18.234 & 0.251 & 16 & 19\\
 P>5d [Fe/H]>$-$0.5   &  -6.013 & 0.035 & 0.036 & -3.25 & 0.0 & 0.0 & -0.635 & 0.273 & 0.281 & -0.014 & 0.0 & 0.0 & 108 & 18.352 & 0.119 & 71 & 77\\
 P>10d [Fe/H]>$-$0.5  &  -6.0 & 0.059 & 0.061 & -3.25 & 0.0 & 0.0 & -0.717 & 0.508 & 0.484 & -0.014 & 0.0 & 0.0 & 53 & 18.306 & 0.218 & 25 & 28\\
\hline\end{tabular}
\tablefoot{The units of $\alpha,\, \beta,\, \gamma,\,\epsilon$ and of their uncertainties are mag, mag/dex, mag/dex and mas, respectively.}
\end{table*}

\end{appendix}

%
%

\end{document}